Puskar Mondal ✉ ; Shing-Tung Yau

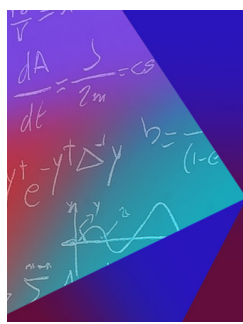

# Global exterior stability of the Minkowski space: Coupled Einstein–Yang–Mills perturbations

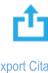

View Online   Export Citation   CrossMark

Puskar Mondal[a])  and Shing-Tung Yau[b])

AFFILIATIONS
Department of Mathematics, Harvard University, Cambridge, Massachusetts 02138, USA and Center of Mathematical Sciences and Applications, Harvard University, Cambridge, Massachusetts 02138, USA

[a])Author to whom correspondence should be addressed: puskar_mondal@fas.harvard.edu
[b])yau@math.harvard.edu

ABSTRACT

Here we prove a global gauge-invariant radiation estimates for the perturbations of the 3 + 1 dimensional Minkowski spacetime in the presence of Yang–Mills sources. In particular, we obtain a novel gauge invariant estimate for the Yang–Mills fields coupled to gravity in a double null framework in the Causal complement of a compact set of a Cauchy slice. A consequence of our result is the global exterior stability of the Minkowski space under coupled Yang–Mills perturbations. A special structure present both in the null Bianchi equations and the null Yang–Mills equations is utilized crucially to obtain the dispersive estimates necessary to conclude the global existence property. Direct use of Bel-Robinson and Yang–Mills stress-energy tensor to obtain the energy estimates is avoided in favor of weighted integration by parts taking advantage of the manifestly symmetric hyperbolic characteristics of null Bianchi and null Yang–Mills equations. Our result holds for any compact semi-simple gauge group. This is the first stability result of Minkowski space including a non-linear source.

Published under an exclusive license by AIP Publishing. https://doi.org/10.1063/5.0185598

## I. INTRODUCTION

A fundamental problem in the context of metric gravity is the existence of a stable ground state. For pure vacuum, the ground state with trivial topology is expected to be the Minkowski space. This is essentially motivated by the fact that the Minkowski space has zero ADM mass as a consequence of the positive mass Theorem.[1,2] Even similar results hold in a quasi-local sense due to Refs. [3–5]. More precisely, any Minkowski ball bounded by a space-like topological two–sphere contains zero gravitational energy. Naively, one would think if one were to perturb the Minkowski space by adding sufficiently small *energy* (defined to control appropriate norms), then such perturbations should disperse by losing energy to infinity through non-compact directions. Such a stability result for pure gravitational perturbations was first obtained by Ref. [6]. In particular, they proved that the Cauchy development of initial data sets that are close in an appropriate sense to the trivial Minkowski data lead to global, smooth, and geodesically complete solutions of the Einstein-vacuum equations that remain close in an appropriate, global sense to the Minkowski spacetime. In this study, the hyperbolic nature of the Bianchi equations was crucially utilized through the use of the Bel-Robinson stress-energy tensor. The special null structure present in these equations allowed the closure of the non-linear estimates. Later on, Ref. [7] provided simpler proof of the stability of Minkowski space in spacetime harmonic gauge. As such, it was believed earlier that the space-time harmonic gauge forms finite time coordinate singularity and therefore not suitable for long-time existence problems (notice that the spacetime harmonic gauge was first used by Ref. [8] to establish a local "in time" wellposedness result for the vacuum Einstein's equations). However, the result of Ref. [9] suggests otherwise in the case of data close to the Minkowski trivial data in an appropriate sense. In particular, they cast Einstein's equations into a system of quasi-linear wave equations and utilized a weak null structure present in the equations in this particular gauge to control the non-linear terms. Suitable energies were constructed and controlled employing the conformal Killing fields of the background.

Reference [10] provided an alternative proof by utilizing the energy estimates obtained from a smaller number of approximate conformal Killing fields (approximate inversion generator, approximate generator of time translation, and approximate scaling vector fields). In particular, she has avoided the use of the approximate rotation generators as commuting vector fields through the use of elliptic estimates. She also utilized a weaker fall-off condition on the data with well defined ADM mass. In addition, this proof also required less regularity of the Weyl curvature (one derivative of the Weyl curvature). In addition to these studies of the classical initial value problem associated with the vacuum Einstein's equations where a combination of maximal slice and a null foliation (constructed utilizing an optical function) is utilized,[11] used a double null foliation to study the radiation problem associated with the vacuum Einstein's equations. In particular, their result implies the global stability of a suitably defined exterior domain of the Minkowski space under pure gravitational perturbations. In the stability problem of an exterior domain, the choice of double null foliation is ideal since one ultimately wants to investigate whether the gravitational radiation can escape to null infinity through the outgoing null direction or whether the associated non-linearities present in the equation can focus the radiation through the incoming null hypersurfaces to form *singularity* in a finite "time." Therefore the question of stability essentially turns into a competition between the geometric dispersion of energy and concentration of energy by the non-linearities present. In order for the dispersion to win at the level of small data, one, unfortunately, can not have an arbitrary quadratic non-linearity present in the relevant equations. In other words, the quadratic non-linearities must verify a particular structure known as the "null" structure. Fortunately, the Bianchi equations while expressed in a double null coordinate precisely verify such a structure thereby affirming the dominance of dispersion phenomena. This in turn yields the stability result.

Moving one step further, one would like to understand the stability issues in the presence of suitable source terms. This is important since at the level of relativistic field theory, the Minkowski space never appears without additional fields (scalar fields, Maxwell fields, or Yang–Mills fields) propagating on it. Therefore it is natural to ask the following question "*is Minkowski space stable under coupled gravitational and matter/radiation perturbations?*" Reference [9] considered the small data perturbations of coupled gravity and mass-less scalar field and proved the global existence of the associated initial value problem in space-time harmonic gauge. The most interesting case of a coupled system is the study of the Einstein–Maxwell system by Ref. [12]. In particular, she extended the vacuum stability method of Ref. [6] including a Maxwell source term. The Maxwell equations share a remarkable similarity with that of the vacuum Bianchi equations and as such a null structure is present in these equations as well. Reference [12] obtained gauge invariant estimate for the $U(1)$ curvature $F$ together with the estimates for the Weyl curvature in a null-maximal framework which then yielded the global existence result. In addition, the initial data she considered included well defined non-zero ADM mass. There have been several other studies as well in the direction of stability of the Minkowski space under coupled gravity-matter perturbations.[13–17]

In this article, we want to include the Yang–Mills source terms in Einstein's equations and study the associated radiation problem. In other words, we study the global evolution of initial data that is specified on the complement of a large ball of an initial Cauchy hypersurface. To this end, we work in a double null framework to investigate the competition between dispersion and non-linearities. A fundamental difference between the $U(1)$ Maxwell theory and a non-abelian Yang–Mills theory is that the latter is no-linear. Therefore, in addition to investigating the non-linearities of the Bianchi equations, one ought to carefully treat the non-linearities of the Yang–Mills theory as well. However, since Yang–Mills theory is a gauge theory, one ought to work in a particular choice of gauge (or equivalently descend to the "orbit space" of the theory). Unfortunately, there does not exist global gauge choice in Yang–Mills theory. The traditional choice of Lorentz gauge is known to develop finite time coordinate singularities in non-abelian theory contrary to the linear Maxwell theory where such a breakdown is absent. One may therefore choose a different gauge such as a "generalized Coulomb gauge" to work with. However, the geometry of the orbit space of the theory (i.e., the space of connections modulo the bundle automorphisms) has a non-trivial consequence in the matter of gauge choice. The positivity of sectional curvature of the orbit space leads to the development of the so-called "Gribov horizon" which essentially indicates the breakdown of the gauge choice. Therefore, we develop a novel gauge invariant framework for the Yang–Mills fields where we write down the Yang–Mills equations in the double null gauge in terms of the fully gauge covariant derivative. By virtue of the compatibility of the fiber metrics with the fully gauge covariant derivative, we are able to obtain the necessary gauge-invariant energy estimates. The true non-linear characteristics of the Yang–Mills theory manifest themselves in the higher-order energy estimates when we commute the full gauge covariant derivatives. The task, therefore, is reduced to establishing that this non-linearity is dominated by linear decay. As we describe in the appropriate section, the linear decay is borderline sufficient to overcome the non-linearities thereby closing the argument. This once again is a natural manifestation of the null structure that is present in the wave equation for the Yang–Mills curvature. Once we obtain the necessary gauge invariant estimates, we can work in a particular collection of gauge choices to yield a global existence result for the coupled theory. The latter part is standard and therefore we omit the discussion of such. Surprisingly there have been very few studies regarding the stability issues associated with the coupled Einstein–Yang–Mills system in the absence of any symmetry. The notable studies include the stability of 3 + 1 dimensional de Sitter spacetime under coupled Einstein–Yang–Mills perturbations by Friedrich,[18] recent global existence and non-linear stability results of de Sitter-like solutions to the Einstein–Yang–Mills system in $n \geq 4$ spacetime dimension by Liu *et al.*,[19] and stability of $n + 1, n \geq 4$ dimensional Milne spacetime by Ref. [20].

Before stating the main theorem that we intend to prove in this article, let us introduce a few notations. The Einstein–Yang–Mills equations on a 3 + 1 dimensional Lorentzian manifold $(M, g)$ reads

$$\text{Ric}[g] - \frac{1}{2} R[g] g = \mathfrak{T}, \quad DF = 0, \quad D\, {}^*F = 0 \qquad (1.1)$$

where Ric and $R[g]$ are the Ricci curvature and the scalar curvature of the manifold $(M, g)$, respectively. $F$ is the curvature of the principle $G$ bundle on $(M, g)$ with $G$ being a compact semisimple Lie group. $D$ is the gauge covariant spacetime covariant derivative compatible with the metric $g$ as well as the metric on the fiber of the bundle $G$. $*$ is the Hodge dual operation on the two–form $F$ and $\mathfrak{T}$ is the stress-energy tensor associated with the Yang–Mill fields that read in a local chart $\mathfrak{T}_{\mu\nu} := \frac{1}{2} \left( F^P{}_{Q\mu\alpha} F^Q{}_{P\nu}{}^\alpha + {}^*F^P{}_{Q\mu\alpha} \, {}^*F^Q{}_{P\nu}{}^\alpha \right)$, $\mu, \nu$ are the spacetime indices where $P, Q$ are the Lie-algebra indices (details are to be provided in Sec. II). Notice that the vacuum Minkowski space ($\mathbb{R}^{1,3}, \eta, F = 0$) with $\eta = \text{diag}(-1, 1, 1, 1)$ in the usual rectangular chart is a trivial solution to the coupled Einstein–Yang–Mills equations (1.1). We want to perturb this trivial solution in an appropriate sense. Firstly, through the decomposition of the Riemann tensor as a sum of the pure gravity (the Weyl tensor) and Ricci curvature [source term through the Einstein's equation (1.1)], Minkowski space can be perturbed by introducing a non-trivial Weyl curvature and a non-trivial stress-energy tensor such the Yang–Mills stress-energy tensor. Small perturbations can be defined as the smallness of an appropriately defined norm of the Weyl curvature and that of the Yang-Mills curvature (gauge-invariant). We denote this norms by $\mathbf{W}_0$ and $\mathbf{Y}_0$, respectively (see the exact definition in Sec. II). To establish a stability theorem, one ought to prove that the appropriately defined norm of the Weyl and Yang–Mills curvature ($\mathbf{W}$ and $\mathbf{Y}$ are defined in Sec. II) remains uniformly bounded in the future "in" direction for any finite time and moreover exhibits decay. In the current context, we consider the initial data on a complement of a ball of a Cauchy hypersurface. We work in a double null gauge and obtain the gauge invariant estimates for the Weyl curvature and the Yang–Mills curvature. The double null gauge is defined as the level sets of the optical functions $u$ and $\underline{u}$ obtained by solving Eikonal equations (see Sec. II for the technical detail). Level sets of $u$ are the outgoing null cones denoted by $H_u$ while that of $\underline{u}$ are the incoming null cones denoted by $\underline{H}_{\underline{u}}$. We state the main theorem as follows.

**Theorem 1.1.** *Let $(M, g)$ be a $3 + 1$ dimensional Lorentzian globally hyperbolic manifold diffeomorphic to $\mathbb{R}^{1+3}$. Let $S_{u_0, \underline{u}_0}$ be a space-like topological two–sphere obtained by the intersection of an incoming and outgoing null cone $\underline{H}_{\underline{u}_0}$ and $H_{u_0}$ such that $u_0 + \underline{u}_0 = 0$. Let $\Sigma_0$ be a space-like Cauchy hypersurface diffeomorphic to $\mathbb{R}^3$ and foliated by the topological two–spheres $S_{u, \underline{u}}$ such that $u + \underline{u} = 0$ and such spheres constitute the initial spheres for the canonical double null foliation of the causal future of $\Sigma_0$. Let us also consider the space-like topological ball $B_{u_0, \underline{u}_0} \subset \Sigma_0$ bounded by the sphere $S_{u_0, \underline{u}_0}$ and denote the exterior initial Cauchy hypersurface $\Sigma_0 - B_{u_0, \underline{u}_0}$ by $\widehat{\Sigma}_0$ ($\widehat{\Sigma}_0$ is diffeomorphic to the complement of a ball in $\mathbb{R}^3$). Under the assumption that $\mathbf{W}_0 + \mathbf{Y}_0 \leq \epsilon_0$ for a sufficiently small $\epsilon_0 > 0$, the following estimate for the Weyl and Yang–Mills curvature holds in the entire causal complement of the ball $B_{u_0, \underline{u}_0}$*

$$\mathbf{W} + \mathbf{Y} \leq \delta(\epsilon_0) \tag{1.2}$$

*uniformly in $u$ and $\underline{u}$ for a $\delta = \delta(\epsilon_0) > 0$ and such causal complement is foliated by two families of incoming and outgoing null hypersurfaces $H$ and $\underline{H}$, respectively. Here $H \equiv H_u$ and $\underline{H} \equiv \underline{H}_{\underline{u}}$.*

*Remark 1.* Notice that we can not choose an initial CMC or maximal slice, construct a double null foliation for the bootstrap region, and match both the foliations since the space-like foliation of the boot-strap domain obtained from the associated canonical double-null foliation may not be CMC or maximal. In such a case, one would need an additional oscillation-type lemma proved in Ref. 11.

This theorem can be treated as the main *a priori* estimate toward proving global exterior stability of the Minkowski space under coupled gravity-Yang–Mills perturbations. We provide a physical intuition behind why controlling the appropriate norm of the Weyl and Yang–Mills curvature is sufficient for the stability statement. First, recall from the geodesic deviation equation that the spacetime Riemann curvature is the mathematical manifestation of gravity, and as such it exhausts all gravitational degrees of freedom. In other words, appropriate norms of the metric perturbations can be obtained from appropriate norms of the Riemann curvature purely employing the elliptic estimates. Due to its irreducible decomposition in terms of Weyl curvature and the Ricci curvature, the full gravitational degrees of freedom split into pure gravity (Weyl curvature) and sourced gravity (Ricci curvature that is fixed by the Yang–Mills source term in the current context). The global decay of the Weyl and Yang–Mills curvature then implies that the gravitational energy contained within any spacelike topological ball bounded by a membrane decays in the direction of future null infinity (the notion of gravitational energy bounded by a spacelike membrane can be defined by the Wang–Yau quasi-local energy[3,4]). If the gravitational energy contained within every spacelike membrane decays in the exterior domain then from from the rigidity of the gravitational energy (i.e., if the Wang–Yau definition of gravitational energy contained within the spacelike ball bounded by a membrane vanishes, then it must be isometric to Minkowski) the exterior approaches Minkowski space. Therefore, there are potential two ways to prove the stability argument: one by standard Cauchy problem argument and another by proving the decay of the Wang–Yau quasi-local energy. Both of these approaches involve standard computations and therefore we do not pursue it here.

The main novelty in the proof of the Theorem 1.1 is the complete gauge invariance. We sketch the main argument behind the Proof of 1.1. As discussed in Ref. 21, the Yang–Mills sourced null Bianchi equations and null Yang-Mills equations are manifestly hyperbolic contrary to the equations $d\Gamma + \Gamma\Gamma = R$ and $dA + [A, A] = F$ which are manifestly non-hyperbolic, where $\Gamma$, $A$, $R$, and $F$ denote connection coefficients on the frame tangent bundle, connection on the principle $G$–bundle (or gauge bundle), the curvature of frame tangent bundle, and curvature of the principle $G$–bundle, respectively. In the double null framework, $d\Gamma + \Gamma\Gamma = R$ reduces to transport and elliptic equations which may be utilized to control the connections in terms of the curvature. However, in order to utilize the equations $dA + [A, A] = F$, we need to make a gauge choice for the Yang–Mills theory. This makes the global analysis tremendously complicated and non-trivial. The gauge choice for Yang–Mills equations is an extremely non-trivial one due to the presence of Gribov phenomena.[22,23] The orbit space of the theory (space of connections modulo the bundle automorphisms or the gauge transformations) has a positive definite sectional curvature and therefore the choice of a well-known Coulomb gauge remains only valid in a small enough neighborhood of a background connection. This geometric property of the orbit space does not allow for any global gauge choice for the Yang–Mills theory.[24] Lorentz gauge for the Yang–Mills theory is also expected to be pathological in the sense that it can develop gauge singularity in finite time (although in the regime of small data, such a problem may be avoided through a careful analysis). Note that none of these problems are present in the Maxwell theory. To avoid these issues, we work in a full gauge covariant framework where we *do not* split the gauge covariant derivative into its spacetime (or spatial) part and a gauge part given by the Yang–Mills connection. The *true* non-linearity of the Yang–Mills theory shows up through the commutation of gauge covariant derivative. To our knowledge, this approach was not utilized previously. Once we have obtained the necessary gauge invariant estimates, the remaining task to obtain a global solution is straightforward and therefore we do not discuss the details here.

First, we assume that the connection and the curvature norms (both Weyl and Yang–Mills) are bounded from above by a small number $\epsilon$. Under this bootstrap assumption, we can integrate the transport equations for the connections as well as utilize the estimates from the elliptic equations to show that the suitable connection norms can be controlled by the Weyl and Yang–Mills curvature norms. Now we utilize the symmetric hyperbolic characteristic of the null Bianchi and null Yang–Mills equations to control the Weyl and Yang–Mills curvature energies. The null structure present in the Bianchi and Yang–Mills equations plays a vital role in controlling the curvature energies along the null cones solely in terms of the initial data uniformly in $u$ and $\underline{u}$ in the exterior region. Roughly speaking the non-linear terms in the null Bianchi and null Yang–Mills equations always appear at least as the product of a term that decays sufficiently fast (i.e., the decay is integrable) and a slowly decaying term. For Yang–Mills equations, this structure persists at the level of higher order energies when the commutation of gauge covariant derivative yields further Yang–Mills curvature terms. Finally, choosing the initial data norm sufficiently small we can make curvature and connection norms (since connections are estimated by curvature) less than $\frac{\epsilon}{2}$ thereby closing the boot-strap. As it turns out, the expected linear decay of the Yang–Mills fields is sufficient to tame the associated non-linearities.

Once these estimates are proven, several issues need to be addressed which we do not pursue here since they have become standard in the language of modern mathematical general relativity. Firstly, a non-trivial task is to solve Einstein's constraint equations for the initial data (Note that the estimates on the Weyl and Yang–Mills curvature can be used to prove the estimates for the data $(\widehat{g}, k, F)$, where $\widehat{g}$ is and $k$ are the Riemannian metric induced on the Cauchy slice $\Sigma$ by the Lorentzian metric $g$ and $k$ is the second fundamental form of the slice. Solving the constraint equation is standard by now. In spatial dimensions $n \geq 2$ there is a well-known technique pioneered by Lichnerowicz, for solving the constraint equations on a constant-mean curvature hypersurface (see Choquet Bruhat,[25] Bartnik and Isenberg[26] for detailed expositions of this "conformal" method). Now since the estimates for spacetime Riemann curvature are obtained by that of Weyl and Yang-Mills curvature, the pair $(\widehat{g}, k)$ can be estimated employing elliptic estimates (see Ref. [27] for estimates of $(\widehat{g}, k)$ in terms of the estimates of the Weyl curvature in vacuum. This can be generalized to the case with source terms in a straightforward way). One can work in a particular choice of gauge (e.g., spacetime harmonic gauge is an appropriate choice in the current context) to obtain the global existence of a classical solution. We do not discuss this here rather only concern ourselves with obtaining the gauge-invariant estimates for the Weyl and Yang–Mills curvature components.

The structure of the article is as follows. We start with writing down the Yang–Mills sourced Bianchi and Yang–Mills equations in the double null framework. We discuss the main theorem and its consequences along with a brief idea of the proof. Then we start estimating the relevant connection norms that are required for the energy estimates. This part is similar to the ideas of Ref. [11] and therefore we only estimate the non-trivial Yang–Mills coupling terms. Equipped with the connection estimates, we show that the energy (energy flux to be precise) for the curvature along the two families of the null hypersurfaces remain uniformly bounded in the exterior domain (causal complement of a large ball on the initial Cauchy slice) in terms of the data on the initial Cauchy slice. Through an application of Sobolev trace inequality, we immediately obtain the point-wise behavior of the Weyl and Yang–Mills curvatures. We conclude by discussing a few implications as well as the potential use of quasi-local energies (the one introduced by Refs. [3] and [4]) to study the stability problems.

## II. EINSTEIN–YANG–MILLS EQUATIONS IN DOUBLE NULL FOLIATION

Let the two null hupersurfaces $H_{u_0}$ and $\underline{H}_{\underline{u}_0}$ (to be defined later) intersect at a topological two−sphere $S_{u_0,\underline{u}_0}$ in a globally hyperbolic spacetime $M$ equipped with a Lorentzian metric $g$. The null hypersurfaces $H$ and $\underline{H}$ are described by the level sets of the optical functions $u$ and $\underline{u}$, respectively (Fig. 1). In other words $u$ and $\underline{u}$ satisfy the Eikonal equations

$$g^{\mu\nu}\partial_\mu u\partial_\nu u = 0, \quad g^{\mu\nu}\partial_\mu \underline{u}\partial_\nu \underline{u} = 0. \tag{2.1}$$

Through the variation of $u$ and $\underline{u}$, we can foliate a spacetime slab $D_{u,\underline{u}}$ by these two families of null hypersurfaces. The geodesic generators of the double null foliation are the vector fields $L$ and $\bar{L}$ given by

$$L := -g^{\mu\rho}\partial_\rho u\partial_\mu, \quad \bar{L} := -g^{\mu\rho}\partial_\rho \underline{u}\partial_\mu \tag{2.2}$$

and manifestly they satisfy

$$\nabla_L L = 0 = \nabla_{\bar{L}} \bar{L}. \tag{2.3}$$

Whenever, we say $H$ (resp. $\underline{H}$) we will always mean $H_u$ (resp. $\underline{H}_{\underline{u}}$). In this notation, $H_{u_0}$ and $\underline{H}_{\underline{u}_0}$ are the two initial null hypersurfaces corresponding to $u = u_0$ and $\underline{u} = \underline{u}_0$, respectively. Intersection of $H$ and $\underline{H}$ is a topological two−sphere $S_{u,\underline{u}}$. The spacetime metric in the double null gauge may be written as (in a local chart $(u, \underline{u}, \theta^1, \theta^2)$)

$$g := -2\Omega^2(du \otimes d\underline{u} + d\underline{u} \otimes du) + \gamma_{AB}(d\theta^A - b^A d\underline{u}) \otimes (d\theta^B - b^B d\underline{u}), \tag{2.4}$$

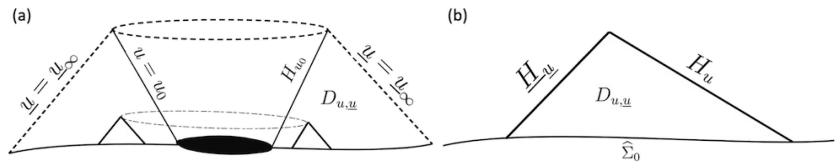

**FIG. 1.** (a) The Cauchy data is provided on the compliment of the topological ball $B$ (shaded) in $\Sigma_0$ i.e., $\widehat{\Sigma}_0 := \Sigma_0 - B$. Therefore the globally hyperbolic portion $D_{u,\underline{u}}$ is the compliment of the causal future of the ball $B$. Therefore, one would like to understand the future completeness of this globally hyperbolic portion of the spacetime. This portion is entitled as the exterior domain. (b) A portion of the exterior domain after projecting out the topological two-spheres i.e., each point on the diagram is a sphere. This exterior domain is foliated by two families of null hypersurfaces $H_u$ and $\underline{H}_{\underline{u}}$, respectively.

where $\Omega$ is the null lapse function and $b := b^A \frac{\partial}{\partial \theta^A}$ is the null shift vector field. $\{\theta^A\}_{A=1}^2$ are the coordinates on the topological sphere $S_{u,\underline{u}}$. The induced metric on $S_{u,\underline{u}}$ is $\gamma_{AB}$. We can identify a normalized null frame $(e_4, e_3, e_1, e_2)$ such that $g(e_4, e_4) = g(e_3, e_3) = 0 = g(e_A, e_4) = g(e_A, e_3)$ = 0 and $g(e_4, e_3) = -2$, where $(e_1, e_2)$ is an arbitrary frame on $S_{u,\underline{u}}$. We may identify $e_4$ and $e_3$ as follows

$$e_4 = \Omega^{-1} \frac{\partial}{\partial \underline{u}}, \quad e_3 = \Omega^{-1} \left( \frac{\partial}{\partial u} + b^A \frac{\partial}{\partial \theta^A} \right). \tag{2.5}$$

For the coordinate system, we may first define a coordinate chart $\mathcal{A}$ on $S_{u_0,\underline{u}_0}$ then drag it by the flow of the vector field $e_4$ along $H_{u_0}$ and then drag it by the flow of $e_3$ to fill out the entire slab $D_{u,\underline{u}}$. For more detailed information about the double null foliation of a spacetime, see Refs. 11 and 28. We define the areal radius of the topological two-sphere $S_{u,\underline{u}}$ as follows

$$r(u, \underline{u}) := \sqrt{\frac{\text{Area}(S_{u,\underline{u}})}{4\pi}}. \tag{2.6}$$

In addition we also define $\tau_+ := \sqrt{1 + \underline{u}^2}$ and $\tau_- := \sqrt{1 + u^2}$. In the exterior region, $|u| \approx \tau_-$ and $r \approx \tau_+$. The exterior domain is defined by the condition that $\underline{u}$ is sufficiently large.

Now we define a Yang–Mills theory on $(M, g)$. We denote by $\mathfrak{P}$ a $C^\infty$ principal bundle with base a $3 + 1$ dimensional Lorentzian manifold $M$ and a Lie group $G$. We assume that $G$ is compact semi-simple (for physical purposes) and therefore admits a positive definite non-degenerate bi-invariant metric. Its Lie algebra $\mathfrak{g}$ by construction admits an adjoint invariant, positive definite scalar product denoted by $\langle , \rangle$ which enjoys the property: for $A, B, C \in \mathfrak{g}$,

$$\langle [A, B], C \rangle = \langle A, [B, C] \rangle. \tag{2.7}$$

as a consequence of adjoint invariance. A Yang-Mills connection is defined as a one-form $w$ on $\mathfrak{P}$ with values in $\mathfrak{g}$ endowed with compatible properties. It's representative in a local trivialization of $\mathfrak{P}$ over $U \subset M$

$$\varphi : p \mapsto (x, a), \quad p \in \mathfrak{P}, \; x \in U, \; a \in G \tag{2.8}$$

is the one-form $s^*w$ on $U$, where $s$ is the local section of $\mathfrak{P}$ corresponding canonically to the local trivialization $s(x) = \varphi^{-1}(x, e)$, called a *gauge*. Let $A_1$ and $A_2$ be representatives of $w$ in gauges $s_1$ and $s_2$ over $U_1 \subset M$ and $U_2 \in M$. In $U_1 \cap U_2$, one has

$$A_1 = Ad(u_{12}^{-1})A_2 + u_{12}\Theta_{MC}, \tag{2.9}$$

where $\Theta_{MC}$ is the Maurer-Cartan form on $G$, and $u_{12} : U_1 \cap U_2 \to G$ generates the transformation between the two local trivializations:

$$s_1 = R_{u_{12}} s_2, \tag{2.10}$$

$R_{u_{12}}$ is the right translation on $\mathfrak{P}$ by $u_{12}$. Given the principal bundle $\mathfrak{P} \to M$, a Yang–Mills potential $A$ on $M$ is a section of the fibered tensor product $T^*M \otimes_M \mathfrak{P}_{Affine,\mathfrak{g}}$ where $\mathfrak{P}_{Affine,\mathfrak{g}}$ is the affine bundle with base $M$ and typical fiber $\mathfrak{g}$ associated to $\mathfrak{P}$ via relation (2.9). If $\widehat{A}$ is another Yang–Mills potential on $M$, then $A - \widehat{A}$ is a section of the tensor product of vector bundles $T^*M \otimes_M P_{Ad,\mathfrak{g}}$, where $\mathfrak{P}_{Ad,\mathfrak{g}} := \mathfrak{P} \times_{Ad} \mathfrak{g}$ is the vector bundle associated to $\mathfrak{P}$ by the adjoint representation of $G$ on $\mathfrak{g}$. There is an inner product in the fibers of $\mathfrak{P}_{Ad,\mathfrak{g}}$, deduced from that on $\mathfrak{g}$. The curvature $\mathbf{C}$ of the connection $w$ considered as a one-form on $\mathfrak{P}$ is a $\mathfrak{g}$-valued two-form on $\mathfrak{P}$. Its representative in a gauge where $w$ is represented by $A$ is given by

$$F := dA + [A, A], \tag{2.11}$$

and relation between two representatives $F_1$ and $F_2$ on $U_1 \cap U_2$ is $F_1 = Ad(u_{12}^{-1})F_2$ and therefore $F$ is a section of the vector bundle $\Lambda^2 T^*M \otimes_M \mathfrak{P}_{Ad,\mathfrak{g}}$. For a section $\mathfrak{D}$ of the vector bundle $\otimes^k T^*M \otimes_M \mathfrak{P}_{Ad,\mathfrak{g}}$, a natural covariant derivative is defined as follows

$$\widehat{D}\mathfrak{D} := D\mathfrak{D} + [A, \mathfrak{D}], \tag{2.12}$$

where $D$ is the usual covariant derivative induced by the Lorentzian structure of $M$ and by construction $\widehat{D}\mathfrak{O}$ is a section of the vector bundle $\otimes^{k+1} T^*M \otimes_M \mathfrak{P}_{Ad,\mathfrak{g}}$. The Yang–Mills coupling constant $g_{YM}$ is set to 1.

The classical Yang–Mills equations (in the absence of sources) correspond to setting the natural [spacetime and gauge as defined in (2.12)] covariant divergence of this curvature two-form $F$ to zero. By virtue of its definition in terms of the connection, this curvature also satisfies the Bianchi identity that asserts the vanishing of its gauge covariant exterior derivative. Taken together these equations provide a geometrically natural nonlinear generalization of Maxwell's equations (when the latter are written in terms of a "vector potential") and of course play a fundamental role in modern elementary particle physics. If nontrivial bundles are considered or nontrivial spacetime topologies are involved, then the foregoing so-called "local trivializations" of the bundles in question must be patched together to give global descriptions but, by virtue of the covariance of the formalism, there is a natural way of carrying out this patching procedure at least over those regions of spacetime where the connections are well-defined.

Now that we have defined a Yang–Mills theory over the Lorentzian manifold $M$ (which is to be sufficiently close to Minkowski spacetime in an appropriate sense), we may proceed to couple it with Einstein's equations and write down the full set of equations in the double null coordinates. Before doing so we need to project the covariant derivatives (spacetime and gauge) onto the topological two–spheres that are the leaves of the double null foliation. In this section, we explicitly define all the entities associated with the double null foliation and write down structure equations. The spacetime covariant derivative $D$ admits the following usual decomposition in terms of its components parallel and orthogonal to the topological two–sphere $S_{u,\underline{u}}$

$$D_{e_a} e_b = \nabla_{e_a} e_b - \frac{1}{2} \langle D_{e_a} e_b, e_3 \rangle e_4 - \frac{1}{2} \langle D_{e_a} e_b, e_4 \rangle e_3, \qquad (2.13)$$

where $\nabla$ is the $S_{u,\underline{u}}$-parallel covariant derivative. Similar to the spacetime covariant derivative, the spacetime gauge covariant derivative $\widehat{D}$ admits the following decomposition

$$\widehat{D}_{e_a} e_b = \widehat{\nabla}_{e_a} e_b - \frac{1}{2} \langle \widehat{D}_{e_a} e_b, e_3 \rangle e_4 - \frac{1}{2} \langle \widehat{D}_{e_a} e_b, e_4 \rangle e_3$$
$$= \nabla_{e_a} e_b - \frac{1}{2} \langle D_{e_a} e_b, e_3 \rangle e_4 - \frac{1}{2} \langle D_{e_a} e_b, e_4 \rangle e_3 \qquad (2.14)$$

simply because the basis $(e_4, e_3, e_a)_{a=1}^2$ are not sections of the gauge bundle and therefore gauge covariant derivative acts as ordinary covariant derivative. Now recall the following definitions of the outgoing and incoming null second fundamental form of $S_{u,v}$

$$\chi_{ab} := \langle D_{e_a} e_4, e_b \rangle, \qquad \underline{\chi}_{ab} := \langle D_{e_a} e_3, e_b \rangle, \qquad (2.15)$$

the spacetime covariant (also gauge covariant) derivative satisfies

$$D_{e_a} e_b = \nabla_{e_a} e_b + \frac{1}{2} \underline{\chi}_{ab} e_4 + \frac{1}{2} \chi_{ab} e_3. \qquad (2.16)$$

Now let $K \in sections\{TS_{u,\underline{u}}\}$ and $Z$ be a $\mathfrak{g}$– valued frame vector field on $S_{u,\underline{u}}$ that transforms as a tensor under gauge transformation, then using the definition of the gauge covariant derivative (2.9), the following holds for the gauge covariant derivative

$$\widehat{D}_K Z = \widehat{\nabla}_K Z + \frac{1}{2} \underline{\chi}(K, Z) e_4 + \frac{1}{2} \chi(K, Z) e_3, \qquad (2.17)$$

where $\chi(K, Z) := \chi_{ab} K^a Z^P{}_{Qb}$ and $\underline{\chi}(K, Z) := \underline{\chi}_{ab} K^a Z^P{}_{Qb}$. Now we recall the definitions of the remaining connection coefficients adapted to the double null framework

$$\eta_a := -\frac{1}{2} \langle D_{e_3} e_a, e_4 \rangle, \quad \omega := -\frac{1}{4} \langle D_{e_4} e_3, e_4 \rangle = -\frac{1}{2} D_{e_4} \ln \Omega,$$
$$\underline{\eta}_a := -\frac{1}{2} \langle D_{e_4} e_a, e_3 \rangle, \quad \underline{\omega} := -\frac{1}{4} \langle D_{e_3} e_4, e_3 \rangle = -\frac{1}{2} D_{e_3} \ln \Omega \qquad (2.18)$$

and the torsion $\zeta_a := \frac{1}{2} \langle D_{e_a} e_4, e_3 \rangle = \frac{1}{2} (\eta - \underline{\eta})$. Utilizing these definitions, let us write down the kinematical set of structural equations

$$D_{e_a} e_3 = \underline{\chi}_{ab} e_b + \zeta_a e_3, \quad D_{e_3} e_a = \nabla_{e_3} e_a + \eta_a e_3, \qquad (2.19)$$

$$D_{e_4} e_a = \nabla_{e_4} e_a + \underline{\eta}_a e_4, \quad \nabla_{e_3} e_3 = -2\underline{\omega} e_3, \qquad (2.20)$$

$$D_{e_4} e_4 = -2\omega e_4, \quad \nabla_{e_4} e_3 = 2\omega e_3 + 2\underline{\eta}_a e_a, \qquad (2.21)$$

$$D_{e_3} e_4 = 2\underline{\omega} e_4 + 2\eta_a e_a, \quad D_{e_4} e_a = \nabla_{e_4} e_a + \underline{\eta}_a e_4, \qquad (2.22)$$

$$D_{e_a} e_4 = \chi_{ab} e_b - \zeta_a e_4. \tag{2.23}$$

Also note $\eta_a = \zeta_a + \nabla_{e_a} \ln \Omega$, $\bar{\eta}_a = -\zeta_a + \nabla_{e_a} \ln \Omega$. Utilizing these kinematical structure equations, we obtain the dynamical set of structural equations suitable trace of which are nothing but Einstein's equations sourced by Yang–Mills stress-energy tensor and expressed in the double null framework. Before writing down such equations, let us recall the following decomposition of the spacetime Riemann curvature tensor

$$R_{\alpha\beta\gamma\delta} = W_{\alpha\beta\gamma\delta} + \frac{1}{2}(g_{\alpha\gamma}R_{\beta\delta} + g_{\beta\delta}R_{\alpha\gamma} - g_{\beta\gamma}R_{\alpha\delta} - g_{\alpha\delta}R_{\beta\gamma}) + \frac{1}{6}R(g_{\alpha\delta}g_{\beta\gamma} - g_{\alpha\gamma}g_{\beta\delta}), \tag{2.24}$$

where $W$ is the Weyl curvature tensor describing pure gravitational degrees of freedom. $W$ is trace-free and enjoys the same algebraic symmetry of the Riemann curvature. Since the Ricci and the scalar curvatures are fixed by Einstein's field equations, we ought to write down the equation for the null components of the Weyl curvature. These equations will constitute the null Bianchi equations. The components of the Weyl curvature are defined as follows

$$\alpha_{ab} := W(e_a, e_4, e_b, e_4), \quad \bar{\alpha}_{ab} := W(e_a, e_3, e_b, e_3), \tag{2.25}$$

$$2\beta_a := W(e_4, e_3, e_4, e_a), \quad 2\bar{\beta}_a := W(e_3, e_4, e_3, e_a), \tag{2.26}$$

$$\rho := \frac{1}{4} W(e_4, e_3, e_4, e_3), \quad \sigma := \frac{1}{4} {}^*W(e_4, e_3, e_4, e_3), \tag{2.27}$$

where ${}^*W$ is the left Hodge dual of $W$ defined as follows

$${}^*W_{\alpha\beta\gamma\delta} := \frac{1}{2} \epsilon_{\alpha\beta\mu\nu} W^{\mu\nu}{}_{\gamma\delta}, \tag{2.28}$$

where $\epsilon_{\alpha\beta\mu\nu}$ is the volume form on the spacetime $M$. In addition to the Weyl field, we also have the Yang–Mills curvature $F := \frac{1}{2} F^P{}_{Q\mu\nu} dx^\mu \wedge dx^\nu \in \Lambda^2 T^*M \otimes_M \mathfrak{P}_{Ad,\mathfrak{g}}$, $P, Q = 1, 2, 3, \ldots, \dim(V)$, where $V$ is the representation of the Lie algebra $\mathfrak{g}$. The components of the Yang–Mills curvature $F$ are defined as follows

$$\alpha^F_a := F^P{}_Q(e_a, e_4), \quad \bar{\alpha}^F_a := F^P{}_Q(e_a, e_3), \quad \rho^F := \frac{1}{2} F^P{}_Q(e_3, e_4),$$

$$\sigma^F = \frac{1}{2} {}^*F^P{}_Q(e_3, e_4) = F^P{}_Q(e_1, e_2). \tag{2.29}$$

Also decompose the null second fundamental forms $\chi, \bar{\chi}$ into their trace and trace-free components

$$\chi = \hat{\chi} + \frac{1}{2} tr\chi \gamma, \quad \underline{\chi} = \hat{\underline{\chi}} + \frac{1}{2} tr\underline{\chi} \gamma. \tag{2.30}$$

The Einstein's equations (with the choice of unit $8\pi G = c = 1$)

$$R_{\mu\nu} - \frac{1}{2} R g_{\mu\nu} = \mathfrak{T}_{\mu\nu} \tag{2.31}$$

in the double null framework reads

$$\nabla_4 tr\chi + \frac{1}{2}(tr\chi)^2 = -|\hat{\chi}|^2_\gamma - 2\omega tr\chi - \mathfrak{T}_{44} \tag{2.32}$$

$$\nabla_4 \hat{\chi} + tr\chi \hat{\chi} = -2\omega \hat{\chi} - \alpha \tag{2.33}$$

$$\nabla_3 tr\underline{\chi} + \frac{1}{2}(tr\underline{\chi})^2 = -|\hat{\underline{\chi}}|^2_\gamma - 2\underline{\omega} tr\underline{\chi} - \mathfrak{T}_{33} \tag{2.34}$$

$$\nabla_3 \hat{\underline{\chi}} + tr\underline{\chi} \hat{\underline{\chi}} = -2\underline{\omega} \hat{\underline{\chi}} - \underline{\alpha} \tag{2.35}$$

$$\nabla_4 \eta_a = -\chi \cdot (\eta - \underline{\eta}) - \beta - \frac{1}{2} \mathfrak{T}_{a4} \tag{2.36}$$

$$\nabla_3 \underline{\eta}_a = -\underline{\chi} \cdot (\underline{\eta} - \eta) + \underline{\beta} + \frac{1}{2} \mathfrak{T}_{a3} \tag{2.37}$$

$$\nabla_4 \underline{\omega} = 2\omega\underline{\omega} + \frac{3}{4}|\eta - \underline{\eta}|^2 - \frac{1}{4}(\eta - \underline{\eta})\cdot(\eta + \underline{\eta}) - \frac{1}{8}|\eta + \underline{\eta}|^2 + \frac{1}{2}\rho + \frac{1}{4}\mathfrak{T}_{43}$$

$$\nabla_3 \omega = 2\omega\underline{\omega} + \frac{3}{4}|\eta - \underline{\eta}|^2 + \frac{1}{4}(\eta - \underline{\eta})\cdot(\eta + \underline{\eta}) - \frac{1}{8}|\eta + \underline{\eta}|^2 + \frac{1}{2}\rho + \frac{1}{4}\mathfrak{T}_{43}$$

$$\nabla_4 tr\underline{\chi} + \frac{1}{2}tr\chi tr\underline{\chi} = 2\omega tr\underline{\chi} + 2\mathrm{div}\underline{\eta} + 2|\underline{\eta}|_\gamma^2 + 2\rho - \hat{\chi}\cdot\hat{\underline{\chi}} \tag{2.38}$$

$$\nabla_3 tr\chi + \frac{1}{2}tr\underline{\chi}tr\chi = 2\underline{\omega}tr\chi + 2\mathrm{div}\eta + 2|\eta|^2 + 2\rho - \hat{\chi}\cdot\hat{\underline{\chi}} \tag{2.39}$$

$$\nabla_4 \hat{\underline{\chi}} + \frac{1}{2}tr\chi\hat{\underline{\chi}} = \nabla\hat{\otimes}\underline{\eta} + 2\omega\hat{\underline{\chi}} - \frac{1}{2}tr\underline{\chi}\hat{\chi} + \underline{\eta}\hat{\otimes}\underline{\eta} + \hat{\mathfrak{T}}_{ab} \tag{2.40}$$

$$\nabla_3 \hat{\chi} + \frac{1}{2}tr\underline{\chi}\hat{\chi} = \nabla\hat{\otimes}\eta + 2\underline{\omega}\hat{\chi} - \frac{1}{2}tr\chi\hat{\underline{\chi}} + \eta\hat{\otimes}\eta + \hat{\mathfrak{T}}_{ab} \tag{2.41}$$

$$\mathrm{div}\hat{\chi} = \frac{1}{2}\nabla tr\chi - \frac{1}{2}(\eta - \underline{\eta})\cdot\left(\hat{\chi} - \frac{1}{2}tr\chi\gamma\right) - \beta + \frac{1}{2}\mathfrak{T}(e_4,\cdot) \tag{2.42}$$

$$\mathrm{div}\hat{\underline{\chi}} = \frac{1}{2}\nabla tr\underline{\chi} - \frac{1}{2}(\underline{\eta} - \eta)\cdot\left(\hat{\underline{\chi}} - \frac{1}{2}tr\underline{\chi}\gamma\right) - \underline{\beta} + \frac{1}{2}\mathfrak{T}(e_3,\cdot) \tag{2.43}$$

$$\mathrm{curl}\,\eta = \hat{\underline{\chi}}\wedge\hat{\chi} + \sigma\epsilon = -\mathrm{curl}\,\underline{\eta} \tag{2.44}$$

$$K - \frac{1}{2}\hat{\chi}\cdot\hat{\underline{\chi}} + \frac{1}{4}tr\chi tr\underline{\chi} = -\rho + \frac{1}{4}\mathfrak{T}_{43}. \tag{2.45}$$

where $\mathfrak{T}_{\mu\nu} := \frac{1}{2}\left(F^P{}_{Q\mu\alpha}F^Q{}_{P\nu}{}^\alpha + {}^*F^P{}_{Q\mu\alpha}{}^*F^Q{}_{P\nu}{}^\alpha\right)$ is the Yang–Mills stress-energy tensor and $K$ is the sectional curvature of the topological two–sphere $S_{u,\underline{u}}$ (or a constant multiple of Gauss curvature). Here Eqs. (2.42)–(2.45) are the null-constraint equations. Now recall the Bianchi identities

$$D_\mu R_{\alpha\beta\nu\lambda} + D_\nu R_{\alpha\beta\lambda\mu} + D_\lambda R_{\alpha\beta\mu\nu} = 0 \tag{2.46}$$

which together with the decomposition (2.24) and Einstein's equations yields the following Yang–Mills type equations for the Weyl curvature

$$D^\alpha W_{\alpha\beta\gamma\delta} = J[\mathfrak{T}]_{\beta\gamma\delta}, \quad D_{[\mu} W_{\gamma\delta]\alpha\beta} = \frac{1}{3}\epsilon_{\nu\mu\gamma\delta}J[\mathfrak{T}]^{*\nu}{}_{\alpha\beta}. \tag{2.47}$$

Here $J[\mathfrak{T}]$ is the source term determined fully by the Yang–Mills stress-energy tensor $\mathfrak{T}$. After elementary algebraic manipulations, the differential equations for the Weyl curvature may be cast into the following double-null form[29]

$$\nabla_3 \alpha_{ab} + \frac{1}{2}tr\underline{\chi}\alpha_{ab} = (\nabla\hat{\otimes}\beta)_{ab} + 4\underline{\omega}\alpha_{ab} - 3(\hat{\chi}\rho + {}^*\hat{\chi}\sigma)_{ab} + ((\zeta + 4\eta)\hat{\otimes}\beta)_{ab} \tag{2.48}$$

$$+ \frac{1}{2}(D_3\mathfrak{T}_{44} - D_4\mathfrak{T}_{43})\gamma_{ab} - D_4\mathfrak{T}_{ab} + \frac{1}{2}(D_b\mathfrak{T}_{4a} + D_a\mathfrak{T}_{4b})$$

$$\nabla_4 \beta_a + 2tr\chi\beta_a = (\mathrm{div}\alpha)_a - 2\omega\beta_a + (\eta\cdot\alpha)_a - \frac{1}{2}(D_a\mathfrak{T}_{44} - D_4\mathfrak{T}_{4a}) \tag{2.49}$$

$$\nabla_3 \beta_a + tr\underline{\chi}\beta_a = \nabla_a\rho + {}^*\nabla_a\sigma + 2\underline{\omega}\beta_a + 2(\hat{\chi}\cdot\underline{\beta})_a + 3(\eta\rho + {}^*\eta\sigma)_a$$
$$+ \frac{1}{2}(D_a\mathfrak{T}_{34} - D_4\mathfrak{T}_{3a})$$

$$\nabla_4 \sigma + \frac{3}{2}tr\chi\sigma = -\mathrm{div}\,{}^*\beta + \frac{1}{2}\hat{\underline{\chi}}\cdot{}^*\alpha - \zeta\cdot{}^*\beta - 2\underline{\eta}\cdot{}^*\beta - \frac{1}{4}(D_\mu\mathfrak{T}_{4\nu} - D_\nu\mathfrak{T}_{4\mu})\epsilon^{\mu\nu}{}_{34}$$

$$\nabla_3 \sigma + \frac{3}{2}tr\underline{\chi}\sigma = -\mathrm{div}\,{}^*\underline{\beta} + \frac{1}{2}\hat{\chi}\cdot{}^*\underline{\alpha} - \zeta\cdot{}^*\underline{\beta} - 2\eta\cdot{}^*\underline{\beta} + \frac{1}{4}(D_\mu\mathfrak{T}_{3\nu} - D_\nu\mathfrak{T}_{3\mu})\epsilon^{\mu\nu}{}_{34}$$

$$\nabla_4 \rho + \frac{3}{2}tr\chi\rho = \mathrm{div}\beta - \frac{1}{2}\hat{\underline{\chi}}\cdot\alpha + \zeta\cdot\beta + 2\underline{\eta}\cdot\beta - \frac{1}{4}(D_3\mathfrak{T}_{44} - D_4\mathfrak{T}_{34}) \tag{2.50}$$

$$\nabla_3 \rho + \frac{3}{2}tr\underline{\chi}\rho = -\mathrm{div}\underline{\beta} - \frac{1}{2}\hat{\chi}\cdot\underline{\alpha} + \zeta\cdot\underline{\beta} - 2\eta\cdot\underline{\beta} + \frac{1}{4}(D_3\mathfrak{T}_{34} - D_4\mathfrak{T}_{33}) \tag{2.51}$$

$$\nabla_4 \underline{\beta}_a + tr\chi \underline{\beta}_a = -\nabla_a \rho + {}^*\nabla_a \sigma + 2\omega \underline{\beta}_a + 2(\hat{\underline{\chi}} \cdot \beta)_a - 3(\underline{\eta}\rho - {}^*\underline{\eta}\sigma)_a$$
$$- \frac{1}{2}(D_a \mathfrak{T}_{43} - D_3 \mathfrak{T}_{4a})$$

$$\nabla_3 \underline{\beta}_a + 2tr\underline{\chi} \underline{\beta}_a = -(\text{div}\underline{\alpha})_a - 2\underline{\omega}\underline{\beta}_a + (\underline{\eta} \cdot \underline{\alpha})_a + \frac{1}{2}(D_a \mathfrak{T}_{33} - D_3 \mathfrak{T}_{3a}) \qquad (2.52)$$

$$\nabla_4 \underline{\alpha}_{ab} + \frac{1}{2} tr\chi \underline{\alpha} = -(\nabla \hat{\otimes} \underline{\beta})_{ab} + 4\omega \underline{\alpha}_{ab} - 3(\hat{\underline{\chi}}_{ab}\rho - {}^*\hat{\underline{\chi}}_{ab}\sigma) + ((\zeta - 4\underline{\eta}) \hat{\otimes} \underline{\beta})_{ab} \qquad (2.53)$$
$$+ \frac{1}{2}(D_4 \mathfrak{T}_{33} - D_3 \mathfrak{T}_{34})\gamma_{ab} - D_3 \mathfrak{T}_{ab} + \frac{1}{2}(D_a \mathfrak{T}_{3b} + D_b \mathfrak{T}_{3a}).$$

The null Yang–Mills equations are written as

$$\widehat{\nabla}_4 \underline{\alpha}^F + \frac{1}{2} tr\chi \underline{\alpha}^F = -\widehat{\nabla}\underline{\rho}^F - {}^*\widehat{\nabla}\underline{\sigma}^F - 2\,{}^*\underline{\eta}\underline{\sigma}^F - 2\underline{\eta}\underline{\rho}^F + 2\omega\underline{\alpha}^F - \hat{\underline{\chi}} \cdot \alpha^F \qquad (2.54)$$

$$\widehat{\nabla}_3 \alpha^F + \frac{1}{2} tr\underline{\chi} \alpha^F = -\widehat{\nabla}\rho^F + {}^*\widehat{\nabla}\sigma^F - 2\,{}^*\eta\sigma^F + 2\eta\rho^F + 2\underline{\omega}\alpha^F - \hat{\underline{\chi}} \cdot \underline{\alpha}^F \qquad (2.55)$$

$$\widehat{\nabla}_4 \rho^F + tr\chi \rho^F = -\widehat{\text{div}}\alpha^F - (\eta - \underline{\eta}) \cdot \alpha^F \qquad (2.56)$$

$$\widehat{\nabla}_4 \sigma^F + tr\chi \sigma^F = -\widehat{\text{curl}}\alpha^F + (\eta - \underline{\eta}) \cdot {}^*\alpha^F \qquad (2.57)$$

$$\widehat{\nabla}_3 \rho^F + tr\underline{\chi} \rho^F = -\widehat{\text{div}}\underline{\alpha}^F + (\eta - \underline{\eta}) \cdot \underline{\alpha}^F \qquad (2.58)$$

$$\widehat{\nabla}_3 \sigma^F + tr\underline{\chi} \sigma^F = -\widehat{\text{curl}}\underline{\alpha}^F + (\eta - \underline{\eta}) \cdot {}^*\underline{\alpha}^F. \qquad (2.59)$$

We may write down explicitly the null components of the Yang–Mills stress-energy tensor

$$\mathfrak{T}_{43} = \rho^F \cdot \rho^F + \sigma^F \cdot \sigma^F, \quad \mathfrak{T}_{4A} = \alpha_A^F \cdot \rho^F + \epsilon_A{}^B \alpha_B^F \cdot \sigma^F, \qquad (2.60)$$

$$\mathfrak{T}_{3A} = -\underline{\alpha}_A^F \cdot \rho^F + \epsilon_A{}^B \underline{\alpha}_B^F \cdot \sigma^F, \quad \mathfrak{T}_{44} = \alpha_A^F \cdot \alpha_B^F \gamma^{AB}, \quad \mathfrak{T}_{33} = \underline{\alpha}_A^F \cdot \underline{\alpha}_B^F \gamma^{AB}, \qquad (2.61)$$

$$\mathfrak{T}_{ab} = \frac{1}{2}(\rho^F \cdot \rho^F + \sigma^F \cdot \sigma^F)\gamma_{AB} - (\alpha_A^F \cdot \underline{\alpha}_B^F + \underline{\alpha}_A^F \cdot \alpha_B^F - \alpha_C^F \cdot \underline{\alpha}_D^F \gamma^{CD} \gamma_{AB}), \qquad (2.62)$$

The Yang–Mills source terms that appear in the null Bianchi equations are explicitly computed as follows. Since we do not need the exact expressions for the source terms, we write these in the schematic form for convenience.

$$D_a \mathfrak{T}_{44} - D_4 \mathfrak{T}_{4a} \sim \langle \alpha^F, \widehat{\nabla}_b \alpha^F \rangle - \chi_{bc} \mathfrak{T}_{c4} + \eta_b \mathfrak{T}_{44} - 2\omega \mathfrak{T}_{4b} - \widehat{\nabla}_4(\alpha_b^F \rho^F - \alpha_b^F \sigma^F),$$
$$D_b \mathfrak{T}_{34} - D_4 \mathfrak{T}_{3b} \sim (\langle \widehat{\nabla}\rho^F, \rho^F \rangle + \langle \widehat{\nabla}\sigma^F, \sigma^F \rangle) - \underline{\chi}\alpha^F(\sigma^F + \rho^F) - \widehat{\nabla}_4 \underline{\alpha}^F(\rho^F + \sigma^F)$$
$$- \underline{\alpha}^F(\widehat{\nabla}_4 \rho^F + \widehat{\nabla}_4 \sigma^F), \qquad (2.63)$$

$$(D_\mu \mathfrak{T}_{4\nu} - D_\nu \mathfrak{T}_{4\mu})\epsilon^{\mu\nu}{}_{34} \sim \widehat{\nabla}(\alpha^F \rho^F - \alpha^F \sigma^F) - \chi(\rho^F \rho^F + \alpha^F \underline{\alpha}^F + \sigma^F \sigma^F)$$
$$+ (\eta - \underline{\eta})\alpha^F(\rho^F + \sigma^F) \qquad (2.64)$$

$$D_3 \mathfrak{T}_{44} - D_4 \mathfrak{T}_{34} \sim \langle \alpha^F, \widehat{\nabla}_3 \alpha^F \rangle + \underline{\omega}|\alpha^F|^2 + \eta_A(\alpha^F \cdot \sigma^F + \alpha^F \cdot \rho^F) + \langle \rho^F, \widehat{\nabla}_4 \rho^F \rangle$$
$$+ \langle \sigma^F, \widehat{\nabla}_4 \sigma^F \rangle + \underline{\eta}(\rho^F \underline{\alpha}^F + \rho^F \cdot \sigma^F),$$

$$D_b \mathfrak{T}_{43} - D_3 \mathfrak{T}_{4b} \sim \langle \rho^F, \widehat{\nabla}\rho^F \rangle + \langle \sigma^F, \widehat{\nabla}\sigma^F \rangle - \underline{\chi}\alpha^F(\rho^F + \sigma^F) + \chi \underline{\alpha}^F(\rho^F + \sigma^F), \qquad (2.65)$$

$$(D_\mu R_{3\nu} - D_\nu R_{3\mu})\epsilon^{\mu\nu}{}_{34} \sim \widehat{\nabla}(\underline{\alpha}^F \rho^F - \underline{\alpha}^F \sigma^F) - \underline{\chi}(\rho^F \rho^F + \underline{\alpha}^F \alpha^F + \sigma^F \sigma^F)$$
$$+ (\eta + \underline{\eta})\underline{\alpha}^F(\rho^F + \sigma^F), \qquad (2.66)$$

$$D_3 \mathfrak{T}_{43} - D_4 \mathfrak{T}_{33} \sim 2\langle \rho^F, -\hat{\text{div}}\underline{\alpha}^F + tr\underline{\chi}\rho^F + (\eta - \underline{\eta})\underline{\alpha}^F \rangle + 2\langle \sigma_F, -\hat{\text{curl}}\underline{\alpha}^F - tr\underline{\chi}\sigma^F$$
$$+ (\eta - \underline{\eta}) *\underline{\alpha}^F \rangle - 2\eta(\rho^F \underline{\alpha}^F + \rho^F \sigma^F) - \langle \underline{\alpha}^F, -\frac{1}{2}tr\chi\underline{\alpha}^F - \widehat{\nabla}\rho^F - {}^*\widehat{\nabla}\sigma^F - 2 \, {}^*\underline{\eta}\sigma^F$$
$$- 2\underline{\eta}\rho^F + 2\omega\underline{\alpha}^F - \hat{\underline{\chi}}\alpha^F \rangle + 2\omega|\underline{\alpha}^F|^2 - 4\underline{\eta}_a(\underline{\alpha}^F \rho^F - \underline{\alpha}^F \sigma^F),$$
(2.67)

$$D_b \mathfrak{T}_{33} - D_3 \mathfrak{T}_{3b} \sim \langle \underline{\alpha}^F, \widehat{\nabla}\underline{\alpha}^F \rangle - \chi \underline{\alpha}^F (\rho^F + \sigma^F) + \underline{\eta}|\underline{\alpha}^F|^2 - 2\underline{\omega}\underline{\alpha}^F (\rho^F + \sigma^F)$$
$$- \widehat{\nabla}_3 (\underline{\alpha}^F \rho^F + \underline{\alpha}^F \sigma^F).$$

Note the most important point: the Yang–Mills curvatures appear in the source term for Einstein's equations in a gauge-invariant fashion. Let us make this more precise. The source terms in the Bianchi equations involve Derivatives of the Yang–Mills stress-energy tensor. The stress-energy tensor in its covariant form is a section of the co-tangent bundle of the spacetime and not a section of the gauge bundle. Therefore, under Yang–Mills gauge transformation, it remains invariant. In particular, the null components of the stress-energy tensor $\mathfrak{T}$ verifies schematically $\mathfrak{T} \sim \langle A, B \rangle$ where $\langle \cdot, \cdot \rangle$ denotes the inner product on the fiber of the gauge bundle and $A, B$ are Lie-algebra valued Yang–Mills curvature components. Now, in obtaining the estimates when we act $\langle A, B \rangle$ by the ordinary derivative $\nabla$, using metric compatibility of the gauge covariant derivative this reads $\nabla \langle A, B \rangle = \langle \widehat{\nabla} A, B \rangle + \langle A, \widehat{\nabla} B \rangle$. Therefore, we never have to use the split $\widehat{\nabla} = \nabla + [A, \cdot]$ instead obtain all the necessary estimates for the Yang–Mills curvature components in terms of the fully gauge covariant derivative $\widehat{\nabla}$. This allows us to obtain gauge invariant estimates for the Yang–Mills curvature.

Now we define the norm of the initial data (the Weyl curvature and the Yang–Mills curvature).

In this section, Let us now write down the main theorem of the article. First, let us define the generalized Sobolev norm for the data on the initial slice that we shall consider. (One could use one less derivative to close the argument. See Ref. 30 where a characteristic initial value problem is solved in one order lower regularity.)

$$\mathbf{W}_0 := \int_{\widehat{\Sigma}_0} r^4(|\alpha|^2 + |\beta|^2) + \int_{\widehat{\Sigma}_0} \tau_-^4 |\underline{\beta}|^2 + \int_{\widehat{\Sigma}_0} \tau_-^2 r^2 (\rho^2 + \sigma^2)$$
(2.68)
$$+ \int_{\widehat{\Sigma}_0} (\tau_-^4 |\underline{\alpha}|^2 + r^4 |\beta|^2 + r^2 \tau_-^2 |\underline{\beta}|^2 + r^4 |\rho|^2 + r^4 |\sigma|^2),$$
$$+ \int_{\widehat{\Sigma}_0} r^4 (|r\nabla\alpha|^2 + |r\nabla\beta|^2) + \int_{\widehat{\Sigma}_0} \tau_-^4 |r\nabla\underline{\beta}|^2 + \int_{\widehat{\Sigma}_0} \tau_-^2 r^2 (|r\nabla\rho|^2 + |r\nabla\sigma|^2)$$
$$+ \int_{\widehat{\Sigma}_0} (\tau_-^4 |r\nabla\underline{\alpha}|^2 + r^4 |r\nabla\beta|^2 + r^2 \tau_-^2 |r\nabla\underline{\beta}|^2 + r^4 |r\nabla\rho|^2 + r^4 |r\nabla\sigma|^2)$$
$$+ \int_{\widehat{\Sigma}_0} r^4 |rD_4 \alpha|^2 + \int_{\widehat{\Sigma}_0} \tau_-^4 |\tau_- D_3 \underline{\alpha}|^2$$
$$+ \int_{\widehat{\Sigma}_0} r^4 (|r^2 \nabla^2 \alpha|^2 + |r^2 \nabla^2 \beta|^2) + \int_{\widehat{\Sigma}_0} \tau_-^4 |r^2 \nabla^2 \underline{\beta}|^2 + \int_{\widehat{\Sigma}_0} \tau_-^2 r^2 (|r^2 \nabla^2 \rho|^2 + |r^2 \nabla^2 \sigma|^2)$$
$$+ \int_{\widehat{\Sigma}_0} \left( \tau_-^4 |r^2 \nabla^2 \underline{\alpha}|^2 + r^4 |r^2 \nabla^2 \beta|^2 + r^2 \tau_-^2 |r^2 \nabla^2 \underline{\beta}|^2 + r^4 |r^2 \nabla^2 \rho|^2 \right.$$
$$\left. + r^4 |r^2 \nabla^2 \sigma|^2 \right) + \int_{\widehat{\Sigma}_0} r^4 |r^2 \nabla D_4 \alpha|^2 + \int_{\widehat{\Sigma}_0} \tau_-^4 |\tau_- r \nabla D_3 \underline{\alpha}|^2$$

and

$$\mathbf{Y}_0 := \int_{\widehat{\Sigma}_0} (r^2 |\alpha^F|^2 + \tau_-^2 |\rho^F|^2 + \tau_-^2 |\sigma^F|^2)$$
$$+ \int_{\widehat{\Sigma}_0} (\tau_-^2 |\underline{\alpha}^F|^2 + r^2 |\rho^F|^2 + r^2 |\sigma^F|^2),$$
$$+ \sum_{I=1}^{3} \int_{\widehat{\Sigma}_0} (r^2 |r^I \widehat{\nabla}^I \alpha^F|^2 + \tau_-^2 |r^I \widehat{\nabla}^I \rho^F|^2 + \tau_-^2 |r^I \widehat{\nabla}^I \sigma^F|^2)$$
$$+ \int_{\widehat{\Sigma}_0} (\tau_-^2 |r^I \widehat{\nabla}^I \underline{\alpha}^F|^2 + r^2 |r^I \widehat{\nabla}^I \rho^F|^2 + r^2 |r^I \widehat{\nabla}^I \sigma^F|^2)$$
$$+ \int_{\widehat{\Sigma}_0} r^2 |r^I \widehat{\nabla}^{I-1} \widehat{D}_4 \alpha^F|^2 + \int_{\widehat{\Sigma}_0} \tau_-^2 |r^{I-1} \tau_- \widehat{\nabla}^{I-1} \widehat{D}_3 \underline{\alpha}^F|^2$$
$$+ \int_{\widehat{\Sigma}_0} r^6 |\widehat{D}_4^2 \alpha^F|^2 + \int_{\widehat{\Sigma}_0} \tau_-^6 |\widehat{D}_3^2 \underline{\alpha}^F|^2.$$
(2.69)

The energy (the generalized weighted Sobolev norm) that we shall control in the bulk $D_{u,\underline{u}}$ (i.e., the causal globally hyperbolic portion of the spacetime) is defined as follows

$$\mathbf{W} := \int_H r^4(|\alpha|^2 + |\beta|^2) + \int_H \tau_-^4 |\underline{\beta}|^2 + \int_H \tau_-^2 r^2 (\rho^2 + \sigma^2) \tag{2.70}$$

$$+ \int_{\underline{H}} (\tau_-^4 |\underline{\alpha}|^2 + r^4 |\beta|^2 + r^2 \tau_-^2 |\underline{\beta}|^2 + r^4 |\rho|^2 + r^4 |\sigma|^2),$$

$$+ \int_H r^4(|r\nabla \alpha|^2 + |r\nabla \beta|^2) + \int_H \tau_-^4 |r\nabla \underline{\beta}|^2 + \int_H \tau_-^2 r^2 (|r\nabla \rho|^2 + |r\nabla \sigma|^2)$$

$$+ \int_{\underline{H}} (\tau_-^4 |r\nabla \underline{\alpha}|^2 + r^4 |r\nabla \beta|^2 + r^2 \tau_-^2 |r\nabla \underline{\beta}|^2 + r^4 |r\nabla \rho|^2 + r^4 |r\nabla \sigma|^2)$$

$$+ \int_H r^4 |rD_4 \alpha|^2 + \int_{\underline{H}} \tau_-^4 |\tau_- D_3 \underline{\alpha}|^2 +$$

$$+ \int_H r^4 (|r^2 \nabla^2 \alpha|^2 + |r^2 \nabla^2 \beta|^2) + \int_H \tau_-^4 |r^2 \nabla^2 \underline{\beta}|^2 + \int_H \tau_-^2 r^2 (|r^2 \nabla^2 \rho|^2 + |r^2 \nabla^2 \sigma|^2)$$

$$+ \int_{\underline{H}} (\tau_-^4 |r^2 \nabla^2 \underline{\alpha}|^2 + r^4 |r^2 \nabla^2 \beta|^2 + r^2 \tau_-^2 |r^2 \nabla^2 \underline{\beta}|^2 + r^4 |r^2 \nabla^2 \rho|^2 + r^4 |r^2 \nabla^2 \sigma|^2)$$

$$+ \int_H r^k |r^2 \nabla D_4 \alpha|^2 + \int_{\underline{H}} \tau_-^4 |\tau_- r \nabla D_3 \underline{\alpha}|^2$$

and

$$\mathbf{Y} := \int_H (r^2 |\alpha^F|^2 + \tau_-^2 |\rho^F|^2 + \tau_-^2 |\sigma^F|^2) \tag{2.71}$$

$$+ \int_{\underline{H}} (\tau_-^2 |\underline{\alpha}^F|^2 + r^2 |\rho^F|^2 + r^2 |\sigma^F|^2),$$

$$+ \sum_{I=1}^{3} \int_H (r^2 |r^I \widehat{\nabla}^I \alpha^F|^2 + \tau_-^2 |r^I \widehat{\nabla}^I \rho^F|^2 + \tau_-^2 |r^I \widehat{\nabla}^I \sigma^F|^2) \tag{2.72}$$

$$+ \int_{\underline{H}} (\tau_-^2 |r^I \widehat{\nabla}^I \underline{\alpha}^F|^2 + r^2 |r^I \widehat{\nabla}^I \rho^F|^2 + r^2 |r^I \widehat{\nabla}^I \sigma^F|^2)$$

$$+ \int_H r^2 |r^I \widehat{\nabla}^{I-1} \widehat{D}_4 \alpha^F|^2 + \int_{\underline{H}} \tau_-^2 |r^{I-1} \tau_- \widehat{\nabla}^{I-1} \widehat{D}_3 \underline{\alpha}^F|^2$$

$$+ \int_H r^6 |\widehat{D}_4^2 \alpha^F|^2 + \int_{\underline{H}} \tau_-^6 |\widehat{D}_3^2 \underline{\alpha}^F|^2.$$

As a consequence, we immediately note that a global exterior stability theorem for the Minkowski space under coupled Einstein–Yang–Mills perturbations follow.

## III. IMPORTANT INEQUALITIES

One of the most important features of our study is the utilization of the fully gauge-covariant structure of the Yang–Mills equations. In other words, we would like to work with the Yang–Mills curvature and the fully gauge covariant derivative directly instead of the working with Yang–Mills connection. To that end, we need all the necessary inequalities in a fully gauge-invariant form. Let us now extend all the Sobolev and trace inequalities to the fibers of the vector bundle $\otimes^k T^* M \otimes_M \mathfrak{P}_{Ad,\mathfrak{g}}$. Let us write both the degenerate and the non-degenerate (no loss in decay along $\underline{u}$) versions. Notice that these inequalities are nothing but exhibitions of the scaling properties of the associated physical entities. Assume $\sup_{D_{u,\underline{u}}} |tr\chi - \frac{2}{r}| \leq \epsilon$, $\sup_{D_{u,\underline{u}}} |tr\underline{\chi} + \frac{2}{r}| \leq \epsilon$ for sufficiently small $\epsilon$.

*Lemma 3.1.* Let $F$ is an arbitrary $S_{u,\underline{u}}$– tangent section of the associated bundle $k \otimes T^* S \otimes \mathfrak{P}_{Ad,\rho}$. The following Sobolev embedding holds for $F$

$$\sup_{S_{u,\underline{u}}} r|F|_{\gamma,\delta} \leq C_{sob} \left( \int_{S_{u,\underline{u}}} |F|^2_{\gamma,\delta} + r^2 |\widehat{\nabla} F|^2_{\gamma,\delta} + r^4 |\widehat{\nabla}^2 F|^2_{\gamma,\delta} \right)^{\frac{1}{2}} \tag{3.1}$$

*Proof.* Write $f = \sqrt{F^a_{bIJ} F^b_{aKL} \gamma^{IK} \gamma^{JL} + \delta}$ and since $f$ is gauge invariant $\nabla f = \widehat{\nabla} f = \frac{1}{2f} \langle F, \widehat{\nabla} F \rangle_{\gamma,\delta}$. Now use the standard Sobolev embedding for $f$ and take the limit $\delta \to 0$ to yield the result. □

*Proposition 3.1.* The following gauge-invariant Sobolev inequality holds in the context of Cauchy-double-null frame-work

$$\sup_{S_{u,\underline{u}}} r^{\frac{3}{2}}|F| \leq C\left[\left(\int_{S_{u,\underline{u}_0}} r^4|F|^4_{\gamma,\delta}\right)^{\frac{1}{4}} + \left(\int_{S_{u,\underline{u}_0}} r^4|r\widehat{\nabla}F|^4_{\gamma,\delta}\right)^{\frac{1}{4}}\right.$$
$$\left. + \left(\int_{H_u \cap D_{u,\underline{u}}} |F|^2_{\gamma,\delta} + r^2|\widehat{\nabla}F|^2_{\gamma,\delta} + r^2|\widehat{\nabla}_4 F|^2_{\gamma,\delta} + r^4|\widehat{\nabla}^2 F|^2_{\gamma,\delta} + r^4|\widehat{\nabla}\widehat{\nabla}_4 F|^2_{\gamma,\delta}\right)^{\frac{1}{2}}\right]. \quad (3.2)$$

*The following degenerate version holds too*

$$\sup_{S_{u,\underline{u}}} \left(r\tau_-^{\frac{1}{2}}|F|\right) \leq C\left[\left(\int_{S_{u,\underline{u}_0}} r^2\tau_-^2|F|^4_{\gamma,\delta}\right)^{\frac{1}{4}} + \left(\int_{S_{u,\underline{u}_0}} r^2\tau_-^2|r\widehat{\nabla}F|^4_{\gamma,\delta}\right)^{\frac{1}{4}}\right.$$
$$\left. + \left(\int_{H_u \cap D_{u,\underline{u}}} |F|^2_{\gamma,\delta} + r^2|\widehat{\nabla}F|^2_{\gamma,\delta} + \tau_-^2|\widehat{\nabla}_4 F|^2 + r^4|\widehat{\nabla}^2 F|^2_{\gamma,\delta} + r^2\tau_-^2|\widehat{\nabla}\widehat{\nabla}_4 F|^2_{\gamma,\delta}\right)^{\frac{1}{2}}\right]. \quad (3.3)$$

*Analogous inequality holds on $\underline{H}$ too i.e.,*

$$\sup_{S_{u,\underline{u}}} r^{\frac{3}{2}}|F| \leq C\left[\left(\int_{S_{u_0,\underline{u}}} r^4|F|^4_{\gamma,\delta}\right)^{\frac{1}{4}} + \left(\int_{S_{u_0,\underline{u}}} r^4|r\widehat{\nabla}F|^4_{\gamma,\delta}\right)^{\frac{1}{4}}\right.$$
$$\left. + \left(\int_{\underline{H}_{\underline{u}} \cap D_{u,\underline{u}}} |F|^2_{\gamma,\delta} + r^2|\widehat{\nabla}F|^2_{\gamma,\delta} + r^2|\widehat{\nabla}_3 F|^2_{\gamma,\delta} + r^4|\widehat{\nabla}^2 F|^2_{\gamma,\delta} + r^4|\widehat{\nabla}\widehat{\nabla}_3 F|^2_{\gamma,\delta}\right)^{\frac{1}{2}}\right], \quad (3.4)$$

*and*

$$\sup_{S_{u,\underline{u}}} \left(r\tau_-^{\frac{1}{2}}|F|\right) \leq C\left[\left(\int_{S_{u_0,\underline{u}}} r^2\tau_-^2|F|^4_{\gamma,\delta}\right)^{\frac{1}{4}} + \left(\int_{S_{u_0,\underline{u}}} r^2\tau_-^2|r\widehat{\nabla}F|^4_{\gamma,\delta}\right)^{\frac{1}{4}}\right.$$
$$\left. + \left(\int_{\underline{H}_{\underline{u}} \cap D_{u,\underline{u}}} |F|^2_{\gamma,\delta} + r^2|\widehat{\nabla}F|^2_{\gamma,\delta} + \tau_-^2|\widehat{\nabla}_3 F|^2 + r^4|\widehat{\nabla}^2 F|^2_{\gamma,\delta} + r^2\tau_-^2|\widehat{\nabla}\widehat{\nabla}_3 F|^2_{\gamma,\delta}\right)^{\frac{1}{2}}\right]. \quad (3.5)$$

*Proof.* The proof relies on the following two inequalities

$$\sup_{S_{u,\underline{u}}} |G| \leq Cr^{-\frac{1}{2}}\left(\int_{S_{u,\underline{u}}} |G|^4 + r^4|\nabla G|^4\right)^{\frac{1}{4}},$$
$$\int_{S_{u,\underline{u}}} r^4|G|^4 \leq \int_{S_{u,\underline{u}_0}} r^4|G|^4 + C\left(\int_{H_u \cap D_{u,\underline{u}}} r^6|G|^6\right)^{\frac{1}{2}}\left(\int_{H_u \cap D_{u,\underline{u}}} r^2|\nabla_4 G|^2\right)^{\frac{1}{2}} \quad (3.6)$$

Now for a section of the associated bundle, both of these two inequalities hold with $\nabla$ replaced by $\widehat{\nabla}$ and $\nabla_4$ replaced by $\widehat{\nabla}_4$. This follows in an exact similar way as the previous case i.e., set $G = \sqrt{F^a_{bA_1A_2A_3\cdots A_n}F^b_{aB_1B_2B_3\cdots B_n}\gamma^{A_1B_1}\gamma^{A_2B_2}\gamma^{A_3B_3}\cdots\gamma^{A_nB_n} + \delta}$, apply the standard Sobolev inequality, and pass to the limit $\delta \to 0$ i.e.,

$$|\nabla G|^2 = |\nabla\sqrt{F^a_{bA_1A_2A_3\cdots A_n}F^b_{aB_1B_2B_3\cdots B_n}\gamma^{A_1B_1}\gamma^{A_2B_2}\gamma^{A_3B_3}\cdots\gamma^{A_nB_n} + \delta}|^2$$
$$= |\frac{\langle F, \widehat{\nabla}F\rangle}{2G}|^2 \leq \frac{1}{4}|\widehat{\nabla}F|^2_{\gamma,\delta}.$$

Similar argument yields

$$|\nabla_4 F|^2 \leq \frac{1}{4}|\widehat{\nabla}_4 F|^2_{\gamma,\delta}. \quad (3.7)$$

Therefore first we obtain

$$\sup_{S_{u,\underline{u}}} |F| \leq Cr^{-\frac{1}{2}}\left(\int_{S_{u,\underline{u}}} |F|^4_{\gamma,\delta} + r^4|\widehat{\nabla}F|^4_{\gamma,\delta}\right)^{\frac{1}{4}}. \quad (3.8)$$

Similarly we obtain

$$\int_{S_{u,\underline{u}}} r^4 |G|^4_{\gamma,\delta} \leq \int_{S_{u,\underline{u}_0}} r^4 |G|^4_{\gamma,\delta} + C \left( \int_{H_u \cap D_{u,\underline{u}}} r^6 |G|^6_{\gamma,\delta} \right)^{\frac{1}{2}} \left( \int_{H_u \cap D_{u,\underline{u}}} r^2 |\widehat{\nabla}_4 G|^2_{\gamma,\delta} \right)^{\frac{1}{2}}$$

Now we can treat $\widehat{\nabla} F$ as a new section $J$ of the bundle $^{k+1} \otimes T^* S \otimes \mathfrak{P}_{Ad,\mathfrak{g}}$ and apply the previous two inequalities to conclude the result. □

In addition to the gauge-invariant Sobolev inequalities in null cones, we also have the co-dimension-1 trace inequalities and usual Sobolev embedding on the topological two–spheres $S_{u,\underline{u}}$ (see Ref. 30, Sec. 5, inequalities 2 and 5 after appropriate scaling). In addition, one also has the following gauge-invariant transport inequalities that are useful for estimating connections in terms of the Weyl and Yang–Mills curvature.

*Proposition 3.2.* Consider a globally hyperbolic portion of the spacetime $D_{u,\underline{u}}$ foliated by a double null foliation. Let us assume that $|\Omega tr\chi - \overline{\Omega tr\chi}| \leq \frac{\epsilon}{r^2}$ for sufficiently small $\epsilon > 0$. Also an $S$–tangential tensor field $\mathcal{G}$ satisfies the following transport equation

$$\widehat{\nabla}_4 \mathcal{G} + \lambda_0 tr\chi \mathcal{G} = \Omega^{-1} \mathcal{H}. \tag{3.9}$$

The following gauge-invariant inequality holds for $\mathcal{G}$

$$|r^{\lambda_1} \mathcal{G}|_{p,S}(u,\underline{u}) \lesssim \left( |r^{\lambda_1} \mathcal{G}|_{p,S}(u,\underline{u}_0) + \int_{\underline{u}_0}^{\underline{u}} |r^{\lambda_1} \mathcal{H}|_{p,S}(u,\underline{u}')d\underline{u}' \right) \tag{3.10}$$

for $\lambda_1 = 2(\lambda_0 - \frac{1}{p})$.

*Proposition 3.3.* Assume that for $\epsilon > 0$ sufficiently small, $|\Omega tr\underline{\chi} - \overline{\Omega tr\underline{\chi}}| \lesssim \epsilon r^{-1} \tau_-^{-1}$ and let an $S$–tangential tensor field $\mathcal{G}$ satisfies the following transport equation

$$\widehat{\nabla}_3 \mathcal{G} + \lambda_0 tr\underline{\chi} \mathcal{G} = \Omega^{-1} \underline{\mathcal{H}}. \tag{3.11}$$

The following gauge-invariant inequality holds for $\mathcal{G}$

$$|r^{\lambda_1} \mathcal{G}|_{p,S}(u,\underline{u}) \lesssim \left( |r^{\lambda_1} \mathcal{G}|_{p,S}(u_0,\underline{u}) + \int_{u_0}^{u} |r^{\lambda_1} \underline{\mathcal{H}}|_{p,S}(u',\underline{u})du' \right) \tag{3.12}$$

for $\lambda_1 = 2(\lambda_0 - \frac{1}{p})$.

## IV. COMMUTATION FORMULAE

In order to obtain the higher order energy estimates, we need the commutation formulae $[\widehat{\nabla}^I, \widehat{\nabla}_4]$ and $[\widehat{\nabla}^I, \widehat{\nabla}_3]$. For a $S$–tangential tensor field that is a section of the bundle $^k \otimes T^* S$, the gauge covariant derivative $\widehat{\nabla}$ and the ordinary covariant derivative $\nabla$ coincide. These commutators are important since, the true non-linear characteristics of the Yang–Mills fields manifest itself through these commutators in the higher order energy estimates.

*Lemma 4.1.* Suppose $\mathcal{G}$ is a section of the product vector bundle $^k \otimes T^* S \otimes \mathfrak{P}_{Ad,\rho}$, $k \geq 1$, that satisfies $\widehat{\nabla}_4 \mathcal{G} = \mathcal{F}_1$ and $\widehat{\nabla}_4 \widehat{\nabla}^I \mathcal{G} = \mathcal{F}^I_1$, then $\mathcal{F}^I_1$ verifies the following schematic expression

$$\mathcal{F}^I_1 \sim \sum_{J_1+J_2+J_3+J_4=I-1} \nabla^{J_1}(\eta+\underline{\eta})^{J_2} \nabla^{J_3} \beta \widehat{\nabla}^{J_4} \mathcal{G}$$

$$+ \sum_{J_1+J_2+J_3+J_4+J_5=I-1} \nabla^{J_1}(\eta+\underline{\eta})^{J_2} \widehat{\nabla}^{J_3} \alpha^F \widehat{\nabla}^{J_4}(\rho^F, \sigma^F) \widehat{\nabla}^{J_5} \mathcal{G}$$

$$+ \underbrace{\sum_{J_1+J_2+J_3+J_4=I-1} \nabla^{J_1}(\eta+\underline{\eta})^{J_2} \widehat{\nabla}^{J_3} \alpha^F \widehat{\nabla}^{J_4} \mathcal{G}}_{C_1} + \sum_{J_1+J_2+J_3=I} \nabla^{J_1}(\eta+\underline{\eta})^{J_2} \widehat{\nabla}^{J_3} \mathcal{F}_1$$

$$+ \sum_{J_1+J_2+J_3+J_4=I} \nabla^{J_1}(\eta+\underline{\eta})^{J_2} \widehat{\nabla}^{J_3} \chi \widehat{\nabla}^{J_4} \mathcal{G}. \tag{4.1}$$

Similarly, for $\widehat{\nabla}_3 \mathcal{G} = \mathcal{F}_2$, and $\widehat{\nabla}_3 \widehat{\nabla}^I \mathcal{G} = \mathcal{F}^I_2$,

$$\mathcal{F}_2^I \sim \sum_{J_1+J_2+J_3+J_4=I-1} \nabla^{J_1}(\eta+\underline{\eta})^{J_2} \nabla^{J_3} \underline{\beta} \widehat{\nabla}^{J_4} \mathcal{G}$$

$$+ \sum_{J_1+J_2+J_3+J_4+J_5=I-1} \nabla^{J_1}(\eta+\underline{\eta})^{J_2} \widehat{\nabla}^{J_3} \underline{\alpha}^F \widehat{\nabla}^{J_4}(\rho^F, \sigma^F) \widehat{\nabla}^{J_5} \mathcal{G} \quad (4.2)$$

$$+ \underbrace{\sum_{J_1+J_2+J_3+J_4=I-1} \nabla^{J_1}(\eta+\underline{\eta})^{J_2} \widehat{\nabla}^{J_3} \underline{\alpha}^F \widehat{\nabla}^{J_4} \mathcal{G}}_{C_2} + \sum_{J_1+J_2+J_3=I} \nabla^{J_1}(\eta+\underline{\eta})^{J_2} \widehat{\nabla}^{J_3} \mathcal{F}_2$$

$$+ \sum_{J_1+J_2+J_3+J_4=I} \nabla^{J_1}(\eta+\underline{\eta})^{J_2} \widehat{\nabla}^{J_3} \underline{\chi} \widehat{\nabla}^{J_4} \mathcal{G}. \quad (4.2)$$

By moving the top derivatives of $\mathcal{G}$ multiplied by $tr\chi$ and $tr\underline{\chi}$ from the right hand side to the left hand side, one may also obtain

$$\mathcal{F}_1^I + \frac{I}{2} tr\chi \widehat{\nabla}^I \mathcal{G} \sim \sum_{J_1+J_2+J_3+J_4=I-1} \nabla^{J_1}(\eta+\underline{\eta})^{J_2} \nabla^{J_3} \beta \widehat{\nabla}^{J_4} \mathcal{G}$$

$$+ \sum_{J_1+J_2+J_3+J_4+J_5=I-1} \nabla^{J_1}(\eta+\underline{\eta})^{J_2} \widehat{\nabla}^{J_3} \alpha^F \widehat{\nabla}^{J_4}(\rho^F, \sigma^F) \widehat{\nabla}^{J_5} \mathcal{G} \quad (4.3)$$

$$+ \sum_{J_1+J_2+J_3+J_4=I-1} \nabla^{J_1}(\eta+\underline{\eta})^{J_2} \widehat{\nabla}^{J_3} \alpha^F \widehat{\nabla}^{J_4} \mathcal{G} + \sum_{J_1+J_2+J_3=I} \nabla^{J_1}(\eta+\underline{\eta})^{J_2} \widehat{\nabla}^{J_3} \mathcal{F}_1$$

$$+ \sum_{J_1+J_2+J_3+J_4=I} \nabla^{J_1}(\eta+\underline{\eta})^{J_2} \widehat{\nabla}^{J_3} \hat{\chi} \widehat{\nabla}^{J_4} \mathcal{G} + \sum_{J_1+J_2+J_3+J_4=I-1} \nabla^{J_1}(\eta+\underline{\eta})^{J_2+1} \widehat{\nabla}^{J_3} tr\chi \widehat{\nabla}^{J_4} \mathcal{G}. \quad (4.3)$$

$$\mathcal{F}_2^I + \frac{I}{2} tr\underline{\chi} \widehat{\nabla}^I \mathcal{G} \sim \sum_{J_1+J_2+J_3+J_4=I-1} \nabla^{J_1}(\eta+\underline{\eta})^{J_2} \nabla^{J_3} \underline{\beta}^R \widehat{\nabla}^{J_4} \mathcal{G}$$

$$+ \sum_{J_1+J_2+J_3+J_4=I-1} \nabla^{J_1}(\eta+\underline{\eta})^{J_2} \widehat{\nabla}^{J_3} \underline{\alpha}^F \widehat{\nabla}^{J_4} \mathcal{G} \quad (4.4)$$

$$+ \sum_{J_1+J_2+J_3=I} \nabla^{J_1}(\eta+\underline{\eta})^{J_2} \widehat{\nabla}^{J_3} \mathcal{F}_2 + \sum_{J_1+J_2+J_3+J_4=I} \nabla^{J_1}(\eta+\underline{\eta})^{J_2} \widehat{\nabla}^{J_3} \underline{\hat{\chi}} \widehat{\nabla}^{J_4} \mathcal{G}$$

$$+ \sum_{J_1+J_2+J_3+J_4=I-1} \nabla^{J_1}(\eta+\underline{\eta})^{J_2+1} \widehat{\nabla}^{J_3} tr\underline{\chi} \widehat{\nabla}^{J_4} \mathcal{G}.$$

This form of the commutation formulae will be very useful since we need to keep track of $tr\chi$ and $tr\underline{\chi}$ which behave like $2/r$ and $-2/r$ that are not integrable.

*Proof.* Recall the following gauge-covariant commutation rule

$$[\widehat{\nabla}_4, \widehat{\nabla}_B] \mathcal{G}^P_{QA_1A_2A_3\cdots A_n} = [\widehat{D}_4, \widehat{D}_B] \mathcal{G}^P_{QA_1A_2A_3\cdots A_n} + (\nabla_B \log \Omega) \widehat{\nabla}_4 \mathcal{G}^P_{QA_1A_2A_3\cdots A_n}$$

$$- \gamma^{CD} \chi_{BD} \widehat{\nabla}_C \mathcal{G}^P_{QA_1A_2A_3\cdots A_n} - \sum_{i=1}^n \gamma^{CD} \chi_{BD} \bar{\eta}_{A_i} \mathcal{G}^P_{QA_1A_2A_3\cdots \widehat{A}_i C \cdots A_n}$$

$$+ \sum_{i=1}^n \gamma^{CD} \chi_{A_i B} \bar{\eta}_D \mathcal{G}^P_{QA_1A_2A_3\cdots \hat{A}_i C \cdots A_n}$$

and

$$[\widehat{D}_4, \widehat{D}_A] \mathcal{G}^P_{QA_1A_2\cdots A_n} = (\nabla_A \log \Omega) \widehat{\nabla}_4 \mathcal{G}^P_{QA_1A_2\cdots A_n} - \sum_i R(e_C, e_{A_i}, e_4, e_A) \mathcal{G}^P_{QA_1\cdots \hat{A}_i,\cdots A_n}$$

$$+ F^P_{R4A} \mathcal{G}^R_{QA_1A_2\cdots A_n} - F^R_{Q4A} \mathcal{G}^P_{RA_1A_2\cdots A_n}. \quad (4.5)$$

In schematic form the commutation formula reads

$$[\widehat{\nabla}_4, \widehat{\nabla}_B] \mathcal{G} \sim (\beta + \alpha^F \cdot (\rho^F + \sigma^F)) \mathcal{G} + \alpha^F \mathcal{G} + (\eta + \bar{\eta}) \hat{\nabla}_4 \mathcal{G} - \chi \hat{\nabla} \mathcal{G} + \chi \bar{\eta} \mathcal{G}. \quad (4.6)$$

Similarly for $[\widehat{\nabla}_3, \widehat{\nabla}_B]$, we have

$$[\widehat{\nabla}_3, \widehat{\nabla}_B] \mathcal{G} \sim (\bar{\beta} + \tilde{\alpha}^F \cdot (\rho^F + \sigma^F)) \mathcal{G} + \tilde{\alpha}^F \mathcal{G} + (\eta + \bar{\eta}) \hat{\nabla}_3 \mathcal{G} - \tilde{\chi} \nabla \mathcal{G} + \tilde{\chi} \eta \mathcal{G}. \quad (4.7)$$

The commutation formula for general $\widehat{\nabla}^I$ may be obtained through an iteration argument (see Ref. 28). □

*Remark 2.* Note that the commutation terms $C_1$ and $C_2$ are produced because of the non-abelian (and hence non-linear) characteristic of the Yang–Mills theory. For any section of the bundle $\otimes^k T^*S$, $C_1$ and $C_2$ drop out.

## V. NORMS

3 + 1 dimensional Minkowski space is equipped with a 15 dimensional conformal group. Out of these 15 Killing and conformal Killing vector fields, one may utilize the time-like ones to construct conserved positive definite energies for a theory that posses a stress-energy tensor. However, if we perturb the Minkowski space, the isometry is no longer present and one can not hope to construct the conserved energies. Nevertheless, since we are interested in the small data problem, one might hope that the perturbed space-time does have vector fields that are approximately Killing or conforml Killing in a suitable sense. In other words, appropriate norms of the deformation tensors (or the trace-free part of the deformation tensor) are sufficiently small. In the regime of such smallness, one may hope to utilize the usual energy argument to establish a stability result. With this hope, we define the approximate Killing and conformal vector field in the same forms as their Minkowski ones

$$\hat{\mathbf{n}} := \frac{1}{2}(e_4 + e_3), \quad K = \frac{1}{2}(\underline{u}^2 e_4 + u^2 e_3), \quad S = \frac{1}{2}(\underline{u} e_4 + u e_3). \tag{5.1}$$

On the Minkowski space, $\hat{\mathbf{n}}$, $K$, and $S$ are the generators of time translation, inversion, and scaling vector field, respectively. In this current context, we endow them with the characteristics of "approximate" Killing and conformal Killing vector fields. We need to construct suitable scaled norms for the Weyl curvature components as well as that of Yang–Mills curvature components. Since the Weyl curvature scales as length$^{-2}$ and Yang-Mills curvature scales as length$^{-1}$, an appropriate norm would scale as length$^2$ ∗ weyl + length ∗ Yang – Mills. This simple scaling essentially tells us that in order to establish the existence of a solution of the coupled Einstein–Yang–Mills theory, we need one extra regularity for the Yang–Mills curvature. In the current context, however, there are two length scales present $\underline{u}$ ($\sim r(u,\underline{u})$) and $u$ and it is crucial to separate out these length scales since fields do not decay in an isotropic way. In order to determine the scaling behavior of different Weyl and Yang–Mills curvature components, we contract the respective stress tensors with the generator of approximate inversion and that of approximate time translations. The connection norms follow from the null transport equations (while estimating energies for the curvature we need the connection norms that can be estimated *a priori* by means of the null evolution and constraint equations).

### A. Curvature norms

The "true" gravitational degrees of freedom are encoded in the Weyl curvature of the spacetime metric. Therefore a natural expectation that the total energy of the Einstein–Yang–Mill system associated with characteristics may be described in terms of suitable norms of the Weyl and Yang–Mills curvature (notice that one does not have the concept of local energy density associated with pure gravity due to equivalence principle as such one may define a quasi-local notion[3,4]). One may utilize the Bel-Robinson tensor associated with a Weyl field to construct suitable norms for the pure gravitational sector. Let us recall the Bel-Robinson tensor associated with the Weyl tensor

$$Q(W)_{\alpha\beta\gamma\delta} = W_{\alpha\rho\gamma\sigma} W_\beta{}^\sigma{}_\delta{}^\rho + {}^*W_{\alpha\rho\gamma\sigma} {}^*W_\beta{}^\sigma{}_\delta{}^\rho. \tag{5.2}$$

In addition, we will also have the Yang–Mills stress-energy tensor

$$\mathfrak{T}_{\alpha\beta} = F^a_{b\alpha\mu} F^b_{a\beta}{}^\mu + {}^*F^a_{b\alpha\mu} {}^*F^b_{a\beta}{}^\mu. \tag{5.3}$$

Here ${}^*W_{\alpha\beta\gamma\delta} := \frac{1}{2} \epsilon_{\alpha\beta}{}^{\mu\nu} W_{\mu\nu\gamma\delta}$ and ${}^*F_{\alpha\beta} := \frac{1}{2} \epsilon_{\alpha\beta}{}^{\mu\nu} F_{\mu\nu}$. We note ${}^*({}^*F,W) = -F,W$ and therefore $\langle F,F\rangle_{g,\delta} + \langle{}^*F, {}^*F\rangle_{g,\delta} = 0$. Let us define the Bel-Robinson energy associated with the exterior Cauchy hypersurface $\widehat{\Sigma}$

$$\mathcal{Q}_0^{in} := \int_{\widehat{\Sigma}_0} Q(W)(K, K, \hat{\mathbf{n}}, \hat{\mathbf{n}}) \tag{5.4}$$

At the level of curvature flux through the null hyper-surface, this would correspond to the following entities that we are primarily interested in controlling

$$\mathcal{F}_0^W = \int_H Q(W)(K, K, \hat{\mathbf{n}}, e_4) + \int_{\underline{H}} Q(W)(K, K, \hat{\mathbf{n}}, e_3). \tag{5.5}$$

For the Yang–Mills fields, we define the energy employing the Yang–Mills stress-energy tensor

$$\mathcal{T}_0^{in} = \int_{\widehat{\Sigma}_0} \mathfrak{T}[F](K, \hat{\mathbf{n}}). \tag{5.6}$$

and the associated flux

$$\mathcal{F}_0^F = \int_H \mathfrak{T}[F](K, e_4) + \int_{\underline{H}} \mathfrak{T}[F](K, e_3). \tag{5.7}$$

Now utilizing the relations below

$$Q(W)(e_4, e_4, e_4, e_4) = 2|\alpha|^2, \quad Q(W)(e_4, e_4, e_4, e_3) = 4|\beta(W)|^2, \tag{5.8}$$

$$Q(W)(e_4, e_4, e_3, e_3) = 4(\rho(\mathcal{W})^2 + \sigma(W)^2), \quad Q(W)(e_4, e_3, e_3, e_3) = 4|\underline{\beta}(W)|^2,$$

$$Q(W)(e_3, e_3, e_3, e_3) = 2|\underline{\alpha}(W)|^2, \tag{5.9}$$

we compute

$$Q(K, K, \hat{\mathbf{n}}, e_4) = \frac{1}{4}\underline{u}^4|\alpha|^2 + \frac{1}{2}u^4|\beta|^2 + \frac{(u^4 + 2u^2\underline{u}^2)}{2}|\beta|^2 + \frac{(u^4 + 2u^2\underline{u}^2)}{2}(\rho^2 + \sigma^2), \tag{5.10}$$

$$Q(K, K, \hat{\mathbf{n}}, e_3) = \frac{1}{4}u^4|\underline{\alpha}|^2 + \frac{1}{2}\underline{u}^4|\beta|^2 + \frac{\underline{u}^4 + 2\underline{u}^2 u^2}{2}(\rho^2 + \sigma^2) + \frac{u^4 + 2u^2\underline{u}^2}{2}|\underline{\beta}|^2, \tag{5.11}$$

$$\mathfrak{I}(\mathcal{F})(K, e_4) = \frac{1}{2}\underline{u}^2|\alpha^F|^2 + \frac{1}{2}u^2(|\rho^F|^2 + |\sigma^F|^2), \tag{5.12}$$

$$\mathfrak{I}(\mathcal{F})(K, e_3) = \frac{1}{2}u^2|\underline{\alpha}^F|^2 + \frac{1}{2}\underline{u}^2(|\rho^F|^2 + |\sigma^F|^2) \tag{5.13}$$

which automatically provides us with the scaling behavior. With the help of the scaling behavior (5.10)–(5.13) dictated by the approximate Killing fields, we define the appropriate norms of the Weyl and Yang–Mills curvature that need control. Notice that sphere derivative $\nabla$ (and its gauge covariant counterpart $\widehat{\nabla}$) and $e_4$ null derivative $\nabla_4$ (and its gauge covariant counterpart) scale as $r^{-1}$ while the null derivative $\nabla_3$ scales as $|u|^{-1}$.

$$\mathcal{W}_0 := \int_H r^4(|\alpha|^2 + |\beta|^2) + \int_H \tau_-^4|\underline{\beta}|^2 + \int_H \tau_-^2 r^2(\rho^2 + \sigma^2) \tag{5.14}$$

$$+ \int_{\underline{H}} (\tau_-^4|\underline{\alpha}|^2 + r^4|\beta|^2 + r^2\tau_-^2|\underline{\beta}|^2 + r^4|\rho|^2 + r^4|\sigma|^2),$$

$$\mathcal{W}_1 = \int_H r^4(|r\nabla\alpha|^2 + |r\nabla\beta|^2) + \int_H \tau_-^4|r\nabla\underline{\beta}|^2 + \int_H \tau_-^2 r^2(|r\nabla\rho|^2 + |r\nabla\sigma|^2)$$

$$+ \int_{\underline{H}} (\tau_-^4|r\nabla\underline{\alpha}|^2 + r^4|r\nabla\beta|^2 + r^2\tau_-^2|r\nabla\underline{\beta}|^2 + r^4|r\nabla\rho|^2 + r^4|r\nabla\sigma|^2),$$

$$+ \int_H r^4|rD_4\alpha|^2 + \int_{\underline{H}} \tau_-^4|\tau_- D_3\underline{\alpha}|^2 + \int_H r^6|D_4\beta|^2 + \int_{\underline{H}} \tau_-^6|D_3\underline{\beta}|^2$$

$$\mathcal{W}_2 = \int_H r^4(|r^2\nabla^2\alpha|^2 + |r^2\nabla^2\beta|^2) + \int_H \tau_-^4|r^2\nabla^2\underline{\beta}|^2 + \int_H \tau_-^2 r^2(|r^2\nabla^2\rho|^2 + |r^2\nabla^2\sigma|^2)$$

$$+ \int_{\underline{H}} (\tau_-^4|r^2\nabla^2\underline{\alpha}|^2 + r^4|r^2\nabla^2\beta|^2 + r^2\tau_-^2|r^2\nabla^2\underline{\beta}|^2 + r^4|r^2\nabla^2\rho|^2 + r^4|r^2\nabla^2\sigma|^2)$$

$$+ \int_H r^4|r^2\nabla D_4\alpha|^2 + \int_{\underline{H}} \tau_-^4|\tau_- r\nabla D_3\underline{\alpha}|^2 + \int_H r^8|\nabla D_4\beta|^2 + \int_{\underline{H}} u^6 r^2|\nabla D_3\underline{\beta}|^2$$

and

$$\mathcal{Y}_0 := \int_H (r^2|\alpha^F|^2 + \tau_-^2|\rho^F|^2 + \tau_-^2|\sigma^F|^2) + \int_{\underline{H}} (\tau_-^2|\underline{\alpha}^F|^2 + r^2|\rho^F|^2 + r^2|\sigma^F|^2), \tag{5.15}$$

$$\mathcal{Y}_{1,2,3} = \sum_{I=1}^{3} \int_H (r^2|r^I\widehat{\nabla}^I\alpha^F|^2 + \tau_-^2|r^I\widehat{\nabla}^I\rho^F|^2 + \tau_-^2|r^I\widehat{\nabla}^I\sigma^F|^2) \tag{5.16}$$

$$+ \int_{\underline{H}} (\tau_-^2|r^I\widehat{\nabla}^I\underline{\alpha}^F|^2 + r^2|r^I\widehat{\nabla}^I\rho^F|^2 + r^2|r^I\widehat{\nabla}^I\sigma^F|^2)$$

$$+ \int_H r^2|r^I\widehat{\nabla}^{I-1}\widehat{D}_4\alpha^F|^2 + \int_{\underline{H}} \tau_-^2|r^{I-1}\tau_-\widehat{\nabla}^{I-1}\widehat{D}_3\underline{\alpha}^F|^2 + \int_H r^6|\widehat{D}_4^2\alpha^F|^2 + \int_{\underline{H}} \tau_-^6|\widehat{D}_3^2\underline{\alpha}^F|^2$$

### B. Connection norms

Let us now define the norm of the spin connections. Once again we need to weight by suitable factors of $r$ and $\tau_-$ to respect the scaling

$$\mathcal{O}_{p,q}(u,\underline{u}) := \|r^{2+q-\frac{2}{p}}\nabla^q(tr\chi - \underline{tr\chi})\|_{L^p(S_{u,\underline{u}})} + \|r^{2+q-\frac{2}{p}}\nabla^q(\underline{tr\chi} - \underline{tr\chi})\|_{L^p(S_{u,\underline{u}})} \quad (5.17)$$
$$+ \|r^{2+q-\frac{2}{p}}\nabla^q(\eta,\underline{\eta})\|_{L^p(S_{u,\underline{u}})} + \|r^{2+q-\frac{2}{p}}\nabla^q\omega\|_{L^p(S_{u,\underline{u}})} + \|r^{1+q-\frac{2}{p}}\tau_-\nabla^q\underline{\omega}\|_{L^p(S_{u,\underline{u}})}$$
$$+ \|r^{2+q-\frac{2}{p}}\nabla^q\hat{\chi}\|_{L^p(S_{u,\underline{u}})} + \|r^{1+q-\frac{2}{p}}\tau_-\nabla^q\underline{\hat{\chi}}\|_{L^p(S_{u,\underline{u}})} + \|r^{3+q-\frac{2}{p}}\nabla^q\nabla_4\underline{\omega}\|_{L^p(S_{u,\underline{u}})}$$
$$+ \|r^{1+q-\frac{2}{p}}\nabla^q\nabla_3\underline{\omega}\|_{L^p(S_{u,\underline{u}})}$$

and we also need the following top-order null norm (estimated through elliptic analysis)

$$\mathcal{O}_{\underline{H}} := r^{\frac{1}{2}}(u,\underline{u})\|r^3\nabla^3\hat{\chi}\|_{L^2(\underline{H})} + r^{\frac{1}{2}}(u,\underline{u})\|r^3\nabla^3 tr\chi\|_{L^2(\underline{H})} r^{\frac{1}{2}}(u,\underline{u})\|r^3\nabla^3\eta\|_{L^2(\underline{H})} \quad (5.18)$$
$$+ r^{\frac{1}{2}}(u,\underline{u})\|r^3\nabla^3\underline{\eta}\|_{L^2(\underline{H})} + r^{\frac{1}{2}}(u,\underline{u})\|r^3\nabla^3\underline{\hat{\chi}}\|_{L^2(\underline{H})} + r^{\frac{1}{2}}(u,\underline{u})\|r^3\nabla^3 \underline{tr\chi}\|_{L^2(\underline{H})}$$
$$+ r^{\frac{1}{2}}(u,\underline{u})\|r^3\nabla^3\omega\|_{L^2(\underline{H})} + r^{\frac{1}{2}}(u,\underline{u})\|r^3\nabla^3\underline{\omega}\|_{L^2(\underline{H})}.$$

The total connection norm up to the second derivative is defined as follows

$$\mathcal{O} := \sup_{u,\underline{u}} \mathcal{O}_{p,q} + \sup_{D_{u,\underline{u}}} O_{\underline{H}} \quad (5.19)$$

We are interested in $0 \leq q \leq 2$ and $2 \leq p \leq 4$. Notice that the point-wise norms of the connection coefficients are controlled by $\mathcal{O}_{p,q}(u,\underline{u})$ employing the Sobolev inequalities on the topological two–sphere.

*Remark 3.* Bootstrap Assumption: Assume that $\mathcal{O} \leq \epsilon$ and $\mathbf{W} + \mathbf{Y} := \mathcal{W}_0 + \mathcal{W}_1 + \mathcal{W}_2 + \mathcal{Y}_0 + \mathcal{Y}_1 + \mathcal{Y}_2 + \mathcal{Y}_3 \leq \epsilon$ *for a sufficiently small* $\epsilon > 0$.

## VI. NORM ESTIMATES USING THE EQUATIONS OF MOTIONS

In this section, we want to estimate the connection coefficients in terms of the Weyl and Yang–Mills curvatures utilizing the available null-transport and constraint equations. Null transport equations enable us to make use of the weighted transport inequalities whereas the constraint equations yield elliptic estimates. The technology of such coupled estimates was described in full detail by Ref. 11 for the vacuum case. In the current context, we have the additional Yang-Mills terms. Since the procedure is similar to Ref. 11, we only present estimating one connection component in detail. In addition, at the level of connection estimates, the true nonlinear characteristics of the Yang–Mills theory (the feature that separates it from the usual $U(1)$–Maxwell theory) do not manifest themselves in a crucial way (contrary to the energy estimates for the Yang–Mills curvature where it does).

### A. Connection estimates

*Proposition 6.1.* Assume that the boot-strap assumption (Remark 3) holds. The Ricci coefficient $\hat{\chi}$ verifies the following estimate

$$\|r^{3-\frac{2}{p}}\nabla\hat{\chi}\|_{L^p(S_{u,\underline{u}})} \lesssim (\mathcal{I}_0 + \mathcal{W} + \mathcal{Y}^2), \quad \|r^{2-\frac{2}{p}}\hat{\chi}\|_{L^p(S_{u,\underline{u}})} \lesssim (\mathcal{I}_0 + \mathcal{W} + \mathcal{Y}^2), \quad (6.1)$$

*for* $p \in [2,4]$, *where* $\mathcal{I}_0$ *is the connection norm on a leaf of the initial slice* $\widehat{\Sigma}$ *i.e., the initial norm of the connection coefficients.*

*Proof.* Following Ref. 11, we introduce the tensor $U := \Omega^{-1}\nabla tr\chi + \Omega^{-1} tr\chi \zeta$ and write down its evolution equation

$$\nabla_4 U + \frac{3}{2} tr\chi U = \mathcal{F}, \quad (6.2)$$

where $\mathcal{F}$ is obtained through the use of the null-transport equation after commuting with $\nabla$. To obtain this evolution equation, first we need to obtain an expression for $\nabla_4 \nabla tr\chi$. Recall the equation for $tr\chi$

$$\nabla_4 tr\chi + \frac{1}{2}(tr\chi)^2 = -|\hat{\chi}|^2 - 2\omega tr\chi - \mathfrak{T}_{44}, \quad (6.3)$$

where $\mathfrak{T}_{44} = \alpha^F \cdot \alpha^F$. Explicit computation yields

$$\nabla_4 \nabla tr\chi = -tr\chi \nabla tr\chi - 2\hat{\chi}\nabla\hat{\chi} - 2\nabla\omega tr\chi - 2\omega\nabla tr\chi - 2\alpha^F \cdot \widehat{\nabla}\alpha^F + [\nabla_4, \nabla]tr\chi,$$

where we have utilized the fact that $\mathfrak{T}_{44}$ is a gauge-invariant object together with the metric compatibility of the gauge covariant projected connection $\widehat{\nabla}$. Now compute $[\nabla, \nabla_4]tr\chi$

$$[\nabla, \nabla_4]tr\chi = \chi\nabla tr\chi - (\zeta + \eta)\nabla_4 tr\chi = \chi\nabla tr\chi - (\zeta + \eta)\left(-\frac{1}{2}(tr\chi)^2 - |\hat{\chi}|^2_\gamma - 2\omega tr\chi - \alpha^F \cdot \alpha^F\right).$$

Explicit calculations yield

$$\nabla_4 U = 2\Omega^{-1}\omega\nabla tr\chi + \Omega^{-1}\nabla_4\nabla tr\chi + 2\Omega^{-1}\omega tr\chi\zeta + \Omega^{-1}\zeta\nabla_4 tr\chi + \Omega^{-1}tr\chi\nabla_4\zeta \quad (6.4)$$

which upon substitution of the transport equation for $\zeta$

$$\nabla_4\zeta = 2\nabla\omega + \chi(\underline{\eta} - \zeta) + 2\omega(\zeta + \underline{\eta}) - \beta - \sigma^F \epsilon \cdot \alpha^F - \rho^F \cdot \alpha^F \quad (6.5)$$

yields

$$\mathcal{F} \sim -\hat{\chi}U - \Omega^{-1}\hat{\chi}\nabla\hat{\chi} - \Omega^{-1}\eta|\hat{\chi}|^2 + \Omega^{-1}tr\chi\hat{\chi}\underline{\eta} - \Omega^{-1}tr\chi\beta + \Omega^{-1}\alpha^F \cdot \widehat{\nabla}\alpha^F \\ + \Omega^{-1}tr\chi(\sigma^F, \rho^F) \cdot \underline{\alpha}^F + \Omega^{-1}(\zeta + \eta)\alpha^F \cdot \alpha^F \quad (6.6)$$

An application of the transport inequality yields

$$\|r^{3-\frac{2}{p}}U\|_{L^p(S_{u,\underline{u}})} \lesssim \|r^{3-\frac{2}{p}}U\|_{L^p(S_{u,\underline{u}_0})} + \int_{\underline{u}_0}^{\underline{u}}\|r^{3-\frac{2}{p}}\Omega\mathcal{F}\|_{L^p(S_{u,\underline{u}'})}d\underline{u}'. \quad (6.7)$$

Now we explicitly estimate each term of the error term $\mathcal{F}$

$$\|r^{3-\frac{2}{p}}\Omega\mathcal{F}\|_{L^p(S_{u,\underline{u}'})} \lesssim \|r^{3-\frac{2}{p}}\Omega\hat{\chi}U\|_{L^p(S_{u,\underline{u}'})} + \|r^{3-\frac{2}{p}}\hat{\chi}\nabla\hat{\chi}\|_{L^p(S_{u,\underline{u}'})} + \|r^{3-\frac{2}{p}}\eta|\hat{\chi}|^2\|_{L^p(S_{u,\underline{u}'})} \\ + \|r^{3-\frac{2}{p}}\underline{\eta} \cdot \hat{\chi}tr\chi\|_{L^p(S_{u,\underline{u}'})} + \|r^{3-\frac{2}{p}}tr\chi\beta\|_{L^p(S_{u,\underline{u}'})} + \|r^{3-\frac{2}{p}}\alpha^F \cdot \widehat{\nabla}\alpha^F\|_{L^p(S_{u,\underline{u}'})} \\ + \|r^{3-\frac{2}{p}}tr\chi(\rho^F, \sigma^F) \cdot \alpha^F\|_{L^p(S_{u,\underline{u}'})} + \|r^{3-\frac{2}{p}}(\zeta + \eta)\alpha^F \cdot \alpha^F\|_{L^p(S_{u,\underline{u}'})}. \quad (6.8)$$

We can estimate each term separately

$$\|r^{3-\frac{2}{p}}\Omega\hat{\chi}U\|_{L^p(S_{u,\underline{u}'})} \lesssim \epsilon\frac{1}{r^2}\|r^{3-\frac{2}{p}}U\|_{L^p(S_{u,\underline{u}'})}, \quad \|r^{3-\frac{2}{p}}\hat{\chi}\nabla\hat{\chi}\|_{L^p(S_{u,\underline{u}'})} \lesssim \epsilon\frac{1}{r^2}\|r^{3-\frac{2}{p}}\nabla\hat{\chi}\|_{L^p(S_{u,\underline{u}'})},$$

$$\|r^{3-\frac{2}{p}}\eta|\hat{\chi}|^2\|_{L^p(S_{u,\underline{u}'})} \lesssim \epsilon^3\frac{1}{r^3}, \quad \|r^{3-\frac{2}{p}}\underline{\eta}\cdot\hat{\chi}tr\chi\|_{L^p(S_{u,\underline{u}'})} \lesssim \epsilon^2\frac{1}{r^2}, \quad \|r^{3-\frac{2}{p}}tr\chi\beta\|_{L^p(S_{u,\underline{u}'})} \lesssim \mathcal{W}_0\frac{1}{r^{\frac{3}{2}}},$$

$$\|r^{3-\frac{2}{p}}\alpha^F \cdot \widehat{\nabla}\alpha^F\|_{L^p(S_{u,\underline{u}'})} \lesssim \mathcal{Y}^2\frac{1}{r^3}, \quad \|r^{3-\frac{2}{p}}tr\chi(\rho^F, \sigma^F) \cdot \alpha^F\|_{L^p(S_{u,\underline{u}'})} \lesssim \mathcal{Y}^2\frac{|u|^{-\frac{1}{2}}}{r^{\frac{5}{2}}},$$

$$\|r^{3-\frac{2}{p}}(\zeta + \eta)\alpha^F \cdot \alpha^F\|_{L^p(S_{u,\underline{u}'})} \lesssim \epsilon\mathcal{Y}^2\frac{1}{r^4}. \quad (6.9)$$

Collecting all the terms together we obtain

$$\|r^{3-\frac{2}{p}}U\|_{L^p(S_{u,\underline{u}})} \lesssim \|r^{3-\frac{2}{p}}U\|_{L^p(S_{u,\underline{u}_0})} + \int_{\underline{u}_0}^{\underline{u}}\left(\epsilon\frac{1}{r^2}\|r^{3-\frac{2}{p}}U\|_{L^p(S_{u,\underline{u}'})} + \epsilon\frac{1}{r^2}\|r^{3-\frac{2}{p}}\nabla\hat{\chi}\|_{L^p(S_{u,\underline{u}'})}\right. \\ \left. + \epsilon^3\frac{1}{r^3} + \epsilon^2\frac{1}{r^2} + \mathcal{W}\frac{1}{r^{\frac{3}{2}}} + \mathcal{Y}^2\frac{1}{r^3} + \mathcal{Y}^2\frac{|u|^{-\frac{1}{2}}}{r^{\frac{5}{2}}} + \epsilon\mathcal{Y}^2\frac{1}{r^4}\right)d\underline{u}' \\ \lesssim \mathcal{I}_0 + (\mathcal{W} + \mathcal{Y}^2) + \epsilon\int_{\underline{u}_0}^{\underline{u}}\frac{1}{r^2}\|r^{3-\frac{2}{p}}U\|_{L^p(S_{u,\underline{u}'})}d\underline{u}' + \int_{\underline{u}_0}^{\underline{u}}\frac{1}{r^2}\|r^{3-\frac{2}{p}}\nabla\hat{\chi}\|_{L^p(S_{u,\underline{u}'})}d\underline{u}'.$$

Now in order to estimate $\|r^{3-\frac{2}{p}}\nabla\hat{\chi}\|_{L^p(S_{u,\underline{u}'})}$, we invoke the following constraint equation

$$\text{div}\hat{\chi} = \frac{1}{2}\nabla tr\chi - \frac{1}{2}(\eta - \underline{\eta}) \cdot \left(\hat{\chi} - \frac{1}{2}tr\chi\gamma\right) - \beta + \frac{1}{2}\mathfrak{I}(e_4, \cdot)$$

$$= -\zeta\hat{\chi} + \frac{1}{2}\Omega U - \beta + \frac{1}{2}\alpha^F \cdot (\rho^F + \sigma^F). \tag{6.10}$$

An application of an elliptic estimate yields

$$\|r^{3-\frac{2}{p}}\nabla\hat{\chi}\|_{L^p(S_{u,\underline{u}'})} \lesssim \|r^{3-\frac{2}{p}}\Omega U\|_{L^p(S_{u,\underline{u}'})} + \|r^{3-\frac{2}{p}}(\eta,\underline{\eta})\hat{\chi}\|_{L^p(S_{u,\underline{u}'})} + \|r^{3-\frac{2}{p}}\beta\|_{L^p(S_{u,\underline{u}'})} \tag{6.11}$$

$$+ \|r^{3-\frac{2}{p}}\alpha^F \cdot (\rho^F, \sigma^F)\|_{L^p(S_{u,\underline{u}'})}$$

$$\lesssim \|r^{3-\frac{2}{p}}U\|_{L^p(S_{u,\underline{u}'})} + \frac{\epsilon^2}{r} + \frac{\mathcal{W}}{r^{\frac{1}{2}}} + \frac{\mathcal{Y}^2}{|u|^{\frac{1}{2}}r^{\frac{3}{2}}}.$$

Therefore

$$\|r^{3-\frac{2}{p}}U\|_{L^p(S_{u,\underline{u}})} \lesssim \mathcal{I}_0 + \mathcal{W} + \mathcal{Y}^2 + \epsilon\int_{\underline{u}_0}^{\underline{u}} \frac{1}{r^2}\|r^{3-\frac{2}{p}}U\|_{L^p(S_{u,\underline{u}'})}d\underline{u}' \tag{6.12}$$

and an application of Grönwall's inequality yields

$$\|r^{3-\frac{2}{p}}U\|_{L^p(S_{u,\underline{u}})} \lesssim \mathcal{I}_0 + \mathcal{W} + \mathcal{Y}^2 \tag{6.13}$$

and therefore

$$\|r^{3-\frac{2}{p}}\nabla\hat{\chi}\|_{L^p(S_{u,\underline{u}})} \lesssim \mathcal{I}_0 + \mathcal{W} + \mathcal{Y}^2 \tag{6.14}$$

as well as

$$\|r^{2-\frac{2}{p}}\hat{\chi}\|_{L^p(S_{u,\underline{u}})} \lesssim \mathcal{I}_0 + \mathcal{W} + \mathcal{Y}^2. \tag{6.15}$$

This concludes the proof. □

*Remark 4.* Notice that the Yang–Mills source terms appear as (gauge-invariant) components of the stress-energy tensor that is quadratic in the Yang–Mills field strength. In addition, note that the evolution equation for $tr\chi$ does not contain a Weyl curvature component which allows us to obtain estimates without losing a derivative of Weyl curvature. Since Yang–Mills field strength has one higher order of differentiability, we could close the argument.

*Proposition 6.2.* Assume that the boot-strap assumption (Remark 3) holds. The Ricci coefficient $\underline{\hat{\chi}}$ verifies the following estimate

$$\|r^{2-\frac{2}{p}}|u|\nabla\underline{\hat{\chi}}\|_{L^p(S_{u,\underline{u}})} \lesssim \mathcal{I}_0 + \mathcal{W} + \mathcal{Y}^2, \quad \|r^{1-\frac{2}{p}}|u|\underline{\hat{\chi}}\|_{L^p(S_{u,\underline{u}})} \lesssim \mathcal{I}_0 + \mathcal{W} + \mathcal{Y}^2 \tag{6.16}$$

*for* $p \in [2, 4]$.

*Proof.* The proof follows exactly the same way as that of $\hat{\chi}$. Construct a new entity $\underline{U} := \Omega^{-1}\nabla tr\underline{\chi} - \Omega^{-1}tr\underline{\chi}\zeta$ and apply the transport inequality in the $e_3$ direction. This together with the elliptic estimate from the equation

$$\text{div}\underline{\hat{\chi}} = \frac{1}{2}\nabla tr\underline{\chi} - \frac{1}{2}(\underline{\eta} - \eta) \cdot \left(\underline{\hat{\chi}} - \frac{1}{2}tr\underline{\chi}\gamma\right) - \underline{\beta} + \frac{1}{2}\mathfrak{I}(e_3, \cdot) \tag{6.17}$$

yields the result. Here we encounter the Yang–Mills source terms of the following type

$$T_1 = \int_{u_0}^{u} \|r^{3-\frac{2}{p}}\underline{\alpha}^F \cdot \widehat{\nabla}\underline{\alpha}^F\|_{L^p(S_{u',\underline{u}})}du' \lesssim \mathcal{Y}^2 \int_{u_0}^{u} |u'|^{-3}du' \lesssim \mathcal{Y}^2, \tag{6.18}$$

$$T_2 = \int_{u_0}^{u} \|r^{3-\frac{2}{p}}tr\underline{\chi}(\rho^F, \sigma^F) \cdot \underline{\alpha}^F\|_{L^p(S_{u',\underline{u}})}du' \lesssim \mathcal{Y}^2 \int_{u_0}^{u} r^{-1}|u'|^{-2}du' \lesssim \mathcal{Y}^2, \tag{6.19}$$

$$T_3 = \int_{u_0}^{u} \|r^{3-\frac{2}{p}}(\zeta - \underline{\eta})\underline{\alpha}^F \cdot \underline{\alpha}^F\|_{L^p(S_{u',\underline{u}})}du' \lesssim \epsilon\mathcal{Y}^2 \int_{u_0}^{u} r^{-1}|u'|^{-3}du' \lesssim \mathcal{Y}^2, \tag{6.20}$$

$$T_4 = \int_{u_0}^{u} \|r^{3-\frac{2}{p}}\underline{\alpha}^F \cdot (\rho^F, \sigma^F)\|_{L^p(S_{u',\underline{u}})}du' \lesssim \mathcal{Y}^2 \int_{u_0}^{u} r^{-1}|u'|^{-1}du' \lesssim \mathcal{Y}^2. \tag{6.21}$$

Here the Yang–Mills source terms are controlled through global Sobolev inequalities on null hypersurfaces. Collecting all the terms and applications of Grönwall's inequality yields the result. □

*Proposition 6.3.* Assume that the boot-strap assumption (Remark 3) holds. The Ricci coefficients $\eta, \underline{\eta}$ verify the following estimate

$$\|r^{3-\frac{2}{p}}\nabla\eta\|_{L^p(S_{u,\underline{u}})} \lesssim \mathcal{I}_0 + \mathcal{W} + \mathcal{Y}^2, \|r^{2-\frac{2}{p}}\eta\|_{L^p(S_{u,\underline{u}})} \lesssim \mathcal{I}_0 + \mathcal{W} + \mathcal{Y}^2, \tag{6.22}$$

$$\|r^{3-\frac{2}{p}}\nabla\underline{\eta}\|_{L^p(S_{u,\underline{u}})} \lesssim \mathcal{I}_0 + \mathcal{W} + \mathcal{Y}^2, \|r^{2-\frac{2}{p}}\underline{\eta}\|_{L^p(S_{u,\underline{u}})} \lesssim \mathcal{I}_0 + \mathcal{W} + \mathcal{Y}^2 \tag{6.23}$$

*Proof.* The proof is obtained by introducing the mass aspect functions[11]

$$\mu = -\text{div}\,\eta + \frac{1}{2}\hat{\chi}\cdot\underline{\hat{\chi}} - \rho, \quad \underline{\mu} = -\text{div}\,\underline{\eta} + \frac{1}{2}\underline{\hat{\chi}}\cdot\hat{\chi} - \rho \tag{6.24}$$

together with the Hodge system

$$\text{div}\,\eta = -\mu + \frac{1}{2}\hat{\chi}\cdot\underline{\hat{\chi}} - \rho, \text{curl}\,\eta = \hat{\chi}\wedge\underline{\hat{\chi}} + \sigma\epsilon, \tag{6.25}$$

$$\text{div}\,\underline{\eta} = -\underline{\mu} + \frac{1}{2}\underline{\hat{\chi}}\cdot\hat{\chi} - \rho, \text{curl}\,\underline{\eta} = \underline{\hat{\chi}}\wedge\hat{\chi} - \sigma\epsilon. \tag{6.26}$$

These mass aspect functions are introduced to avoid he loss of derivatives of the Weyl curvature components $\beta$ and $\underline{\beta}$ since the null evolution equations for $\eta$ and $\underline{\eta}$ contain $\beta$ and $\underline{\beta}$, respectively. Therefore, a naïve application of the transport inequality on $\nabla_4\nabla\eta$ ($\nabla_3\nabla\underline{\eta}$) would cause a regularity non-closure problem. In our previous article[21] of semi-global estimates for the characteristic initial value problem associated with the Einstein–Yang–Mills equations, we have utilized slightly different mass aspect functions ($\mu = -\text{div}\,\eta - \rho, \underline{\mu} = -\text{div}\,\underline{\eta} - \rho$). Here once again, we focus on controlling the Yang–Mills coupling terms that appear since the remaining procedure is exactly similar to Ref. 11. The Yang–Mills terms that appear in the evolution equation for $\mu$ are as follows

$$YM \sim \alpha^F \cdot \widehat{\nabla}(\rho^F, \sigma^F) + \widehat{\nabla}\alpha^F \cdot (\rho^F, \sigma^F) + tr\underline{\chi}|\alpha^F|^2 + \underline{\omega}|\alpha^F|^2 + \eta(\alpha^F \cdot \sigma^F + \alpha^F \cdot \rho^F) \tag{6.27}$$
$$+ tr\chi\rho^F \cdot \sigma^F + (\eta,\underline{\eta})\rho^F \cdot \sigma^F + \hat{\chi}\alpha \cdot \underline{\alpha}^F.$$

Following the method of Ref. 11, the norms that we need to control are the following

$$\int_{\underline{u}_0}^{\underline{u}} \|r^{2-\frac{2}{p}}\alpha^F\cdot\widehat{\nabla}(\rho^F,\sigma^F)\|_{L^p(S_{u,\underline{u}'})}d\underline{u}' \lesssim \mathcal{Y}^2\int_{\underline{u}_0}^{\underline{u}}|u|^{-\frac{1}{2}}r^{-\frac{7}{2}}d\underline{u}' \lesssim \mathcal{Y}^2\frac{1}{r^{\frac{5}{2}}}, \tag{6.28}$$

$$\int_{\underline{u}_0}^{\underline{u}} \|r^{2-\frac{2}{p}}\widehat{\nabla}\alpha^F\cdot(\rho^F,\sigma^F)\|_{L^p(S_{u,\underline{u}'})}d\underline{u}' \lesssim \mathcal{Y}^2\int_{\underline{u}_0}^{\underline{u}}|u|^{-\frac{1}{2}}r^{-\frac{7}{2}}d\underline{u}' \lesssim \mathcal{Y}^2\frac{1}{r^{\frac{5}{2}}}, \tag{6.29}$$

$$\int_{\underline{u}_0}^{\underline{u}} \|r^{2-\frac{2}{p}}tr\underline{\chi}|\alpha^F|^2\|_{L^p(S_{u,\underline{u}'})}d\underline{u}' \lesssim \mathcal{Y}^2\int_{\underline{u}_0}^{\underline{u}}r^{-4}d\underline{u}' \lesssim \mathcal{Y}^2\frac{1}{r^3}, \tag{6.30}$$

$$\int_{\underline{u}_0}^{\underline{u}} \|r^{2-\frac{2}{p}}\underline{\omega}|\alpha^F|^2\|_{L^p(S_{u,\underline{u}'})}d\underline{u}' \lesssim \epsilon\mathcal{Y}^2\int_{\underline{u}_0}^{\underline{u}}|u|^{-1}r^{-4}d\underline{u}' \lesssim \epsilon\mathcal{Y}^2\frac{1}{r^3}, \tag{6.31}$$

$$\int_{\underline{u}_0}^{\underline{u}} \|r^{2-\frac{2}{p}}\eta\alpha^F\cdot(\rho^F,\sigma^F)\|_{L^p(S_{u,\underline{u}'})}d\underline{u}' \lesssim \epsilon\mathcal{Y}^2\int_{\underline{u}_0}^{\underline{u}}|u|^{-\frac{1}{2}}r^{-\frac{9}{2}}d\underline{u}' \lesssim \epsilon\mathcal{Y}^2\frac{1}{r^{\frac{7}{2}}}, \tag{6.32}$$

$$\int_{\underline{u}_0}^{\underline{u}} \|r^{2-\frac{2}{p}}tr\chi\rho^F\cdot\sigma^F\|_{L^p(S_{u,\underline{u}'})}d\underline{u}' \lesssim \epsilon\mathcal{Y}^2\int_{\underline{u}_0}^{\underline{u}}|u|^{-1}r^{-3}d\underline{u}' \lesssim \mathcal{Y}^2\frac{1}{r^2}, \tag{6.33}$$

$$\int_{\underline{u}_0}^{\underline{u}} \|r^{2-\frac{2}{p}}(\eta,\underline{\eta})\rho^F\cdot\sigma^F\|_{L^p(S_{u,\underline{u}'})}d\underline{u}' \lesssim \epsilon\mathcal{Y}^2\int_{\underline{u}_0}^{\underline{u}}|u|^{-1}r^{-4}d\underline{u}' \lesssim \epsilon\mathcal{Y}^2\frac{1}{r^3}, \tag{6.34}$$

$$\int_{\underline{u}_0}^{\underline{u}} \|r^{2-\frac{2}{p}}\hat{\chi}\alpha^F\cdot\underline{\alpha}^F\|_{L^p(S_{u,\underline{u}'})}d\underline{u}' \lesssim \epsilon\mathcal{Y}^2\int_{\underline{u}_0}^{\underline{u}}|u|^{-\frac{3}{2}}r^{-\frac{7}{2}}d\underline{u}' \lesssim \epsilon\mathcal{Y}^2\frac{1}{r^{\frac{5}{2}}}. \tag{6.35}$$

Collecting all the terms and following the steps of Ref. 11, one obtains the result. A similar procedure follows for $\underline{\eta}$ as well. □

The remaining connection estimates follow similarly. Once again, the only modification one needs to worry about is the Yang–Mills source terms.

*Proposition 6.4.* Assume that the boot-strap assumption (Remark 3) holds. The Ricci coefficients verify the following estimate

$$\|r^{3-\frac{2}{p}}\nabla tr\chi\|_{L^p(S_{u,\underline{u}})} \lesssim \mathcal{I}_0 + \mathcal{W} + \mathcal{Y}^2, \|r^{2-\frac{2}{p}}(tr\chi - \underline{tr\chi})\|_{L^p(S_{u,\underline{u}})} \lesssim \mathcal{I}_0 + \mathcal{W} + \mathcal{Y}^2, \tag{6.36}$$

$$\|r^{3-\frac{2}{p}}\nabla tr\underline{\chi}\|_{L^p(S_{u,\underline{u}})} \lesssim \mathcal{I}_0 + \mathcal{W} + \mathcal{Y}^2, \|r^{2-\frac{2}{p}}(tr\underline{\chi} - \underline{tr\underline{\chi}})\|_{L^p(S_{u,\underline{u}})} \lesssim \mathcal{I}_0 + \mathcal{W} + \mathcal{Y}^2, \tag{6.37}$$

$$\|r^{2-\frac{2}{p}}\omega\|_{L^p(S_{u,\underline{u}})} \lesssim \mathcal{I}_0 + \mathcal{W} + \mathcal{Y}^2, \quad \|r^{1-\frac{2}{p}}|u|\underline{\omega}\|_{L^p(S_{u,\underline{u}})} \lesssim \mathcal{I}_0 + \mathcal{W} + \mathcal{Y}^2, \tag{6.38}$$

$$\|r^{3-\frac{2}{p}}\nabla\omega\|_{L^p(S_{u,\underline{u}})} \lesssim \mathcal{I}_0 + \mathcal{W} + \mathcal{Y}^2, \quad \|r^{2-\frac{2}{p}}|u|\nabla\underline{\omega}\|_{L^p(S_{u,\underline{u}})} \lesssim \mathcal{I}_0 + \mathcal{W} + \mathcal{Y}^2. \tag{6.39}$$

*Proof.* Once again, we only sketch the estimates for the Yang–Mills coupling terms. The Yang–Mills coupling terms that appear while executing the procedure of Ref. 11 to obtain the relevant estimates are as follows

$$\int_{\underline{u}_0}^{\underline{u}} \|r^{3-\frac{2}{p}}\alpha^F \cdot \widehat{\nabla}\alpha^F\|_{L^p(S_{u,\underline{u}'})} d\underline{u}' \lesssim \mathcal{Y}^2 \int_{\underline{u}_0}^{\underline{u}} r^{-3} d\underline{u}' \lesssim \mathcal{Y}^2, \tag{6.40}$$

$$\int_{u_0}^{u} \|r^{3-\frac{2}{p}}\underline{\alpha}^F \cdot \widehat{\nabla}\underline{\alpha}^F\|_{L^p(S_{u',\underline{u}})} du' \lesssim \mathcal{Y}^2 \int_{u_0}^{u} |u'|^{-3} du' \lesssim \mathcal{Y}^2, \tag{6.41}$$

$$\int_{\underline{u}_0}^{\underline{u}} \|r^{-\frac{2}{p}}\rho^F \cdot \sigma^F\|_{L^p(S_{u,\underline{u}'})} d\underline{u}' \lesssim \mathcal{Y}^2 \int_{\underline{u}_0}^{\underline{u}} |u|^{-1} r^{-4} d\underline{u}' \lesssim \mathcal{Y}^2 \frac{1}{r^3}, \tag{6.42}$$

$$\int_{u_0}^{u} \|r^{-\frac{2}{p}}\rho^F \cdot \sigma^F\|_{L^p(S_{u',\underline{u}})} du' \lesssim \mathcal{Y}^2 \int_{u_0}^{u} |u|^{-1} r^{-4} du' \lesssim \mathcal{Y}^2 \frac{1}{r^2}, \tag{6.43}$$

$$\int_{\underline{u}_0}^{\underline{u}} \|r^{1-\frac{2}{p}}\widehat{\nabla}(\rho^F, \sigma^F) \cdot (\rho^F, \sigma^F)\|_{L^p(S_{u,\underline{u}'})} d\underline{u}' \lesssim \mathcal{Y}^2 \int_{\underline{u}_0}^{\underline{u}} |u|^{-1} r^{-4} d\underline{u}' \lesssim \mathcal{Y}^3 \frac{1}{r^3}, \tag{6.44}$$

$$\int_{u_0}^{u} \|r^{1-\frac{2}{p}}\widehat{\nabla}(\rho^F, \sigma^F) \cdot (\rho^F, \sigma^F)\|_{L^p(S_{u',\underline{u}})} du' \lesssim \mathcal{Y}^2 \int_{u_0}^{u} |u|^{-1} r^{-4} du' \lesssim \mathcal{Y}^2 \frac{1}{r^2} \tag{6.45}$$

□.

The second derivative estimates follow exactly in a similar way and we omit the proof. We refer the reader to Ref. 11 for the exact procedure.

*Proposition 6.5.* Assume that the boot-strap assumption (Remark 3) holds. The Ricci coefficients verify the following second derivative estimates

$$\|r^{4-\frac{2}{p}}\nabla^2\hat{\chi}\|_{L^p(S_{u,\underline{u}})} \lesssim (\mathcal{I}_0 + \mathcal{W}_0 + \mathcal{Y}^2), \quad \|r^{3-\frac{2}{p}}|u|\nabla^2\hat{\underline{\chi}}\|_{L^p(S_{u,\underline{u}})} \lesssim \mathcal{I}_0 + \mathcal{W} + \mathcal{Y}^2, \tag{6.46}$$

$$\|r^{4-\frac{2}{p}}\nabla^2\eta\|_{L^p(S_{u,\underline{u}})} \lesssim \mathcal{I}_0 + \mathcal{W} + \mathcal{Y}^2, \quad \|r^{4-\frac{2}{p}}\nabla^2\underline{\eta}\|_{L^p(S_{u,\underline{u}})} \lesssim \mathcal{I}_0 + \mathcal{W} + \mathcal{Y}^2, \tag{6.47}$$

$$\|r^{4-\frac{2}{p}}\nabla^2 tr\chi\|_{L^p(S_{u,\underline{u}})} \lesssim \mathcal{I}_0 + \mathcal{W} + \mathcal{Y}^2, \quad \|r^{4-\frac{2}{p}}\nabla^2 tr\underline{\chi}\|_{L^p(S_{u,\underline{u}})} \lesssim \mathcal{I}_0 + \mathcal{W} + \mathcal{Y}^2, \tag{6.48}$$

$$\|r^{4-\frac{2}{p}}\nabla^2\omega\|_{L^p(S_{u,\underline{u}})} \lesssim \mathcal{I}_0 + \mathcal{I}_0 + \mathcal{W} + \mathcal{Y}^2, \quad \|r^{3-\frac{2}{p}}|u|\nabla^2\underline{\omega}\|_{L^p(S_{u,\underline{u}})} \lesssim \mathcal{I}_0 + \mathcal{W} + \mathcal{Y}^2, \tag{6.49}$$

$$\|r^{4-\frac{2}{p}}\nabla\nabla_4\omega\|_{L^p(S_{u,\underline{u}})} \lesssim \mathcal{I}_0 + \mathcal{W} + \mathcal{Y}^2, \quad \|r^{2-\frac{2}{p}}|u|^2\nabla\nabla_3\underline{\omega}\|_{L^p(S_{u,\underline{u}})} \lesssim \mathcal{I}_0 + \mathcal{W} + \mathcal{Y}^2, \tag{6.50}$$

$$\|r^{5-\frac{2}{p}}\nabla^2\nabla_4\omega\|_{L^p(S_{u,\underline{u}})} \lesssim \mathcal{I}_0 + \mathcal{W} + \mathcal{Y}^2, \quad \|r^{3-\frac{2}{p}}|u|^2\nabla^2\nabla_3\underline{\omega}\|_{L^p(S_{u,\underline{u}})} \lesssim \mathcal{I}_0 + \mathcal{W} + \mathcal{Y}^2 \tag{6.51}$$

*Proof.* The proof of this proposition is essentially the same as the first derivative estimates. This is nothing but a consequence of scaling.

□

*Proposition 6.6.* The following third derivative estimates hold for the connection coefficients under the boot-strap assumption in Remark 3

$$r^{\frac{1}{2}}(u,\underline{u})\|r^3\nabla^3\hat{\chi}\|_{L^2(\underline{H})} \lesssim \mathcal{I}_0 + \mathcal{W} + \mathcal{Y}^2, \quad r^{\frac{1}{2}}(u,\underline{u})\|r^3\nabla^3 tr\chi\|_{L^2(\underline{H})} \lesssim \mathcal{I}_0 + \mathcal{W} + \mathcal{Y}^2,$$

$$r^{\frac{1}{2}}(u,\underline{u})\|r^3\nabla^3\eta\|_{L^2(\underline{H})} \lesssim \mathcal{I}_0 + \mathcal{W} + \mathcal{Y}^2, \quad r^{\frac{1}{2}}(u,\underline{u})\|r^3\nabla^3\underline{\eta}\|_{L^2(\underline{H})} \lesssim \mathcal{I}_0 + \mathcal{W} + \mathcal{Y}^2,$$

$$r^{\frac{1}{2}}(u,\underline{u})\|r^3\nabla^3\hat{\underline{\chi}}\|_{L^2(\underline{H})} \lesssim \mathcal{I}_0 + \mathcal{W} + \mathcal{Y}^2, \quad r^{\frac{1}{2}}(u,\underline{u})\|r^3\nabla^3 tr\underline{\chi}\|_{L^2(\underline{H})} \lesssim \mathcal{I}_0 + \mathcal{W} + \mathcal{Y}^2,$$

$$r^{\frac{1}{2}}(u,\underline{u})\|r^3\nabla^3\underline{\omega}\|_{L^2(\underline{H})} \lesssim \mathcal{I}_0 + \mathcal{W} + \mathcal{Y}^2. \tag{6.52}$$

*Proof.* The proof is essentially the same as presented in Ref. 11 for the vacuum case. Here we only estimate the Yang–Mills source terms. Note that the Riemann curvature components in Ref. 11 become Weyl curvature in the current context. See Ref. 21 for the Einstein–Yang–Mills case with the regularity level one lower than the present case. However, such an argument can be extended easily to the higher derivative case since the relative regularity is preserved (i.e., the regularity level of Yang–Mills curvature is one order higher than that of the Weyl curvature), and the addition of derivatives results in the modification of the scale factors in an appropriate way.  □

### B. Energy estimate for the Weyl and Yang–Mills curvature

The true non-linear effect of the Yang–Mills coupling terms appears at the level of energy estimates. Null Bianchi equations and the null Yang–Mills equations are manifestly hyperbolic. This is essentially a consequence of Einstein's and Yang–Mills equations being derivable from a Lagrangian. This can also be interpreted as the existence of the Bel-Robinson tensor for gravity and a canonical Yang–Mills stress-energy tensor (see Ref. 21). We can explicitly utilize this manifestly hyperbolic character to estimate the pure gravitational and Yang–Mills energy. The following integration lemma turns out to be crucial.

*Lemma 6.1.* Let $f$ be a real-valued scalar function on the spacetime that is invariant under the Yang–Mills gauge transformation. The following integration by parts identities holds for $f$.

$$\int_{D_{u\underline{u}}} \nabla_4 f = \int_{\underline{H}_u} f - \int_{\widetilde{\Sigma}_0} f + \int_{D_{u\underline{u}}} (2\omega - tr\chi) f \qquad (6.53)$$

and

$$\int_{D_{u\underline{u}}} \nabla_3 f = \int_{H_u} f - \int_{\widetilde{\Sigma}_0} f + \int_{D_{u\underline{u}}} (2\underline{\omega} - tr\underline{\chi}) f, \qquad (6.54)$$

*Proof.*

$$\int_{D_{u\underline{u}}} \nabla_4 f = \int_{u_0}^{u} du' \int_{\underline{u}_0(u)}^{\underline{u}} \left( \int_{S_{u'\underline{u}'}} \frac{\partial f}{\partial \underline{u}'} \Omega \mu_\gamma \right) d\underline{u}' \qquad (6.55)$$

$$= \int_{u_0}^{u} du' \int_{\underline{u}_0(u')}^{\underline{u}} \left\{ \frac{d}{d\underline{u}'} \int_{S_{u'\underline{u}'}} f \Omega \mu_\gamma - \int_{S_{u'\underline{u}'}} f \left( \frac{\partial \Omega}{\partial \underline{u}'} + \frac{\Omega}{2} tr\chi \partial_u \gamma \right) \mu_\gamma \right\} d\underline{u}'$$

$$= \int_{u_0}^{u} du' \int_{S_{u'\underline{u}}} f \Omega \mu_\gamma - \int_{u_0}^{u} \int_{S_{u'\underline{u}_0}} f \Omega \mu_\gamma + \int_{D_{u\underline{u}}} f (2\omega - tr\chi)$$

$$= \int_{\underline{H}_u} f - \int_{\widetilde{\Sigma}_0} f + \int_{D_{u\underline{u}}} (2\omega - tr\chi) f.$$

The other part follows similarly. □

*Remark 5.* Note that in the Lemma 6.1, $f$ is a scalar function as opposed to a Yang–Mills curvature component treated locally as a scalar from the point of view of the PDEs since the later would not be invariant under a gauge transformation.

Utilizing this integration lemma, we may proceed to obtain the energy estimates in the exterior domain. Let us now identify the Bianchi pairs: $(\alpha, \beta)$, $(\underline{\alpha}, \underline{\beta})$, $(\beta, \{\rho, \sigma\})$, $(\underline{\beta}, \{\rho, \sigma\})$, $(\alpha^F, \{\rho^F, \sigma^F\})$, $(\underline{\alpha}^F, \{\rho^F, \sigma^F\})$.

#### 1. Energy estimates for the gravity sector

Recall the expressions for the variations of the are along incoming and outgoing null directions

$$\frac{d}{du} \int \mu_\gamma = \int tr\underline{\chi} \Omega \mu_\gamma, \quad \frac{d}{d\underline{u}} \int \mu_\gamma = \int tr\chi \Omega \mu_\gamma \qquad (6.56)$$

and evaluate the variation of the areal radius $r(u, \underline{u}) := \sqrt{\frac{\text{Area}(S_{u,\underline{u}})}{4\pi}}$ as follows

$$\frac{dr}{du} = \frac{r}{2} \overline{\Omega tr\underline{\chi}}, \quad \frac{dr}{d\underline{u}} = \frac{r}{2} \overline{\Omega tr\chi}. \qquad (6.57)$$

*Lemma 6.2.* The following estimates hold true in the exterior region

$$\int_{H_u} r^4 |\alpha|^2 + \int_{\underline{H}_u} r^4 |\beta|^2 \lesssim \int_{\widetilde{\Sigma}_0} r^4 |\alpha|^2 + \int_{\widetilde{\Sigma}_0} r^4 |\beta|^2 + \epsilon(\mathbf{W} + \mathbf{Y}), \qquad (6.58)$$

$$\int_{H} r^6 |\nabla \alpha|^2 + \int_{\underline{H}} r^6 |\nabla \beta|^2 \leq \int_{\widetilde{\Sigma}_0} r^6 |\nabla \alpha|^2 + \int_{\widetilde{\Sigma}_0} r^6 |\nabla \beta|^2 + \epsilon(\mathbf{W} + \mathbf{Y}), \qquad (6.59)$$

$$\int_{H} r^8 |\nabla^2 \alpha|^2 + \int_{\underline{H}} r^8 |\nabla^2 \beta|^2 \leq \int_{\widetilde{\Sigma}_0} r^8 |\nabla^2 \alpha|^2 + \int_{\widetilde{\Sigma}_0} r^8 |\nabla^2 \beta|^2 + \epsilon(\mathbf{W} + \mathbf{Y}), \qquad (6.60)$$

*Proof.* Here we present the proof for the topmost derivatives. The lower-order estimates would then follow by scaling. First note that the following identity holds

$$\int_H r^8 |\nabla^2 \alpha|^2 + \int_{\underline{H}} r^8 |\nabla^2 \beta|^2 - \underbrace{\int_{D_{u,\underline{u}}} r^8 \left(4\Omega^{-1}\overline{\Omega tr\underline{\chi}} - 2tr\underline{\chi}\right)|\nabla^2 \alpha|^2}_{<0} \quad (6.61)$$

$$+ \underbrace{\int_{D_{u,\underline{u}}} r^8 (5tr\chi - 4\Omega^{-1}\overline{\Omega tr\chi})|\nabla^2 \beta|^2}_{>0} = \int_{\widehat{\Sigma}_0} r^8 |\nabla^2 \alpha|^2 + \int_{\widehat{\Sigma}_0} r^8 |\nabla^2 \beta|^2 + \mathcal{E}_{\nabla^2 \alpha, \nabla^2 \beta},$$

where the error term $\mathcal{E}_{\nabla^2 \alpha, \nabla^2 \beta}$ may be written in the following schematic form

$$\mathcal{E}_{\nabla^2 \alpha, \nabla^2 \beta} \sim \int_{D_{u,\underline{u}}} r^8 \langle \nabla^2 \alpha, \nabla^2 tr\underline{\chi}\alpha + \nabla tr\underline{\chi}\nabla \alpha \rangle + \int_{D_{u,\underline{u}}} \left[ 2r^8 \left\langle \nabla^2 \alpha, \nabla^2 \left(4\underline{\omega}\alpha - 3(\widehat{\chi}\rho + {}^*\widehat{\chi}\sigma) \right. \right. \right.$$

$$\left. \left. + (\zeta + 4\eta)\hat{\otimes}\beta + \frac{1}{2}(D_3 \mathfrak{T}_{44} - D_4 \mathfrak{T}_{43})\gamma_{ab} - D_4 \mathfrak{T}_{ab} + \frac{1}{2}(D_b \mathfrak{T}_{4a} + D_a \mathfrak{T}_{4b}) \right) \right\rangle$$

$$+ 2r^8 \langle \nabla^2 \alpha, -\widehat{\underline{\chi}} \cdot \nabla^2 \alpha - \nabla \widehat{\underline{\chi}} \cdot \nabla \alpha \rangle + 2r^8 \langle \nabla^2 \beta, -2\nabla^2 (tr\chi)\beta_a - 2\nabla(tr\chi)\nabla \beta_a - 2\nabla^2 (\omega \beta_a)$$

$$+ \nabla^2 (\eta \cdot \alpha)_a - \frac{1}{2}\nabla^2 (D_a \mathfrak{T}_{44} - D_4 \mathfrak{T}_{4a}) \right\rangle + 2r^8 \langle \nabla^2 \beta, -\widehat{\chi} \cdot \nabla^2 \beta - \nabla \widehat{\chi} \cdot \nabla \beta \rangle \right].$$

Before we continue with estimating the error term, we need to address the underlined terms that at first glance appear to be not controllable since $tr\chi \sim 2/r$ and $tr\underline{\chi} \sim -2/r$ are not integrable. However, a closer inspection reveals that these terms do have favorable signs. This follows from the estimates of the Ricci coefficients i.e., $|tr\chi - \frac{2}{r}| \lesssim \frac{\epsilon}{r^2}$ and $|tr\underline{\chi} + \frac{2}{r}| \lesssim \frac{\epsilon}{r^2}$ under the bootstrap assumption (Remark 3). Therefore we can write the following

$$\int_{D_{u,\underline{u}}} r^8 \left(4\Omega^{-1}\overline{\Omega tr\underline{\chi}} - 2tr\underline{\chi}\right)|\nabla^2 \alpha|^2 = \int_{D_{u,\underline{u}}} r^8 \left(4\Omega^{-1}\overline{\Omega tr\underline{\chi}} - 4tr\underline{\chi}\right)|\nabla^2 \alpha|^2 \quad (6.62)$$

$$+ \int_{D_{u,\underline{u}}} r^8 \left(2tr\underline{\chi} + \frac{4}{r}\right)|\nabla^2 \alpha|^2 - \int_{D_{u,\underline{u}}} 4r^7 |\nabla^2 \alpha|^2$$

and

$$\int_{D_{u,\underline{u}}} r^8 (5tr\chi - 4\Omega^{-1}\overline{\Omega tr\chi})|\nabla^2 \beta|^2 = \int_{D_{u,\underline{u}}} r^8 (4tr\chi - 4\Omega^{-1}\overline{\Omega tr\chi})|\nabla^2 \beta|^2 \quad (6.63)$$

$$+ \int_{D_{u,\underline{u}}} r^8 \left(tr\chi - \frac{2}{r}\right)|\nabla^2 \beta|^2 + \int_{D_{u,\underline{u}}} 2r^7 |\nabla^2 \beta|^2.$$

Collecting all the terms together, the energy equality (6.61) becomes

$$\int_H r^8 |\nabla^2 \alpha|^2 + \int_{\underline{H}} r^8 |\nabla^2 \beta|^2 + \int_{D_{u,\underline{u}}} 4r^7 |\nabla^2 \alpha|^2 + \int_{D_{u,\underline{u}}} 2r^7 |\nabla^2 \beta|^2 = \int_{\widehat{\Sigma}_0} r^8 |\nabla^2 \alpha|^2 \quad (6.64)$$

$$+ \int_{\widehat{\Sigma}_0} r^8 |\nabla^2 \beta|^2 - \int_{D_{u,\underline{u}}} r^8 (4\Omega^{-1}\overline{\Omega tr\underline{\chi}} - 4tr\underline{\chi})|\nabla^2 \alpha|^2 - \int_{D_{u,\underline{u}}} r^8 \left(4tr\underline{\chi} + \frac{8}{r}\right)|\nabla^2 \alpha|^2$$

$$\int_{D_{u,\underline{u}}} r^8 (4tr\chi - 4\Omega^{-1}\overline{\Omega tr\chi})|\nabla^2 \beta|^2 + \int_{D_{u,\underline{u}}} r^8 \left(tr\chi - \frac{2}{r}\right)|\nabla^2 \beta|^2 + \mathcal{E}_{\nabla^2 \alpha, \nabla^2 \beta}.$$

The bootstrap assumption on the connection coefficients imply $|(4\Omega^{-1}\overline{\Omega tr\underline{\chi}} - 4tr\underline{\chi})|, |(4tr\underline{\chi} + \frac{8}{r})|, |(4tr\chi - 4\Omega^{-1}\overline{\Omega tr\chi})|, |(tr\chi - \frac{2}{r})| \lesssim \frac{\epsilon}{r^2}$ and therefore

$$\int_H r^8 |\nabla^2 \alpha|^2 + \int_{\underline{H}} r^8 |\nabla^2 \beta|^2 \lesssim \int_{\widehat{\Sigma}_0} r^8 |\nabla^2 \alpha|^2 + \int_{\widehat{\Sigma}_0} r^8 |\nabla^2 \beta|^2 + \epsilon \mathbf{W} + \mathcal{E}_{\nabla^2 \alpha, \nabla^2 \beta}. \quad (6.65)$$

Now the challenging part is controlling the error term $\mathcal{E}_{\nabla^2 \alpha, \nabla^2 \beta}$. Firstly, we note the special structure of Einstein's equations that manifests itself in the null-Bianchi equations.

$$|\int_{D_{u,\underline{u}}} r^8 \nabla^2 \alpha (\nabla tr\underline{\chi}) \nabla \alpha| \lesssim \int_{u,\underline{u}} \|r^4 \nabla^2 \alpha\|_{L^2(S)} \|r^{3-\frac{1}{2}} \nabla tr\underline{\chi}\|_{L^4(S)} \|r^{\frac{7}{2}} \nabla \alpha\|_{L^4(S)} r^{-2} \quad (6.66)$$

$$\lesssim \sup_{u,\underline{u}} \|r^{\frac{5}{2}} \nabla tr\underline{\chi}\|_{L^4(S_{u,\underline{u}})} \left( \int_{u_0}^{u} r^{-2} \right) \int_H (r^8|\nabla^2 \alpha|^2 + r^4|\alpha|^2) \lesssim \epsilon(\mathcal{W}_0 + \mathcal{W}_2).$$

Notice that $\alpha$ is only controlled on $H$. This is beneficial since after controlling $\alpha$ in $H$, we are left with an integral in the $u$ direction (in the bulk integral). Therefore, we can control any non-linear term that involves $\alpha$ and a Ricci coefficient even when the Ricci coefficient exhibits slow decay. The slowly decaying connection coefficient $\underline{\omega}$ appears with $\alpha$ in the following expression

$$|\int_{D_{u,\underline{u}}} r^8 |\nabla^2 \alpha|^2 \underline{\omega}| \lesssim \sup_{D_{u,\underline{u}}} (r|u|\|\underline{\omega}\|) \left( \int_{u_0}^{u} r^{-1}|u'|^{-1} du' \right) \int_H r^8 |\nabla^2 \alpha|^2 \lesssim \epsilon \mathcal{W}_2 \quad (6.67)$$

due to the fact that $r^{-1}|u|^{-1}$ is integrable with respect to $u$ in the exterior domain. This crucial *null* structure is present in the null Yang–Mills equations as well. The remaining vital estimates are the ones arising due to the coupling of the Yang–Mills fields with the gravitational degrees of freedom. The Yang–Mills coupling term with the Weyl curvature component $\alpha$ reads

$$|\int_{D_{u,\underline{u}}} r^8 \langle \nabla^2 \alpha, \nabla^2 D_4 \mathfrak{T}_{AB} \rangle| \quad (6.68)$$

We estimate the top order terms since the remaining terms scale accordingly. We will frequently use the global Sobolev inequality (or simply the co-dimension 1 trace inequality) on the null cones (Lemma 4.3). We outline the following non-linear estimates that arise as the error terms. Under the boot-strap assumption (Remark 3),

$$|\int_{D_{u,\underline{u}}} r^8 \nabla^2 \alpha \widehat{\nabla}^2 \rho^F \cdot \widehat{\nabla} \alpha^F| \lesssim \sup_{u,\underline{u}} \||u|r^{\frac{5}{2}} \widehat{\nabla}^2 \rho^F\|_{L^4(S_{u,\underline{u}})} \left( \int_{u_0}^{u} |u'|^{-1} r^{-1} du' \right)$$

$$\int_H (r^8|\nabla^2 \alpha|^2 + r^4|\widehat{\nabla} \alpha^F|^2 + r^6|\widehat{\nabla}^2 \alpha^F|^2) \lesssim \epsilon(\mathcal{W}_2 + \mathcal{Y}_1 + \mathcal{Y}_2),$$

$$|\int_{D_{u,\underline{u}}} r^8 \nabla^2 \alpha \rho^F \cdot \widehat{\nabla}^3 \alpha^F| \lesssim \sup_{D_{u,\underline{u}}} \left( r^2|u|^{\frac{1}{2}} |\rho^F| \right) \left( \int_{u_0}^{u} r^{-2} |u'|^{-\frac{1}{2}} du' \right) \int_H (r^8|\nabla^2 \alpha|^2 + r^8|\widehat{\nabla}^3 \alpha^F|^2)$$

$$\lesssim \epsilon(\mathcal{W}_2 + \mathcal{Y}_3),$$

$$|\int_{D_{u,\underline{u}}} r^8 \nabla^2 \alpha \nabla^2 tr\chi \rho^F \cdot \rho^F| \lesssim \sup_{D_{u,\underline{u}}} \left( r^2|u|^{\frac{1}{2}} |\rho^F| \right) \sup_{u,\underline{u}} \|r^{\frac{7}{2}} \nabla^2 tr\chi\|_{L^4(S_{u,\underline{u}})} \left( \int_{u_0}^{u} |u'|^{-\frac{1}{2}} r^{-\frac{3}{2}} du' \right)$$

$$\int_H (r^8|\nabla^2 \alpha|^2 + u^2|\rho^F|^2 + u^2 r^2 |\widehat{\nabla} \rho^F|^2) \lesssim \epsilon^2 (\mathcal{W}_2 + \mathcal{Y}_0 + \mathcal{Y}_1)$$

$$|\int_{D_{u,\underline{u}}} r^8 \nabla^2 \alpha \widehat{\nabla}^2 \alpha^F \cdot \widehat{\nabla} \rho^F| \lesssim \sup_{u,\underline{u}} \||u|r^{\frac{3}{2}} \widehat{\nabla} \rho^F\|_{L^4(S_{u,\underline{u}})} \left( \int_{u_0}^{u} |u'|^{-1} r^{-1} du' \right)$$

$$\int_H (r^8|\nabla^2 \alpha|^2 + r^6|\widehat{\nabla}^2 \alpha^F|^2 + r^8|\widehat{\nabla}^3 \alpha^F|^2) \lesssim \epsilon(\mathcal{W}_2 + \mathcal{Y}_1 + \mathcal{Y}_2 + \mathcal{Y}_3),$$

$$|\int_{D_{u,\underline{u}}} r^8 \nabla^2 \alpha \widehat{\nabla} \underline{\alpha}^F \cdot \widehat{\nabla} \alpha^F tr\chi| \lesssim \epsilon \left( \left( \int_{u_0}^{u} |u'|^{-2} du' \right) \int_H r^8|\nabla^2 \alpha|^2 + \left( \int_{\underline{u}_0}^{\underline{u}} r^{-3} d\underline{u}' \right) \int_{\underline{H}} u^2 r^2 |\widehat{\nabla} \underline{\alpha}^F|^2 \right)$$

$$\lesssim \epsilon(\mathcal{W}_2 + \mathcal{Y}_1),$$

$$|\int_{D_{u,\underline{u}}} r^8 \nabla^2 \alpha \widehat{\nabla} \underline{\alpha}^F \cdot \widehat{\nabla} \widehat{\nabla}_4 \alpha^F| \lesssim \sup_{u,\underline{u}} \left( \|r^{\frac{3}{2}} |u|^{\frac{3}{2}} \widehat{\nabla} \underline{\alpha}^F\|_{L^4(S_{u,\underline{u}})} \right) \left( \int_{u_0}^{u} |u'|^{-\frac{3}{2}} r^{-1} du' \right)$$

$$\int_H (r^8 |\nabla^2 \alpha|^2 + r^6 |\nabla \widehat{\nabla}_4 \alpha^F|^2 + r^8 |\widehat{\nabla}^2 \widehat{\nabla}_4 \alpha^F|^2) \lesssim \epsilon(\mathcal{W}_2 + \mathcal{Y}_2 + \mathcal{Y}_3).$$

The remaining terms are straightforward to estimate

$$|\int_{D_{u,\underline{u}}} r^8 \nabla^2 \alpha \nabla^2 tr\chi \alpha^F \cdot \underline{\alpha}^F| \lesssim \epsilon^2 (\mathcal{W}_2 + \mathcal{Y}_0 + \mathcal{Y}_1), \quad (6.69)$$

$$|\int_{D_{u,\underline{u}}} r^8 \nabla^2 \alpha \widehat{\nabla} \underline{\alpha}^F \cdot \widehat{\nabla} \widehat{\nabla}_4 \alpha^F| \lesssim \epsilon(\mathcal{W}_2 + \mathcal{Y}_2 + \mathcal{Y}_3), \quad (6.70)$$

$$|\int_{D_{u,\underline{u}}} r^8 \nabla^2 \alpha(\eta + \underline{\eta})\widehat{\nabla}\alpha^F \cdot \widehat{\nabla}\rho^F| \lesssim \epsilon^2(\mathcal{W}_2 + \mathcal{Y}_1 + \mathcal{Y}_2), \tag{6.71}$$

$$|\int_{D_{u,\underline{u}}} r^8 \nabla^2 \alpha \nabla^2(\eta,\underline{\eta})\alpha^F \cdot \rho^F| \lesssim \epsilon^2(\mathcal{W}_2 + \mathcal{Y}_0 + \mathcal{Y}_1). \tag{}$$

Now we need to estimate

$$|\int_{D_{u,\underline{u}}} r^8 \nabla^2 \alpha_{AB} \nabla^2(D_B \mathfrak{T}_{4A} + D_A \mathfrak{T}_{4B}|. \tag{6.72}$$

We have already estimated most of the terms in the expression of $\nabla^2 \nabla \mathfrak{T}_{4A}$ previously. The remaining terms are controlled in a similar way

$$|\int_{D_{u,\underline{u}}} r^8 \nabla^2 \alpha \nabla^2 tr\chi \rho^F \cdot \rho^F| \lesssim \int_{D_{u,\underline{u}}} \|r^4 \nabla^2 \alpha\|_{L^2(S_{u,\underline{u}})} \|r^{\frac{7}{2}} \nabla^2 tr\chi\|_{L^4(S_{u,\underline{u}})} \|r^{\frac{1}{2}}|u|\rho^F\|_{L^4(S_{u,\underline{u}})} r^{-2}|u|^{-\frac{3}{2}}$$

$$\sup_{D_{u,\underline{u}}} \left(r^2|u|^{\frac{1}{2}}|\rho^F|\right)$$

$$\lesssim \sup_{D_{u,\underline{u}}} \left(r^2|u|^{\frac{1}{2}}|\rho^F|\right) \sup_{u,\underline{u}} \|r^{\frac{7}{2}} \nabla^2 tr\chi\|_{L^4(S_{u,\underline{u}})} \left(\int_{u_0}^{u} r^{-2}|u'|^{-\frac{3}{2}} du'\right)$$

$$\int_H (r^8|\nabla^2\alpha|^2 + u^2|\rho^F|^2 + u^2 r^2|\widehat{\nabla}\rho^F|^2)$$

$$\lesssim \epsilon^2(\mathcal{W}_2 + \mathcal{Y}_0 + \mathcal{Y}_1)$$

Similarly

$$|\int_{D_{u,\underline{u}}} r^8 \nabla^2 \alpha \nabla^2 \hat{\chi} \rho^F \cdot \rho^F| \lesssim \epsilon^2(\mathcal{W}_2 + \mathcal{Y}_0 + \mathcal{Y}_1), \tag{6.73}$$

$$|\int_{D_{u,\underline{u}}} r^8 \nabla^2 \alpha \nabla \hat{\chi} \widehat{\nabla} \rho^F \cdot \rho^F| \lesssim \epsilon^2(\mathcal{W}_2 + \mathcal{Y}_1 + \mathcal{Y}_2), \tag{6.74}$$

$$|\int_{D_{u,\underline{u}}} r^8 \nabla^2 \alpha tr\chi \rho^F \cdot \widehat{\nabla}^2 \rho^F| \lesssim \epsilon(\mathcal{W}_2 + \mathcal{Y}_2), \tag{6.75}$$

$$|\int_{D_{u,\underline{u}}} r^8 \nabla^2 \alpha \nabla(tr\chi,\hat{\chi})\widehat{\nabla}(\rho^F,\sigma^F) \cdot (\rho^F,\sigma^F)| \lesssim \epsilon^2(\mathcal{W}_2 + \mathcal{Y}_1 + \mathcal{Y}_2), \tag{6.76}$$

$$|\int_{D_{u,\underline{u}}} r^8 \nabla^2 \alpha \nabla(tr\chi,\hat{\chi})\widehat{\nabla}\alpha^F \cdot \underline{\alpha}^F| \lesssim \epsilon^2(\mathcal{W}_2 + \mathcal{Y}_1 + \mathcal{Y}_2), \tag{6.77}$$

$$|\int_{D_{u,\underline{u}}} r^8 \nabla^2 \alpha \nabla(tr\chi,\hat{\chi})\alpha^F \cdot \widehat{\nabla}\underline{\alpha}^F| \lesssim \epsilon^2(\mathcal{W}_2 + \mathcal{Y}_1 + \mathcal{Y}_2), \tag{6.78}$$

$$|\int_{D_{u,\underline{u}}} r^8 \nabla^2 \alpha\ tr\chi \alpha^F \cdot \widehat{\nabla}^2 \underline{\alpha}^F| \lesssim \epsilon(\mathcal{W}_2 + \mathcal{Y}_1 + \mathcal{Y}_2) \tag{6.79}$$

Collecting all the terms together, we obtain

$$|\int_{D_{u,\underline{u}}} r^8 \nabla^2 \alpha_{AB} \nabla^2(D_B \mathfrak{T}_{4A} + D_A \mathfrak{T}_{4B})| \lesssim \epsilon(\mathcal{W}_2 + \mathcal{Y}_0 + \mathcal{Y}_1 + \mathcal{Y}_2 + \mathcal{Y}_3). \tag{6.80}$$

Notice the following term with special structure (slowly decaying connection $\hat{\chi}$ appears with $\alpha$)

$$|\int_{D_{u,\underline{u}}} r^8 \nabla^2 \alpha \underline{\hat{\chi}} \nabla^2 \alpha| \lesssim \sup_{D_{u,\underline{u}}} (|u|r|\underline{\hat{\chi}}|) \left(\int_{u_0}^{u} |u'|^{-1} r^{-1} du'\right) \int_H r^8 |\nabla^2 \alpha|^2 \lesssim \epsilon \mathcal{W}_2. \tag{6.81}$$

Now we need to estimate the error term that is generated via commutation of $\nabla^2$ and $\nabla_3$ while acting on $\alpha$. We estimate these terms as follows

$$|\int_{D_{u,\underline{u}}} r^8 \nabla^2 \alpha \nabla((\eta + \underline{\eta})tr\underline{\chi}\alpha)| \lesssim \epsilon(\mathcal{W}_0 + \mathcal{W}_1 + \mathcal{W}_2), \tag{6.82}$$

$$|\int_{D_{u,\underline{u}}} r^8 \nabla^2 \alpha \nabla(\eta\underline{\chi}\alpha)| \lesssim \epsilon(\mathcal{W}_0 + \mathcal{W}_1 + \mathcal{W}_2), \tag{}$$

$$|\int_{D_{u,\underline{u}}} r^8 \nabla^2 \alpha \nabla(\underline{\beta}\alpha)| \lesssim \epsilon(\mathcal{W}_1 + \mathcal{W}_2), \tag{}$$

$$|\int_{D_{u,\underline{u}}} r^8 \nabla^2 \alpha \nabla((\eta + \underline{\eta})\nabla\beta)| \lesssim \epsilon(\mathcal{W}_1 + \mathcal{W}_2), \tag{}$$

$$|\int_{D_{u,\underline{u}}} r^8 \nabla^2 \alpha \nabla(\underline{\omega}\alpha)| \lesssim \epsilon(\mathcal{W}_0 + \mathcal{W}_1 + \mathcal{W}_2),$$

$$|\int_{D_{u,\underline{u}}} r^8 \nabla^2 \alpha \nabla(\chi(\rho,\sigma)(\eta,\underline{\eta}))| \lesssim \epsilon(\mathcal{W}_0 + \mathcal{W}_1 + \mathcal{W}_2),$$

$$|\int_{D_{u,\underline{u}}} r^8 \nabla^2 \alpha \nabla((\eta,\underline{\eta})\beta)| \lesssim \epsilon(\mathcal{W}_0 + \mathcal{W}_1 + \mathcal{W}_2),$$

$$|\int_{D_{u,\underline{u}}} r^8 \nabla^2 \alpha_{AB} \nabla(D_4 \mathfrak{T}_{AB})| \lesssim \epsilon(\mathcal{W}_2 + \mathcal{Y}_0 + \mathcal{Y}_1 + \mathcal{Y}_2),$$

$$|\int_{D_{u,\underline{u}}} r^8 \nabla^2 \alpha_{AB} \nabla D_A \mathfrak{T}_{4B}| \lesssim \epsilon(\mathcal{W}_2 + \mathcal{Y}_0 + \mathcal{Y}_1 + \mathcal{Y}_2).$$

Here notice that $\int_{D_{u,\underline{u}}} r^8 \langle \nabla \alpha, \nabla^2 (D_3 \mathfrak{T}_{44} - D_4 \mathfrak{T}_{43}) \gamma \rangle = 0$ due to the compatibility of $\gamma$ with $\nabla$ as well as the trace-free property of $\alpha$.

$$|\int_{D_{u,\underline{u}}} r^8 \nabla^2 \beta \nabla^2 (tr\chi)\beta| \lesssim \sup_{u,\underline{u}} \|r^{\frac{7}{2}} \nabla^2 tr\chi\|_{L^4(S_{u,\underline{u}})} \int_{u,\underline{u}} \|r^4 \nabla^2 \beta\|_{L^2(S_{u,\underline{u}})} \|r^{\frac{5}{2}} \beta\|_{L^4(S_{u,\underline{u}})} r^{-2} \quad (6.83)$$

$$\lesssim \sup_{u,\underline{u}} \|r^{\frac{7}{2}} \nabla^2 tr\chi\|_{L^4(S_{u,\underline{u}})} \left(\int_{u_0}^{u} r^{-2} du'\right) \int_H (r^8|\nabla^2 \beta|^2 + r^4|\beta|^2)$$

$$\lesssim \epsilon(\mathcal{W}_0 + \mathcal{W}_2)$$

$$|\int_{D_{u,\underline{u}}} r^8 \nabla^2 \beta (\nabla^2 \omega)\beta| \lesssim \sup_{u,\underline{u}} \|r^{\frac{7}{2}} \nabla^2 \omega\|_{L^4(S_{u,\underline{u}})} \int_{u,\underline{u}} \|r^4 \nabla^2 \beta\|_{L^2(S_{u,\underline{u}})} \|r^{\frac{5}{2}} \beta\|_{L^4(S_{u,\underline{u}})} r^{-2} \quad (6.84)$$

$$\lesssim \sup_{u,\underline{u}} \|r^{\frac{7}{2}} \nabla^2 \omega\|_{L^4(S_{u,\underline{u}})} \left(\int_{u_0}^{u} r^{-2} du'\right) \int_H (r^8|\nabla^2 \beta|^2 + r^4|\beta|^2)$$

$$\lesssim \epsilon(\mathcal{W}_0 + \mathcal{W}_2)$$

$$|\int_{D_{u,\underline{u}}} r^8 \omega |\nabla^2 \beta|^2| \lesssim \epsilon \mathcal{W}_2, \quad |\int_{D_{u,\underline{u}}} r^8 \nabla^2 \beta \nabla(\nabla \eta \alpha)| \lesssim \epsilon(\mathcal{W}_0 + \mathcal{W}_1 + \mathcal{W}_2), \quad (6.85)$$

$$|\int_{D_{u,\underline{u}}} r^8 \nabla^2 \beta \nabla(\eta \nabla \alpha)| \lesssim \epsilon(\mathcal{W}_1 + \mathcal{W}_2). \quad (6.86)$$

Now we need to estimate the error term that arises due to the commutation of $\nabla_4$ and $\nabla^2$ i.e.,

$$|\int_{D_{u,\underline{u}}} r^8 \nabla \beta \nabla(\underline{\eta}\chi\beta)| \lesssim \epsilon(\mathcal{W}_0 + \mathcal{W}_1 + \mathcal{W}_2), \quad |\int_{D_{u,\underline{u}}} r^8 \nabla \beta \nabla \beta \beta| \lesssim \epsilon \mathcal{W}_1,$$

$$|\int_{D_{u,\underline{u}}} r^8 \nabla \beta \nabla(\alpha(\eta,\underline{\eta}))| \lesssim \epsilon(\mathcal{W}_0 + \mathcal{W}_1 + \mathcal{W}_2), |\int_{D_{u,\underline{u}}} r^8 \nabla \beta \nabla(\omega \beta)| \lesssim \epsilon(\mathcal{W}_0 + \mathcal{W}_1),$$

$$|\int_{D_{u,\underline{u}}} r^8 \nabla \beta \alpha^F \cdot \widehat{\nabla}^2 \alpha^F| \lesssim \epsilon(\mathcal{W}_1 + \mathcal{Y}_2), |\int_{D_{u,\underline{u}}} r^8 \nabla \beta \nabla(\chi \alpha^F \cdot (\rho^F,\sigma^F))| \lesssim \epsilon(\mathcal{W}_2 + \mathcal{Y}_0 + \mathcal{Y}_1),$$

$$|\int_{D_{u,\underline{u}}} r^8 \nabla \beta (\rho^F, \sigma^F) \cdot \widehat{\nabla} \widehat{\nabla}_4 \alpha^F| \lesssim \epsilon(\mathcal{W}_1 + \mathcal{Y}_1 + \mathcal{Y}_2),$$

$$|\int_{D_{u,\underline{u}}} r^8 \nabla \beta \nabla((\eta,\underline{\eta})\alpha^F \cdot \alpha^F)| \lesssim \epsilon^2(\mathcal{W}_1 + \mathcal{Y}_0 + \mathcal{Y}_1).$$

The last term to estimate is the coupling term

$$|\int_{D_{u,\underline{u}}} r^8 \nabla^2 \beta_A \nabla^2 (D_A \mathfrak{T}_{44} - D_4 \mathfrak{T}_{4A})|. \quad (6.87)$$

Now we estimate the most non-trivial terms

$$|\int_{D_{u,\underline{u}}} r^8 \nabla^2 \beta \widehat{\nabla} \alpha^F \cdot \widehat{\nabla}^2 \alpha^F| \lesssim \sup_{D_{u,\underline{u}}} \left(r^{\frac{7}{2}} |\widehat{\nabla} \alpha^F|\right) \left(\int_{u_0}^{u} r^{-\frac{5}{2}} du'\right) \int_H (r^8|\nabla^2 \beta|^2 + r^6|\widehat{\nabla}^2 \alpha^F|^2)$$

$$\lesssim \epsilon(\mathcal{W}_2 + \mathcal{Y}_2) \quad (6.88)$$

$$|\int_{D_{u,\underline{u}}} r^8 \nabla^2 \beta \alpha^F \cdot \widehat{\nabla}^3 \alpha^F| \lesssim \sup_{D_{u,\underline{u}}} \left( r^{\frac{5}{2}} |\alpha^F| \right) \left( \int_{u_0}^{u} r^{-\frac{5}{2}} du' \right) \int_H (r^8 |\nabla^2 \beta|^2 + r^8 |\widehat{\nabla}^3 \alpha^F|^2)$$

$$\lesssim \epsilon(\mathcal{W}_2 + \mathcal{Y}_3). \tag{6.89}$$

The collection of all the terms yields the desired result

$$|\mathcal{E}_{\nabla^2 \alpha, \nabla^2 \beta}| \lesssim \epsilon(\mathcal{W}_0 + \mathcal{W}_1 + \mathcal{W}_2 + \mathcal{Y}_0 + \mathcal{Y}_1 + \mathcal{Y}_2 + \mathcal{Y}_3). \tag{6.90}$$

Therefore the proof of the lemma follows. □

Now as a consequence of this lemma, we have a trivial estimate of the bulk integral that will be useful in estimating the mixed derivatives.

*Proposition 6.7.* The following estimate holds

$$4 \int_{D_{u,\underline{u}}} r^7 |\nabla^2 \alpha|^2 + 2 \int_{D_{u,\underline{u}}} r^7 |\nabla^2 \beta|^2 \lesssim \int_{\widehat{\Sigma}_0} r^8 |\nabla^2 \alpha|^2 + \int_{\widehat{\Sigma}_0} r^8 |\nabla^2 \beta|^2 + \epsilon(\mathbf{W} + \mathbf{Y}) \tag{6.91}$$

*Proof.* The proof is a simple consequence of the lemma (6.2) since we have

$$\int_H r^8 |\nabla^2 \alpha|^2 + \int_{\underline{H}} r^8 |\nabla^2 \beta|^2 + \int_{D_{u,\underline{u}}} 4r^7 |\nabla^2 \alpha|^2 + \int_{D_{u,\underline{u}}} 2r^7 |\nabla^2 \beta|^2 \lesssim \int_{\widehat{\Sigma}_0} r^8 |\nabla^2 \alpha|^2$$

$$+ \int_{\widehat{\Sigma}_0} r^8 |\nabla^2 \beta|^2 + \epsilon(\mathbf{W} + \mathbf{Y}) \tag{6.92}$$

and

$$\int_H r^8 |\nabla^2 \alpha|^2 + \int_{\underline{H}} r^8 |\nabla^2 \beta|^2 \lesssim \int_{\widehat{\Sigma}_0} r^8 |\nabla^2 \alpha|^2 + \int_{\widehat{\Sigma}_0} r^8 |\nabla^2 \beta|^2 + \epsilon(\mathbf{W} + \mathbf{Y}) \tag{6.93}$$

. □

*Lemma 6.3.* The following estimate hold in the exterior region

$$\int_{H_u} u^4 |\underline{\beta}|^2 + \int_{\underline{H}_u} u^4 |\underline{\alpha}|^2 \lesssim \int_{\widehat{\Sigma}_0} u^4 |\underline{\alpha}|^2 + \int_{\widehat{\Sigma}_0} u^4 |\underline{\beta}|^2 + \epsilon(\mathbf{W} + \mathbf{Y}), \tag{6.94}$$

$$\int_{H_u} u^4 r^2 |\nabla \underline{\beta}|^2 + \int_{\underline{H}_u} u^4 r^2 |\nabla \underline{\alpha}|^2 \leq \int_{\widehat{\Sigma}_0} u^4 r^2 |\nabla \underline{\alpha}|^2 + \int_{\widehat{\Sigma}_0} u^4 r^2 |\nabla \underline{\beta}|^2 + \epsilon(\mathbf{W} + \mathbf{Y}), \tag{6.95}$$

$$\int_{H_u} u^4 r^4 |\nabla^2 \underline{\beta}|^2 + \int_{\underline{H}_u} u^4 r^4 |\nabla^2 \underline{\alpha}|^2 \leq \int_{\widehat{\Sigma}_0} u^4 r^4 |\nabla^2 \underline{\alpha}|^2 + \int_{\widehat{\Sigma}_0} u^4 r^4 |\nabla^2 \underline{\beta}|^2 + \epsilon(\mathbf{W} + \mathbf{Y}).$$

*Proof.* First, we derive the following identity using the integration Lemma (6.1)

$$\int_{H_u} u^4 r^4 |\nabla^2 \underline{\beta}|^2 + \int_{\underline{H}_u} u^4 r^4 |\nabla^2 \underline{\alpha}|^2 - 2 \int_{D_{u,\underline{u}}} u^4 r^4 (\Omega^{-1} \overline{\Omega tr \chi} - tr \chi) |\nabla^2 \underline{\alpha}|^2 \tag{6.96}$$

$$- \underbrace{\int_{D_{u,\underline{u}}} (4\Omega^{-1} u^3 r^4 - 3u^4 r^4 tr\underline{\chi}) |\nabla^2 \underline{\beta}|^2}_{<0} + 2 \int_{D_{u,\underline{u}}} u^4 r^4 (\Omega^{-1} \overline{\Omega tr \underline{\chi}} - tr\underline{\chi}) |\nabla^2 \underline{\beta}|^2$$

$$= \int_{\widehat{\Sigma}_0} u^4 r^4 |\nabla^2 \underline{\alpha}|^2 + \int_{\widehat{\Sigma}_0} u^4 r^4 |\nabla^2 \underline{\beta}|^2 + \mathcal{E}^{\nabla^2 \underline{\alpha}, \nabla^2 \underline{\beta}},$$

where the error term $\mathcal{E}^{\nabla^2 \underline{\alpha}, \nabla^2 \underline{\beta}}$ verifies the following schematic form

$$\mathcal{E}^{\nabla^2 \underline{\alpha}, \nabla^2 \underline{\beta}} \sim \int_{D_{u,\underline{u}}} 2u^4 r^4 \langle \nabla^2 \underline{\alpha}, -\frac{1}{2} \nabla(\nabla(tr\chi)\underline{\alpha}) + \nabla^2 \Big( 4\omega \underline{\alpha}_{ab} - 3(\hat{\underline{\chi}}_{ab} \rho - {}^*\hat{\underline{\chi}}_{ab} \sigma) + ((\zeta - 4\underline{\eta}) \hat{\otimes} \underline{\beta})_{ab}$$

$$+ \frac{1}{2} (D_4 \mathfrak{I}_{33} - D_3 \mathfrak{I}_{34}) \gamma_{ab} - D_3 \mathfrak{I}_{ab} + \frac{1}{2} (D_a \mathfrak{I}_{3b} + D_b \mathfrak{I}_{3a}) \Big) \rangle + 2u^4 r^4 \langle \nabla^2 \underline{\alpha}, -\nabla(\widehat{\chi} \cdot \nabla \underline{\alpha}) \rangle$$

$$+ 2u^4 r^4 \langle \nabla^2 \underline{\beta}, -2\nabla(\nabla(tr\underline{\chi})\underline{\beta}) - \nabla^2 \Big( 2\underline{\omega} \underline{\beta}_a + (\underline{\eta} \cdot \underline{\alpha})_a + \frac{1}{2} (D_a \mathfrak{I}_{33} - D_3 \mathfrak{I}_{3a}) \Big) \rangle$$

$$+ 2u^4 r^4 \langle \nabla^2 \underline{\beta}, -\nabla(\widehat{\underline{\chi}} \cdot \nabla \underline{\beta}) \rangle.$$

As usual, we first need to control the underlined terms. In fact, we show using the estimates for the connection and the Lapse function that these terms either verify higher-order estimates or have favorable signs. First, consider the term

$$\int_{D_{u,\underline{u}}} (4\Omega^{-1} u^3 r^4 - 3u^4 r^4 tr\underline{\chi})|\nabla^2 \underline{\beta}|^2 \tag{6.97}$$

and rewrite it in the following way

$$\int_{D_{u,\underline{u}}} (4\Omega^{-1} u^3 r^4 - 3u^4 r^4 tr\underline{\chi})|\nabla^2 \underline{\beta}|^2 \tag{6.98}$$
$$= \int_{D_{u,\underline{u}}} \left(4(\Omega^{-1} - 2)u^3 r^4 + 8u^3 r^4 - 3u^4 r^4 \left(tr\underline{\chi} + \frac{2}{r}\right) + 6u^4 r^3\right)|\nabla^2 \underline{\beta}|^2$$
$$= \int_{D_{u,\underline{u}}} \left(4(\Omega^{-1} - 2) u^3 r^4 - 3u^4 r^4 \left(tr\underline{\chi} + \frac{2}{r}\right) - 2r^3|u|^3(4r - 3|u|)\right)|\nabla^2 \underline{\beta}|^2.$$

Therefore the energy identity reads

$$\int_{H_u} u^4 r^4 |\nabla^2 \underline{\beta}|^2 + \int_{\underline{H}_{\underline{u}}} u^4 r^4 |\nabla^2 \underline{\alpha}|^2 + \int_{D_{u,\underline{u}}} 2r^3|u|^3(4r - 3|u|)|\nabla^2 \underline{\beta}|^2 = \int_{\widehat{\Sigma}_0} u^4 r^4 |\nabla^2 \underline{\alpha}|^2 + \int_{\widehat{\Sigma}_0} u^4 r^4 |\nabla^2 \underline{\beta}|^2$$
$$+ 2\int_{D_{u,\underline{u}}} u^4 r^4 (\Omega^{-1} \overline{\Omega tr\chi} - tr\chi)|\nabla^2 \underline{\alpha}|^2 + \int_{D_{u,\underline{u}}} \left(4(\Omega^{-1} - 2) u^3 r^4 - 3u^4 r^4 \left(tr\underline{\chi} + \frac{2}{r}\right)\right)|\nabla^2 \underline{\beta}|^2$$
$$+ 2\int_{D_{u,\underline{u}}} u^4 r^4 (\Omega^{-1} \overline{\Omega tr\underline{\chi}} - tr\underline{\chi})|\nabla^2 \underline{\beta}|^2 + \mathcal{E}^{\nabla^2 \underline{\alpha}, \nabla^2 \underline{\beta}}$$

which in light of boot-strap assumption and connection estimates yields

$$\int_{H_u} u^4 r^4 |\nabla^2 \underline{\beta}|^2 + \int_{\underline{H}_{\underline{u}}} u^4 r^4 |\nabla^2 \underline{\alpha}|^2 \lesssim \int_{\widehat{\Sigma}_0} u^4 r^4 |\nabla^2 \underline{\alpha}|^2 + \int_{\widehat{\Sigma}_0} u^4 r^4 |\nabla^2 \underline{\beta}|^2 + \mathcal{E}^{\nabla^2 \underline{\alpha}, \nabla^2 \underline{\beta}} + \epsilon \mathcal{W}_2$$

since in the exterior domain for sufficiently large $\underline{u}_0$, $4r > 3|u|$. The remaining task is estimate the error term $\mathcal{E}^{\nabla^2 \underline{\alpha}, \nabla^2 \underline{\beta}}$. Once again we start each type of term separately.

$$|\int_{D_{u,\underline{u}}} u^4 r^4 \nabla^2 \underline{\alpha} \nabla^2 (tr\chi)\underline{\alpha}| \lesssim \int_{u,\underline{u}} \|u^2 r^2 \nabla^2 \underline{\alpha}\|_{L^2(S_{u,\underline{u}})} \|r^{\frac{7}{2}} \nabla^2 tr\chi\|_{L^4(S_{u,\underline{u}})} \|r^{\frac{1}{2}} |u|^2 \underline{\alpha}\|_{L^4(S_{u,\underline{u}})} r^{-2}$$
$$\lesssim \sup_{u,\underline{u}} \|r^{\frac{7}{2}} \nabla^2 tr\chi\|_{L^4(S_{u,\underline{u}})} \left(\int_{\underline{u}_0}^{\underline{u}} r^{-2} d\underline{u}'\right) \int_{\underline{H}} (u^4|\underline{\alpha}|^2 + u^4 r^2|\nabla \underline{\alpha}|^2 + u^4 r^4|\nabla^2 \underline{\alpha}|^2) \lesssim \epsilon(\mathcal{W}_0 + \mathcal{W}_1 + \mathcal{W}_2)$$

Notice the following term with special structure

$$|\int_{D_{u,\underline{u}}} u^4 r^4 |\nabla^2 \underline{\alpha}|^2 \omega| \lesssim \sup_{D_{u,\underline{u}}} (r^2|\omega|) \left(\int_{\underline{u}_0}^{\underline{u}} r^{-2} d\underline{u}'\right) \int_{\underline{H}} u^4 r^4 |\nabla^2 \underline{\alpha}|^2 \lesssim \epsilon \mathcal{W}_2. \tag{6.99}$$

Here $\omega$ enjoys fast decay (fast enough such that point-wise norm is integrable in $\underline{u}$) and appears multiplied to the topmost derivative of $\underline{\alpha}$. This is essentially an example of the so-called special structure of the null Bianchi equation i.e., faster-decaying connection coefficients always appear multiplied by the slower-decaying curvature components. The remaining estimates are written as follows

$$|\int_{D_{u,\underline{u}}} u^4 r^4 \nabla^2 \underline{\alpha} (\nabla^2 \omega)\underline{\alpha}| \lesssim \int_{u,\underline{u}} \|u^2 r^2 \nabla^2 \underline{\alpha}\|_{L^2(S_{u,\underline{u}})} \|r^{\frac{7}{2}} \nabla^2 \omega\|_{L^4(S_{u,\underline{u}})} \|r^{\frac{1}{2}} |u|^2 \underline{\alpha}\|_{L^4(S_{u,\underline{u}})} r^{-2}$$
$$\lesssim \sup_{u,\underline{u}} \|r^{\frac{7}{2}} \nabla^2 \omega\|_{L^4(S_{u,\underline{u}})} \left(\int_{\underline{u}_0}^{\underline{u}} r^{-2} d\underline{u}'\right) \int_{\underline{H}} (u^4|\underline{\alpha}|^2 + u^4 r^2|\nabla \underline{\alpha}|^2 + u^4 r^4|\nabla^2 \underline{\alpha}|^2)$$
$$\lesssim \epsilon(\mathcal{W}_0 + \mathcal{W}_1 + \mathcal{W}_2),$$
$$|\int_{D_{u,\underline{u}}} u^4 r^4 \nabla^2 \underline{\alpha} \nabla^2 \hat{\underline{\chi}} \rho| \lesssim \int_{u,\underline{u}} \|u^2 r^2 \nabla^2 \underline{\alpha}\|_{L^2(S_{u,\underline{u}})} \|r^{\frac{5}{2}} |u| \nabla^2 \hat{\underline{\chi}}\|_{L^4(S_{u,\underline{u}})} \|r^{\frac{5}{2}} \rho\|_{L^4(S_{u,\underline{u}})} |u| r^{-3}$$
$$\lesssim \sup_{u,\underline{u}} \|r^{\frac{5}{2}} |u| \nabla^2 \hat{\underline{\chi}}\|_{L^4(S_{u,\underline{u}})} \left(\int_{\underline{u}_0}^{\underline{u}} |u| r^{-3} d\underline{u}'\right) \int_{\underline{H}} (r^4 \rho^2 + r^6 |\nabla \rho|^2 + u^4 r^4 |\nabla^2 \underline{\alpha}|^2)$$
$$\lesssim \epsilon(\mathcal{W}_0 + \mathcal{W}_1 + \mathcal{W}_2),$$

$$|\int_{D_{u,\underline{u}}} u^4 r^4 \nabla^2 \underline{\alpha}(\nabla^2(\eta,\underline{\eta}))\underline{\beta}| \lesssim \int_{u,\underline{u}} \|u^2 r^2 \nabla^2 \underline{\alpha}\|_{L^2(S_{u,\underline{u}})} \|r^{\frac{7}{2}} \nabla^2(\eta,\underline{\eta})\|_{L^4(S_{u,\underline{u}})} \|r^{\frac{3}{2}} |u|\underline{\beta}\|_{L^4(S_{u,\underline{u}})} |u|r^{-3}$$

$$\lesssim \sup_{u,\underline{u}} \|r^{\frac{7}{2}} \nabla^2(\eta,\underline{\eta})\|_{L^4(S_{u,\underline{u}})} \left(\int_{\underline{u}_0}^{\underline{u}} |u|r^{-3} d\underline{u}'\right)$$

$$\int_{\underline{H}} (u^4 r^4 |\nabla^2 \underline{\alpha}|^2 + u^2 r^2 |\underline{\beta}|^2 + u^2 r^4 |\nabla \underline{\beta}|^2)$$

$$\lesssim \epsilon(\mathcal{W}_0 + \mathcal{W}_1 + \mathcal{W}_2).$$

The remaining pure gravity terms are of lower order and therefore harmless. The main task is to estimate the Yang–Mills coupling terms. We estimate each coupling term separately. First, consider the following term

$$|\int_{D_{u,\underline{u}}} u^4 r^4 \nabla^2 \underline{\alpha}_{AB} \nabla^2 D_3 \mathfrak{T}_{AB}|. \tag{6.100}$$

Explicit calculations [using the schematic expressions (2.63)–(2.67) for the source terms]

$$|\int_{D_{u,\underline{u}}} u^4 r^4 \nabla^2 \underline{\alpha} \widehat{\nabla} \rho^F \cdot \widehat{\nabla}^2 \underline{\alpha}^F| \tag{6.101}$$

$$\lesssim \sup_{D_{u,\underline{u}}} \left(|u|^{\frac{1}{2}} r^3 |\widehat{\nabla} \rho^F|\right) \left(\int_{\underline{u}_0}^{\underline{u}} |u|^{-\frac{1}{2}} r^{-3} d\underline{u}'\right) \int_{\underline{H}} (u^4 r^4 |\nabla^2 \underline{\alpha}|^2 + u^2 r^4 |\widehat{\nabla}^2 \underline{\alpha}^F|^2)$$

$$\lesssim \epsilon(\mathcal{W}_2 + \mathcal{Y}_2),$$

$$|\int_{D_{u,\underline{u}}} u^4 r^4 \nabla^2 \underline{\alpha} \rho^F \cdot \widehat{\nabla}^2 \underline{\alpha}^F| \tag{6.102}$$

$$\lesssim \sup_{D_{u,\underline{u}}} r^2 |u|^{\frac{1}{2}} |\rho^F| \left(\int_{\underline{u}_0}^{\underline{u}} |u|r^{-3} d\underline{u}'\right) \int_{\underline{H}} (u^4 r^4 |\nabla^2 \underline{\alpha}|^2 + r^4 |u|^2 |\widehat{\nabla}^2 \underline{\alpha}^F|^2)$$

$$\lesssim \epsilon(\mathcal{W}_2 + \mathcal{Y}_2),$$

$$|\int_{D_{u,\underline{u}}} u^4 r^4 \nabla^2 \underline{\alpha} \nabla^2 tr\chi \rho^F \cdot \rho^F| \lesssim \sup_{u,\underline{u}} \left(\|r^{\frac{7}{2}} \nabla^2 tr\underline{\chi}\|_{L^4(S_{u,\underline{u}})}\right) \sup_{D_{u,\underline{u}}} \left(r^2 |u|^{\frac{1}{2}} |\rho^F|\right)$$

$$\int_{u,\underline{u}} \|u^2 r^2 \nabla^2 \underline{\alpha}\|_{L^2(S_{u,\underline{u}})} \|r^{\frac{3}{2}} \rho^F\|_{L^4(S_{u,\underline{u}})} |u|^{\frac{1}{2}} r^{-5}$$

$$\lesssim \epsilon^2 \int_{\underline{H}} (u^4 r^4 |\nabla^2 \underline{\alpha}|^2 + r^2 |\rho^F|^2 + r^4 |\widehat{\nabla} \rho^F|^2)$$

$$\lesssim \epsilon^2(\mathcal{W}_2 + \mathcal{Y}_0 + \mathcal{Y}_1),$$

$$|\int_{D_{u,\underline{u}}} u^4 r^4 \nabla^2 \underline{\alpha} \widehat{\nabla} \alpha^F \cdot \widehat{\nabla} \widehat{\nabla}_3 \underline{\alpha}^F| \lesssim \sup_{D_{u,\underline{u}}} r^{\frac{7}{2}} |\widehat{\nabla} \alpha^F| \left(\int_{\underline{u}_0}^{\underline{u}} r^{-\frac{5}{2}} d\underline{u}'\right) \int_{\underline{H}} (u^4 r^4 |\nabla^2 \underline{\alpha}|^2 + u^4 r^2 |\widehat{\nabla} \widehat{\nabla}_3 \underline{\alpha}^F|^2)$$

$$\lesssim \epsilon(\mathcal{W}_2 + \mathcal{Y}_1 + \mathcal{Y}_2),$$

$$|\int_{D_{u,\underline{u}}} u^4 r^4 \nabla^2 \underline{\alpha} \widehat{\nabla}^2 \alpha^F \cdot \widehat{\nabla}_3 \underline{\alpha}^F| \tag{6.103}$$

$$\lesssim \sup_{u,\underline{u}} \|r^3 \widehat{\nabla}^2 \alpha^F\|_{L^4(S_{u,\underline{u}})} \left(\int_{\underline{u}_0}^{\underline{u}} r^{-\frac{3}{2}} d\underline{u}'\right)$$

$$\int_{\underline{H}} (u^4 r^4 |\nabla^2 \underline{\alpha}|^2 + u^4 |\widehat{\nabla}_3 \underline{\alpha}^F|^2 + u^4 r^2 |\widehat{\nabla} \widehat{\nabla}_3 \underline{\alpha}^F|^2)$$

$$\lesssim \epsilon(\mathcal{W}_2 + \mathcal{Y}_1 + \mathcal{Y}_2).$$

Similarly, we have the following straightforward estimates

$$|\int_{D_{u,\underline{u}}} u^4 r^4 \nabla^2 \underline{\alpha} \nabla(tr\underline{\chi} \widehat{\nabla} \rho^F \cdot \rho^F)| \lesssim \epsilon(\mathcal{W}_2 + \mathcal{Y}_1 + \mathcal{Y}_2), \tag{6.104}$$

$$|\int_{D_{u,\underline{u}}} u^4 r^4 \nabla^2 \underline{\alpha} \nabla(\nabla(\eta,\underline{\eta})\underline{\alpha}^F \cdot \rho^F)| \lesssim \epsilon^2(\mathcal{W}_2 + \mathcal{Y}_0 + \mathcal{Y}_1 + \mathcal{Y}_2), \tag{6.105}$$

$$|\int_{D_{u,\underline{u}}} u^4 r^4 \nabla^2 \underline{\alpha} \nabla((\eta,\underline{\eta})\widehat{\nabla}\underline{\alpha}^F \cdot \rho^F)| \lesssim \epsilon^2 (\mathcal{W}_2 + \mathcal{Y}_1 + \mathcal{Y}_2), \qquad (6.106)$$

$$|\int_{D_{u,\underline{u}}} u^4 r^4 \nabla^2 \underline{\alpha} \nabla(\underline{\alpha}^F \cdot \widehat{\nabla}^2 \rho^F)| \lesssim \epsilon (\mathcal{W}_2 + \mathcal{Y}_2 + \mathcal{Y}_3), \qquad (6.107)$$

$$|\int_{D_{u,\underline{u}}} u^4 r^4 \nabla^2 \underline{\alpha} \nabla(tr\underline{\chi}\widehat{\nabla}\alpha^F \cdot \underline{\alpha}^F)| \lesssim \epsilon (\mathcal{W}_2 + \mathcal{Y}_1 + \mathcal{Y}_2), \qquad (6.108)$$

$$|\int_{D_{u,\underline{u}}} u^4 r^4 \nabla^2 \underline{\alpha} \nabla(\nabla tr\chi \alpha^F \cdot \underline{\alpha}^F)| \lesssim \epsilon^2 (\mathcal{W}_2 + \mathcal{Y}_0 + \mathcal{Y}_1 + \mathcal{Y}_2). \qquad (6.109)$$

Collecting all the terms, we obtain

$$|\int_{D_{u,\underline{u}}} u^4 r^4 \nabla^2 \underline{\alpha}_{AB} \nabla^2 D_3 \mathfrak{T}_{AB}| \lesssim \epsilon (\mathcal{W}_2 + \mathcal{Y}_0 + \mathcal{Y}_1 + \mathcal{Y}_2 + \mathcal{Y}_3). \qquad (6.110)$$

Now we need to estimate the coupling term

$$|\int_{D_{u,\underline{u}}} u^4 r^4 \nabla^2 \underline{\alpha}_{AB} \nabla \nabla_A \mathfrak{T}_{3B}|. \qquad (6.111)$$

We estimate

$$|\int_{D_{u,\underline{u}}} u^4 r^4 \nabla^2 \underline{\alpha} \widehat{\nabla}^3 \underline{\alpha}^F \cdot (\rho^F, \sigma^F)| \lesssim \sup_{D_{u,\underline{u}}} \left( r^2 |u|^{\frac{1}{2}} |(\rho^F, \sigma^F)| \right) \left( \int_{\underline{u}_0}^{\underline{u}} |u|^{\frac{1}{2}} r^{-3} d\underline{u}' \right) \qquad (6.112)$$

$$\int_{\underline{H}} (u^4 r^4 |\nabla \alpha|^2 + u^2 r^6 |\widehat{\nabla}^3 \underline{\alpha}^F|^2) \lesssim \epsilon (\mathcal{W}_2 + \mathcal{Y}_3)$$

$$|\int_{D_{u,\underline{u}}} u^4 r^4 \nabla^2 \underline{\alpha} \widehat{\nabla}^3 (\rho^F, \sigma^F) \cdot \underline{\alpha}^F| \lesssim \sup_{D_{u,\underline{u}}} \left( |u|^{\frac{3}{2}} r |\underline{\alpha}^F| \right) \left( \int_{\underline{u}_0}^{\underline{u}} |u|^{\frac{1}{2}} r^{-3} d\underline{u}' \right) \qquad (6.113)$$

$$\int_{\underline{H}} (u^4 r^4 |\nabla \alpha|^2 + r^8 |\widehat{\nabla}^3 (\rho^F, \sigma^F)|^2) \lesssim \epsilon (\mathcal{W}_2 + \mathcal{Y}_3)$$

Similarly, we obtain

$$|\int_{D_{u,\underline{u}}} u^4 r^4 \nabla^2 \underline{\alpha} \widehat{\nabla}^2 (\rho^F, \sigma^F) \cdot \widehat{\nabla} \underline{\alpha}^F| \lesssim \sup_{D_{u,\underline{u}}} \left( |u|^{\frac{3}{2}} r^2 |\widehat{\nabla} \underline{\alpha}^F| \right) \left( \int_{\underline{u}_0}^{\underline{u}} |u|^{\frac{1}{2}} r^{-3} d\underline{u}' \right) \qquad (6.114)$$

$$\int_{\underline{H}} (u^4 r^4 |\nabla^2 \underline{\alpha}|^2 + r^6 |\widehat{\nabla}^2 (\rho^F, \sigma^F)|^2) \lesssim \epsilon (\mathcal{W}_2 + \mathcal{Y}_2)$$

The remaining terms are of lower order and can be estimated utilizing the boot-strap assumption straightwardly. Collecting all the terms one obtains

$$|\int_{D_{u,\underline{u}}} u^4 r^4 \nabla^2 \underline{\alpha}_{AB} \nabla^2 \nabla_A \mathfrak{T}_{3B}| \lesssim \epsilon (\mathcal{W}_2 + \mathcal{Y}_0 + \mathcal{Y}_1 + \mathcal{Y}_2 + \mathcal{Y}_3). \qquad (6.115)$$

Notice that all three terms are equivalent in terms of the scaling. Now consider

$$|\int_{D_{u,\underline{u}}} u^4 r^4 \nabla \nabla_A \underline{\beta}_B \nabla \nabla_A (D_B \mathfrak{T}_{33} - D_3 \mathfrak{T}_{B3})|. \qquad (6.116)$$

We estimate the top order terms

$$|\int_{D_{u,\underline{u}}} u^4 r^4 \nabla^2 \underline{\beta} \widehat{\nabla}^2 \underline{\alpha}^F \cdot \widehat{\nabla} \underline{\alpha}^F| \qquad (6.117)$$

$$\lesssim \sup_{D_{u,\underline{u}}} \left( |u|^{\frac{3}{2}} r^2 |\widehat{\nabla} \underline{\alpha}^F| \right) \left( \int_{\underline{u}_0}^{\underline{u}} |u|^{-\frac{1}{2}} r^{-2} d\underline{u}' \right) \int_{\underline{H}} (u^4 r^4 |\nabla \underline{\beta}|^2 + u^2 r^4 |\widehat{\nabla}^2 \underline{\alpha}^F|^2)$$

$$\lesssim \epsilon (\mathcal{W}_2 + \mathcal{Y}_2),$$

$$|\int_{D_{u,\underline{u}}} u^4 r^4 \nabla^2 \underline{\beta} \underline{\alpha}^F \cdot \widehat{\nabla}^3 \underline{\alpha}^F| \lesssim \sup_{D_{u,\underline{u}}} |u|^{\frac{3}{2}} r |\underline{\alpha}^F| \left( \int_{\underline{u}_0}^{\underline{u}} |u|^{-\frac{1}{2}} r^{-2} d\underline{u}' \right) \int_{\underline{H}} (u^4 r^4 |\nabla \underline{\beta}|^2 + u^2 r^6 |\widehat{\nabla}^3 \underline{\alpha}^F|^2)$$

$$\lesssim \epsilon (\mathcal{W}_2 + \mathcal{Y}_3),$$

$$|\int_{D_{u,\underline{u}}} u^4 r^4 \nabla^2 \underline{\beta} \widehat{\nabla}^2 \widehat{\nabla}_3 \underline{\alpha}^F \cdot (\rho^F, \sigma^F)| \tag{6.118}$$

$$\lesssim \sup_{D_{u,\underline{u}}} |u|^{\frac{1}{2}} r^2 |\rho^F, \sigma^F| \left( \int_{\underline{u}_0}^{u} |u|^{-\frac{1}{2}} r^{-2} d\underline{u}' \right)$$

$$\int_{\underline{H}} (u^4 r^4 |\nabla \underline{\beta}|^2 + u^4 r^4 |\widehat{\nabla}^2 \widehat{\nabla}_3 \underline{\alpha}^F|^2)$$

$$\lesssim \epsilon(\mathcal{W}_2 + \mathcal{Y}_3).$$

The remaining terms are lower orders and are controlled as follows

$$|\int_{D_{u,\underline{u}}} u^4 r^4 \nabla^2 \underline{\beta} \nabla (\nabla tr\chi \underline{\alpha}^F \cdot (\rho^F, \sigma^F))| \lesssim \epsilon^2 (\mathcal{W}_2 + \mathcal{Y}_0 + \mathcal{Y}_1 + \mathcal{Y}_2), \tag{6.119}$$

$$|\int_{D_{u,\underline{u}}} u^4 r^4 \nabla^2 \underline{\beta} \nabla (\chi \widehat{\nabla} \underline{\alpha}^F \cdot (\rho^F, \sigma^F))| \lesssim \epsilon(\mathcal{W}_2 + \mathcal{Y}_1 + \mathcal{Y}_2), \tag{6.120}$$

$$|\int_{D_{u,\underline{u}}} u^4 r^4 \nabla^2 \underline{\beta} \nabla (\chi \underline{\alpha}^F \cdot \widehat{\nabla} (\rho^F, \sigma^F))| \lesssim \epsilon(\mathcal{W}_2 + \mathcal{Y}_1 + \mathcal{Y}_2), \tag{6.121}$$

$$|\int_{D_{u,\underline{u}}} u^4 r^4 \nabla^2 \underline{\beta} \nabla (\nabla (\eta, \underline{\eta}) \underline{\alpha}^F \cdot \underline{\alpha}^F)| \lesssim \epsilon^2 (\mathcal{W}_2 + \mathcal{Y}_0 + \mathcal{Y}_1 + \mathcal{Y}_2), \tag{6.122}$$

$$|\int_{D_{u,\underline{u}}} u^4 r^4 \nabla^2 \underline{\beta} \nabla ((\eta, \underline{\eta}) \widehat{\nabla} \underline{\alpha}^F \cdot \underline{\alpha}^F)| \lesssim \epsilon^2 (\mathcal{W}_2 + \mathcal{Y}_1 + \mathcal{Y}_2), \tag{6.123}$$

$$|\int_{D_{u,\underline{u}}} u^4 r^4 \nabla^2 \underline{\beta} \nabla (\nabla \underline{\omega} \underline{\alpha}^F \cdot (\rho^F, \sigma^F))| \lesssim \epsilon^2 (\mathcal{W}_2 + \mathcal{Y}_0 + \mathcal{Y}_1 + \mathcal{Y}_2), \tag{6.124}$$

$$|\int_{D_{u,\underline{u}}} u^4 r^4 \nabla^2 \underline{\beta} \nabla (\underline{\omega} \widehat{\nabla} \underline{\alpha}^F \cdot (\rho^F, \sigma^F))| \lesssim \epsilon^2 (\mathcal{W}_2 + \mathcal{Y}_0 + \mathcal{Y}_1 + \mathcal{Y}_2), \tag{6.125}$$

$$|\int_{D_{u,\underline{u}}} u^4 r^4 \nabla^2 \underline{\beta} \nabla (\underline{\omega} \underline{\alpha}^F \cdot \widehat{\nabla} (\rho^F, \sigma^F))| \lesssim \epsilon^2 (\mathcal{W}_2 + \mathcal{Y}_1 + \mathcal{Y}_2), \tag{6.126}$$

$$|\int_{D_{u,\underline{u}}} u^4 r^4 \nabla^2 \underline{\beta} \nabla (\widehat{\nabla}_3 \alpha^F \cdot \widehat{\nabla} (\rho^F, \sigma^F))| \lesssim \epsilon(\mathcal{W}_2 + \mathcal{Y}_1 + \mathcal{Y}_2). \tag{6.127}$$

The collection of all the terms yields the desired estimate

$$|\mathcal{E}_{\nabla^2 \underline{\alpha}, \nabla^2 \underline{\beta}}| \lesssim \epsilon(\mathcal{W}_0 + \mathcal{W}_1 + \mathcal{W}_2 + \mathcal{Y}_0 + \mathcal{Y}_1 + \mathcal{Y}_2 + \mathcal{Y}_3). \tag{6.128}$$

Therefore the proof of the lemma follows. □

Similar to Proposition (6.7), we have the following bulk estimates

*Proposition 6.8.* *The following estimate holds*

$$\int_{D_{u,\underline{u}}} r^3 |u|^3 (4r - 3|u|) |\nabla^2 \underline{\beta}|^2 \lesssim \int_{\widehat{\Sigma}_0} u^4 r^4 |\nabla^2 \underline{\alpha}|^2 + \int_{\widehat{\Sigma}_0} u^4 r^4 |\nabla^2 \underline{\beta}|^2$$

*Proof.* The proof is a consequence of the Lemma (6.3) since the following estimates hold

$$\int_{H_u} u^4 r^4 |\nabla^2 \underline{\beta}|^2 + \int_{\underline{H}_{\underline{u}}} u^4 r^4 |\nabla^2 \underline{\alpha}|^2 + \int_{D_{u,\underline{u}}} 2r^3 |u|^3 (4r - 3|u|) |\nabla^2 \underline{\beta}|^2 \tag{6.129}$$

$$\lesssim \int_{\widehat{\Sigma}_0} u^4 r^4 |\nabla^2 \underline{\alpha}|^2 + \int_{\widehat{\Sigma}_0} u^4 r^4 |\nabla^2 \underline{\beta}|^2 + \epsilon(\mathbf{W} + \mathbf{Y})$$

and

$$\int_{H_u} u^4 r^4 |\nabla^2 \underline{\beta}|^2 + \int_{\underline{H}_{\underline{u}}} u^4 r^4 |\nabla^2 \underline{\alpha}|^2 \lesssim \int_{\widehat{\Sigma}_0} u^4 r^4 |\nabla^2 \underline{\alpha}|^2 + \int_{\widehat{\Sigma}_0} u^4 r^4 |\nabla^2 \underline{\beta}|^2 + \epsilon(\mathbf{W} + \mathbf{Y})$$

. □

*Lemma 6.4.* *The following estimates hold in the exterior region*

$$\int_H r^4|\beta|^2 + \int_{\underline{H}} r^4(|\rho|^2+|\sigma|^2) \leq \int_{\widehat{\Sigma}_0} r^4|\beta|^2 + \int_{\widehat{\Sigma}_0} r^4(|\rho|^2+|\sigma|^2) + \epsilon(\mathbf{W}+\mathbf{Y}),$$

$$\int_H r^6|\nabla\beta|^2 + \int_{\underline{H}} r^6(|\nabla\rho|^2+|\nabla\sigma|^2) \leq \int_{\widehat{\Sigma}_0} r^6|\nabla\beta|^2 + \int_{\widehat{\Sigma}_0} r^6(|\nabla\rho|^2+|\nabla\sigma|^2) + \epsilon(\mathbf{W}+\mathbf{Y}),$$

$$\int_H r^8|\nabla^2\beta|^2 + \int_{\underline{H}} r^8(|\nabla^2\rho|^2+|\nabla^2\sigma|^2) \leq \int_{\widehat{\Sigma}_0} r^8|\nabla^2\beta|^2 + \int_{\widehat{\Sigma}_0} r^8(|\nabla^2\rho|^2+|\nabla^2\sigma|^2) + \epsilon(\mathbf{W}+\mathbf{Y}).$$

*Proof.* Similar to the estimates for the Bianchi pair $(\alpha,\beta)$, we only concentrate on the topmost derivatives. □

*Lemma 6.5.* The following estimates hold in the exterior region

$$\int_{\underline{H}} r^2 u^2|\underline{\beta}|^2 + \int_H r^2 u^2(\rho^2+\sigma^2) \leq \int_{\widehat{\Sigma}_0} r^2 u^2|\underline{\beta}|^2 + \int_{\widehat{\Sigma}_0} r^2 u^2(\rho^2+\sigma^2) + \epsilon(\mathbf{W}+\mathbf{Y}),$$

$$\int_{\underline{H}} r^4 u^2|\nabla\underline{\beta}|^2 + \int_H r^4 u^2(|\nabla\rho|^2+|\nabla\sigma|^2) \leq \int_{\widehat{\Sigma}_0} r^4 u^2|\nabla\underline{\beta}|^2 + \int_{\widehat{\Sigma}_0} r^4 u^2(|\nabla\rho|^2+|\nabla\sigma|^2) + \epsilon(\mathbf{W}+\mathbf{Y}),$$

$$\int_{\underline{H}} r^6 u^2|\nabla^2\underline{\beta}|^2 + \int_H r^6 u^2(|\nabla^2\rho|^2+|\nabla^2\sigma|^2) \lesssim \int_{\widehat{\Sigma}_0} r^6 u^2|\nabla^2\underline{\beta}|^2 + \int_{\widehat{\Sigma}_0} r^6 u^2(|\nabla^2\rho|^2+|\nabla^2\sigma|^2)$$
$$+\epsilon(\mathbf{W}+\mathbf{Y}),$$

*Proof.* First, we note the following identity as a consequence of the integration Lemma (6.1)

$$\int_{\underline{H}} r^6 u^2|\nabla^2\underline{\beta}|^2 + \int_H r^6 u^2(|\nabla^2\rho|^2+|\nabla^2\sigma|^2) - \underbrace{\int_{D_{u,\underline{u}}} 3r^6 u^2(\Omega^{-1}\overline{\Omega tr\chi}-tr\chi)|\nabla^2\underline{\beta}|^2}$$

$$- \underbrace{\int_{D_{u,\underline{u}}}(3r^6 u^2 \Omega^{-1}\overline{\Omega tr\underline{\chi}} + 2r^6 u\Omega^{-1} - 4r^6 u^2 tr\underline{\chi})|\nabla^2(\rho,\sigma)|^2}_{<0}$$

$$= \int_{\widehat{\Sigma}_0} r^6 u^2|\nabla^2\underline{\beta}|^2 + \int_{\widehat{\Sigma}_0} r^6 u^2(|\nabla^2\rho|^2+|\nabla^2\sigma|^2) + \mathcal{E}^{\nabla^2\underline{\beta},\nabla^2(\rho,\sigma)},$$

where the error term $\mathcal{E}$ is explicitly computed as follows

$$\mathcal{E}^{\nabla^2\underline{\beta},\nabla^2(\rho,\sigma)} \sim \int_{D_{u,\underline{u}}} 2r^6 u^2 \Big\langle \nabla^2\underline{\beta}, -\nabla(\nabla(tr\chi)\underline{\beta}) + 2\nabla^2\Big(\omega\underline{\beta}_a$$
$$+ 2(\hat{\underline{\chi}}\cdot\beta)_a - 3(\underline{\eta}\rho - {}^*\underline{\eta}\sigma)_a - \frac{1}{2}(D_a\mathfrak{T}_{43} - D_3\mathfrak{T}_{4a})\Big) - \widehat{\chi}\cdot\nabla^2\underline{\beta}\Big\rangle$$
$$+ \int_{D_{u,\underline{u}}} 2r^6 u^2 \Big(\Big\langle \nabla^2\rho, -\nabla\Big(\nabla\Big(\frac{3}{2}tr\underline{\chi}\Big)\rho\Big) - \nabla^2\Big(\frac{1}{2}\hat{\chi}\cdot\underline{\alpha} + \zeta\cdot\underline{\beta} - 2\eta\cdot\underline{\beta} + \frac{1}{4}(D_3\mathfrak{T}_{34} - D_4\mathfrak{T}_{33})\Big)$$
$$- \widehat{\chi}\cdot\nabla^2\rho\Big)\Big\rangle + \int_{D_{u,\underline{u}}} 2r^6 u^2 \Big\langle \nabla^2\sigma, -\nabla\Big(\nabla\Big(\frac{3}{2}tr\underline{\chi}\Big)\sigma\Big) + \nabla^2\Big(\frac{1}{2}\hat{\chi}\cdot{}^*\underline{\alpha}$$
$$- \zeta\cdot{}^*\underline{\beta} - 2\eta\cdot{}^*\underline{\beta} + \frac{1}{4}(D_\mu\mathfrak{T}_{3\nu} - D_\nu\mathfrak{T}_{3\mu})\epsilon^{\mu\nu}_{\;\;34}\Big) - \widehat{\chi}\cdot\nabla^2\sigma\Big\rangle.$$

Once again, we need to control the underlined terms and prove that they either verify higher-order estimates or have favorable signs. First, we consider $\int_{D_{u,\underline{u}}} 3r^6 u^2(\Omega^{-1}\overline{\Omega tr\chi}-tr\chi)|\nabla^2\underline{\beta}|^2$ which verifies

$$\int_{D_{u,\underline{u}}} 3r^6 u^2(\Omega^{-1}\overline{\Omega tr\chi}-tr\chi)|\nabla^2\underline{\beta}|^2 \lesssim \epsilon \mathcal{W}_2 \qquad (6.130)$$

due to the bootstrap assumption (3) on the connections i.e., $|\Omega^{-1}\overline{\Omega tr\chi}-tr\chi| \lesssim \frac{\epsilon}{r^2}$. The second underlined term is a bit more delicate and needs care

$$\int_{D_{u,\underline{u}}}(3r^6 u^2 \Omega^{-1}\overline{\Omega tr\underline{\chi}} + 2r^6 u\Omega^{-1} - 4r^6 u^2 tr\underline{\chi})|\nabla^2(\rho,\sigma)|^2$$

$$= \int_{D_{u,\underline{u}}}\Big(3r^6 u^2(\Omega^{-1}\overline{\Omega tr\underline{\chi}} - tr\underline{\chi}) - r^6 u^2\Big(tr\underline{\chi} + \frac{2}{r}\Big) + 2r^6 u\Omega^{-1} + 2r^5 u^2\Big)|\nabla^2(\rho,\sigma)|^2$$

$$= \int_{D_{u,\underline{u}}}\Big(3r^6 u^2(\Omega^{-1}\overline{\Omega tr\underline{\chi}} - tr\underline{\chi}) - r^6 u^2\Big(tr\underline{\chi} + \frac{2}{r}\Big) + 2r^6 u(\Omega^{-1}-2) + 4r^6 u + 2r^5 u^2\Big)|\nabla^2(\rho,\sigma)|^2$$

Therefore the energy identity becomes

$$\int_{\underline{H}} r^6 u^2 |\nabla^2 \underline{\beta}|^2 + \int_H r^6 u^2 (|\nabla^2 \rho|^2 + |\nabla^2 \sigma|^2) - \int_{D_{u,\underline{u}}} (4r^6 u + 2r^5 u^2) |\nabla^2(\rho,\sigma)|^2 \quad (6.131)$$

$$= \int_{\widehat{\Sigma}_0} r^6 u^2 |\nabla^2 \underline{\beta}|^2 + \int_{\widehat{\Sigma}_0} r^6 u^2 (|\nabla^2 \rho|^2 + |\nabla^2 \sigma|^2) + \int_{D_{u,\underline{u}}} 3r^6 u^2 (\Omega^{-1}\overline{\Omega tr\chi} - tr\chi)|\nabla^2 \underline{\beta}|^2$$

$$+ \int_{D_{u,\underline{u}}} \left(3r^6 u^2 (\Omega^{-1}\overline{\Omega tr\underline{\chi}} - tr\underline{\chi}) - r^6 u^2 \left(tr\underline{\chi} + \frac{2}{r}\right) + 2r^6 u(\Omega^{-1} - 2)\right)|\nabla^2(\rho,\sigma)|^2 + \mathcal{E}^{\nabla^2 \underline{\beta}, \nabla^2(\rho,\sigma)}.$$

Now $4r^6 u + 2r^5 u^2 = -4r^6|u| + 2r^5 u^2 = 2r^5|u|(-2r + |u|) < 0$ since $u < 0$ and in the exterior domain for sufficiently large $\underline{u}_0$, we have $2r > |u|$. Therefore, in light of the connection estimates, we obtain the following energy inequality

$$\int_{\underline{H}} r^6 u^2 |\nabla^2 \underline{\beta}|^2 + \int_H r^6 u^2 (|\nabla^2 \rho|^2 + |\nabla^2 \sigma|^2)$$

$$\leq \int_{\widehat{\Sigma}_0} r^6 u^2 |\nabla^2 \underline{\beta}|^2 + \int_{\widehat{\Sigma}_0} r^6 u^2 (|\nabla^2 \rho|^2 + |\nabla^2 \sigma|^2) + |\mathcal{E}^{\nabla^2 \underline{\beta}, \nabla^2(\rho,\sigma)}| + \epsilon \mathcal{W}_2.$$

Therefore, the main remaining task is to estimate the error term $|\mathcal{E}^{\nabla^2 \underline{\beta}, \nabla^2(\rho,\sigma)}|$. Once again, we only focus on the potentially dangerous coupling terms since gravitational self-coupling terms are easily controlled under the boot-strap assumption. In particular, first, we ought to estimate

$$|\int_{D_{u,\underline{u}}} u^2 r^6 \nabla \nabla_B \underline{\beta}_A \nabla \nabla_B (D_A \mathfrak{T}_{43} - D_3 \mathfrak{T}_{4A})|. \quad (6.132)$$

We estimate each term separately

$$|\int_{D_{u,\underline{u}}} u^2 r^6 \nabla^2 \underline{\beta}(\rho^F, \sigma^F) \cdot \widehat{\nabla}^3 (\rho^F, \sigma^F)| \quad (6.133)$$

$$\lesssim \sup_{D_{u,\underline{u}}} \left(|u|^{\frac{1}{2}} r^2 |\rho^F, \sigma^F|\right) \left(\int_{\underline{u}_0}^{\underline{u}} |u|^{\frac{1}{2}} r^{-3} d\underline{u}'\right)$$

$$\int_{\underline{H}} (u^2 r^6 |\nabla^2 \underline{\beta}|^2 + r^8 |\widehat{\nabla}^3 (\rho^F, \sigma^F)|^2)$$

$$\lesssim \epsilon (\mathcal{W}_2 + \mathcal{Y}_3),$$

$$\int_{D_{u,\underline{u}}} u^2 r^6 \nabla^2 \underline{\beta} \widehat{\nabla}(\rho^F, \sigma^F) \cdot \widehat{\nabla}^2 (\rho^F, \sigma^F)|$$

$$\lesssim \sup_{u,\underline{u}} \left(\||u|^{\frac{3}{2}} r^{\frac{3}{2}} \widehat{\nabla}(\rho^F, \sigma^F)\|_{L^4(S_{u,\underline{u}})}\right) \left(\int_{\underline{u}_0}^{\underline{u}} |u|^{-\frac{1}{2}} r^{-2} d\underline{u}'\right)$$

$$\int_{\underline{H}} (u^2 r^6 |\nabla^2 \underline{\beta}|^2 + r^6 |\widehat{\nabla}^2 (\rho^F, \sigma^F)|^2 + r^8 |\widehat{\nabla}^3 (\rho^F, \sigma^F)|^2)$$

$$\lesssim \epsilon (\mathcal{W}_2 + \mathcal{Y}_2 + \mathcal{Y}_3).$$

Similarly,

$$|\int_{D_{u,\underline{u}}} u^2 r^6 \nabla^2 \underline{\beta} \nabla (\nabla \chi \alpha^F \cdot (\rho^F, \sigma^F))| \lesssim \epsilon^2 (\mathcal{W}_2 + \mathcal{Y}_0 + \mathcal{Y}_1 + \mathcal{Y}_2), \quad (6.134)$$

$$|\int_{D_{u,\underline{u}}} u^2 r^6 \nabla^2 \underline{\beta} \nabla (\chi \widehat{\nabla} \alpha^F \cdot (\rho^F, \sigma^F))| \lesssim \epsilon (\mathcal{W}_2 + \mathcal{Y}_1 + \mathcal{Y}_2), \quad (6.135)$$

$$|\int_{D_{u,\underline{u}}} u^2 r^6 \nabla^2 \underline{\beta} \nabla (\chi \alpha^F \cdot \widehat{\nabla}(\rho^F, \sigma^F))| \lesssim \epsilon (\mathcal{W}_2 + \mathcal{Y}_1 + \mathcal{Y}_2), \quad (6.136)$$

$$|\int_{D_{u,\underline{u}}} u^2 r^6 \nabla^2 \underline{\beta} \nabla (\nabla \chi \underline{\alpha}^F \cdot (\rho^F, \sigma^F))| \lesssim \epsilon^2 (\mathcal{W}_2 + \mathcal{Y}_0 + \mathcal{Y}_1 + \mathcal{Y}_2), \quad (6.137)$$

$$|\int_{D_{u,\underline{u}}} u^2 r^6 \nabla^2 \underline{\beta} \nabla (\chi \widehat{\nabla} \underline{\alpha}^F \cdot (\rho^F, \sigma^F))| \lesssim \epsilon (\mathcal{W}_2 + \mathcal{Y}_1 + \mathcal{Y}_2), \quad (6.138)$$

$$|\int_{D_{u,\underline{u}}} u^2 r^6 \nabla^2 \underline{\beta} \nabla (\chi \underline{\alpha}^F \cdot \widehat{\nabla}(\rho^F, \sigma^F))| \lesssim \epsilon^2 (\mathcal{W}_2 + \mathcal{Y}_1 + \mathcal{Y}_2). \quad (6.139)$$

Next we need to estimate the following term

$$\left|\int_{D_{u,\underline{u}}} u^2 r^6 \nabla^2 \rho \nabla^2 (D_3 \mathfrak{I}_{34} - D_4 \mathfrak{I}_{33})\right|. \tag{6.140}$$

Through explicit computation, we estimate the top-order derivative terms first.

$$\left|\int_{D_{u,\underline{u}}} u^2 r^6 \nabla^2 \rho \underline{\alpha}^F \cdot \widehat{\nabla}^3 (\rho^F, \sigma^F)\right| \tag{6.141}$$

$$\lesssim \sup_{D_{u,\underline{u}}} \left(|u|^{\frac{3}{2}} r |\underline{\alpha}^F|\right) \left(\int_{u_0}^{u} |u|^{-\frac{3}{2}} r^{-1}\right)$$

$$\int_H \left(u^2 r^6 |\nabla^2 \rho|^2 + u^2 r^6 |\widehat{\nabla}^3 (\rho^F, \sigma^F)|^2\right)$$

$$\lesssim \epsilon (\mathcal{W}_2 + \mathcal{Y}_3),$$

$$\left|\int_{D_{u,\underline{u}}} u^2 r^6 \nabla^2 \rho \widehat{\nabla} \underline{\alpha}^F \cdot \widehat{\nabla}^2 (\rho^F, \sigma^F)\right|$$

$$\lesssim \sup_{u,\underline{u}} \left(\||u|^{\frac{3}{2}} r^{\frac{3}{2}} \widehat{\nabla} \underline{\alpha}^F\|_{L^4(S_{u,\underline{u}})}\right) \left(\int_{u_0}^{u} |u|^{-\frac{3}{2}} r^{-1} du'\right)$$

$$\int_H \left(u^2 r^6 |\nabla^2 \rho|^2 + u^2 r^4 |\widehat{\nabla}^2 (\rho^F, \sigma^F)|^2 + u^2 r^6 |\widehat{\nabla}^3 (\rho^F, \sigma^F)|^2\right)$$

$$\lesssim \epsilon (\mathcal{W}_2 + \mathcal{Y}_2 + \mathcal{Y}_3),$$

$$\left|\int_{D_{u,\underline{u}}} u^2 r^6 \nabla^2 \rho \widehat{\nabla}^3 \underline{\alpha}^F \cdot (\rho^F, \sigma^F)\right|$$

$$\lesssim \sup_{D_{u,\underline{u}}} \left(|u|^{\frac{1}{2}} r^2 |\rho^F, \sigma^F|\right) \left(\int_{\underline{u}_0}^{\underline{u}} |u|^{\frac{1}{2}} r^{-3} d\underline{u}'\right) \int_{\underline{H}} \left(u^2 r^6 |\nabla^2 \rho|^2 + u^2 r^6 |\widehat{\nabla}^3 \underline{\alpha}^F|^2\right)$$

$$\lesssim \epsilon (\mathcal{W}_2 + \mathcal{Y}_2 + \mathcal{Y}_3),$$

Similarly, the lower order terms terms are estimated as follows

$$\left|\int_{D_{u,\underline{u}}} u^2 r^6 \nabla^2 \rho \nabla (tr\chi \underline{\alpha}^F \cdot \widehat{\nabla} \underline{\alpha}^F)\right| \lesssim \epsilon (\mathcal{W}_2 + \mathcal{Y}_1 + \mathcal{Y}_2), \tag{6.142}$$

$$\left|\int_{D_{u,\underline{u}}} u^2 r^6 \nabla^2 \rho \nabla (\nabla tr\chi \underline{\alpha}^F \cdot \underline{\alpha}^F)\right| \lesssim \epsilon (\mathcal{W}_2 + \mathcal{Y}_0 + \mathcal{Y}_1 + \mathcal{Y}_2), \tag{6.143}$$

$$\left|\int_{D_{u,\underline{u}}} u^2 r^6 \nabla^2 \rho \nabla (\nabla \underline{\eta} \underline{\alpha}^F \cdot (\rho^F, \sigma^F))\right| \lesssim \epsilon (\mathcal{W}_2 + \mathcal{Y}_0 + \mathcal{Y}_1 + \mathcal{Y}_2), \tag{6.144}$$

$$\left|\int_{D_{u,\underline{u}}} u^2 r^6 \nabla^2 \rho \nabla (\underline{\eta} \widehat{\nabla} \underline{\alpha}^F \cdot (\rho^F, \sigma^F))\right| \lesssim \epsilon^2 (\mathcal{W}_2 + \mathcal{Y}_1 + \mathcal{Y}_2), \tag{6.145}$$

$$\left|\int_{D_{u,\underline{u}}} u^2 r^6 \nabla^2 \rho \nabla (\underline{\eta} \underline{\alpha}^F \cdot \widehat{\nabla} (\rho^F, \sigma^F))\right| \lesssim \epsilon^2 (\mathcal{W}_2 + \mathcal{Y}_1 + \mathcal{Y}_2), \tag{6.146}$$

$$\left|\int_{D_{u,\underline{u}}} u^2 r^6 \nabla^2 \rho \nabla (\omega \widehat{\nabla} \underline{\alpha}^F \cdot \underline{\alpha}^F)\right| \lesssim \epsilon^2 (\mathcal{W}_2 + \mathcal{Y}_1 + \mathcal{Y}_2), \tag{6.147}$$

$$\left|\int_{D_{u,\underline{u}}} u^2 r^6 \nabla^2 \rho \nabla (\nabla \omega \underline{\alpha}^F \cdot \underline{\alpha}^F)\right| \lesssim \epsilon^2 (\mathcal{W}_2 + \mathcal{Y}_0 + \mathcal{Y}_1 + \mathcal{Y}_2), \tag{6.148}$$

$$\left|\int_{D_{u,\underline{u}}} u^2 r^6 \nabla^2 \rho \nabla (\nabla \hat{\chi} \alpha^F \cdot \underline{\alpha}^F)\right| \lesssim \epsilon^2 (\mathcal{W}_2 + \mathcal{Y}_0 + \mathcal{Y}_1 + \mathcal{Y}_2), \tag{6.149}$$

$$\left|\int_{D_{u,\underline{u}}} u^2 r^6 \nabla^2 \rho \nabla (\hat{\chi} \widehat{\nabla} \alpha^F \cdot \underline{\alpha}^F)\right| \lesssim \epsilon^2 (\mathcal{W}_2 + \mathcal{Y}_1 + \mathcal{Y}_2), \tag{6.150}$$

$$\left|\int_{D_{u,\underline{u}}} u^2 r^6 \nabla^2 \rho \nabla (\hat{\chi} \alpha^F \cdot \widehat{\nabla} \underline{\alpha}^F)\right| \lesssim \epsilon^2 (\mathcal{W}_2 + \mathcal{Y}_1 + \mathcal{Y}_2), \tag{6.151}$$

$$\left|\int_{D_{u,\underline{u}}} u^2 r^6 \nabla^2 \rho \nabla (\eta \rho^F \cdot \widehat{\nabla} \alpha^F)\right| \lesssim \epsilon^2 (\mathcal{W}_2 + \mathcal{Y}_1 + \mathcal{Y}_2), \tag{6.152}$$

$$\left|\int_{D_{u,\underline{u}}} u^2 r^6 \nabla^2 \rho \nabla (\nabla \eta \rho^F \cdot \alpha^F)\right| \lesssim \epsilon^2 (\mathcal{W}_2 + \mathcal{Y}_0 + \mathcal{Y}_1 + \mathcal{Y}_2). \tag{6.153}$$

Collecting all the terms, one obtains

$$|\mathcal{E}_{\nabla^2 \underline{\beta}, \nabla^2(\rho,\sigma)}| \lesssim \epsilon(\mathcal{W}_0 + \mathcal{W}_1 + \mathcal{W}_2 + \mathcal{Y}_0 + \mathcal{Y}_1 + \mathcal{Y}_2 + \mathcal{Y}_3). \quad (6.154)$$

All the terms together complete the proof of the lemma. □

Similar to Propositions 6.7 and 6.8, the following holds.

*Proposition 6.9.* The following bulk estimates hold

$$- \int_{D_{u,\underline{u}}} (4r^6 u + 2r^5 u^2)|\nabla^2(\rho,\sigma)|^2 \quad (6.155)$$

$$\lesssim \int_{\widehat{\Sigma}_0} r^6 u^2 |\nabla^2 \underline{\beta}|^2 + \int_{\widehat{\Sigma}_0} r^6 u^2 (|\nabla^2 \rho|^2 + |\nabla^2 \sigma|^2) + \epsilon(\mathbf{W} + \mathbf{Y})$$

*Proof.* Recall the identity (6.131) derived in the proof of the Lemma 6.5

$$\int_{\underline{H}} r^6 u^2 |\nabla^2 \underline{\beta}|^2 + \int_H r^6 u^2 (|\nabla^2 \rho|^2 + |\nabla^2 \sigma|^2) - \int_{D_{u,\underline{u}}} (4r^6 u + 2r^5 u^2)|\nabla^2(\rho,\sigma)|^2$$

$$= \int_{\widehat{\Sigma}_0} r^6 u^2 |\nabla^2 \underline{\beta}|^2 + \int_{\widehat{\Sigma}_0} r^6 u^2 (|\nabla^2 \rho|^2 + |\nabla^2 \sigma|^2) + \int_{D_{u,\underline{u}}} 3r^6 u^2 (\Omega^{-1}\overline{\Omega tr\chi} - tr\chi)|\nabla^2 \underline{\beta}|^2$$

$$+ \int_{D_{u,\underline{u}}} \left(3r^6 u^2(\Omega^{-1}\overline{\Omega tr\underline{\chi}} - tr\underline{\chi}) - r^6 u^2 \left(tr\underline{\chi} + \frac{2}{r}\right) + 2r^6 u(\Omega^{-1} - 2)\right)|\nabla^2(\rho,\sigma)|^2 + \mathcal{E}^{\nabla^2 \underline{\beta}, \nabla^2(\rho,\sigma)}.$$

and use the estimates Lemma 6.5 to conclude the proof. □

Due to the global Sobolev inequalities on the null cones (3.1), we need to estimate the mixed derivatives of the Weyl curvature components. We have the following lemma regarding the mixed norm of the Weyl curvature.

*Lemma 6.6.* The following estimates hold for the mixed derivatives of the Weyl curvature

$$\int_H r^6 |D_4 \alpha|^2 + \int_{\underline{H}} r^6 |D_4 \beta|^2 \lesssim \int_{\widehat{\Sigma}_0} r^6 |D_4 \alpha|^2 + \int_{\widehat{\Sigma}_0} r^6 |D_4 \beta|^2 + \epsilon(\mathbf{W} + \mathbf{Y}), \quad (6.156)$$

$$\int_{H_u} u^6 |D_3 \underline{\beta}|^2 + \int_{\underline{H}_{\underline{u}}} u^6 |D_3 \underline{\alpha}|^2 \lesssim \int_{\widehat{\Sigma}_0} u^4 r^2 |D_3 \underline{\alpha}|^2 + \int_{\widehat{\Sigma}_0} u^4 r^2 |D_3 \underline{\beta}|^2 + \epsilon(\mathbf{W} + \mathbf{Y}), \quad (6.157)$$

$$\int_H r^8 |\nabla D_4 \alpha|^2 + \int_{\underline{H}} r^8 |\nabla D_4 \beta|^2 \lesssim \int_{\widehat{\Sigma}_0} r^6 |\nabla D_4 \alpha|^2 + \int_{\widehat{\Sigma}_0} r^6 |\nabla D_4 \beta|^2 + \epsilon(\mathbf{W} + \mathbf{Y}), \quad (6.158)$$

$$\int_{H_u} u^6 r^2 |\nabla D_3 \underline{\beta}|^2 + \int_{\underline{H}_{\underline{u}}} u^6 r^2 |\nabla D_3 \underline{\alpha}|^2 \lesssim \int_{\widehat{\Sigma}_0} u^6 r^2 |\nabla D_3 \underline{\alpha}|^2 + \int_{\widehat{\Sigma}_0} u^6 r^2 |\nabla D_3 \underline{\beta}|^2$$

$$+ \epsilon(\mathbf{W} + \mathbf{Y}).$$

*Proof.* The proof goes in an exactly similar fashion as the spatial energy estimates. However, there can be borderline terms due to the Yang–Mills source terms. We focus on estimating the mixed terms. Here we will have terms of the type $\widehat{\nabla}_4 \widehat{\nabla}_4 \alpha^F$ (and $\widehat{\nabla}_3 \widehat{\nabla}_3 \underline{\alpha}^F$ resp.) appearing from the term $D_4 D_4 \mathfrak{T}_{ab}$ ($D_3 D_3 \mathfrak{T}_{ab}$ resp.). This is precisely why we needed to include the term $\int_H r^6 |\widehat{\nabla}_4 \widehat{\nabla}_4 \alpha_4^F|^2$ ($\int_{\underline{H}} |\widehat{\nabla}_3^2 \underline{\alpha}^F|^2$ resp.) in the Yang–Mills energy. Also note that $\nabla_4$ and $\nabla$ are similar in terms of scaling property. Main estimates read

$$\left|\int_{D_{u,\underline{u}}} r^6 D_4 \alpha \widehat{\nabla} \rho^F \cdot \widehat{\nabla}_4 \alpha^F\right| \lesssim \epsilon(\mathcal{W}_1 + \mathcal{Y}_2),$$

$$\left|\int_{D_{u,\underline{u}}} r^6 D_4 \alpha \rho^F \cdot \widehat{\nabla}_4^2 \alpha^F\right| \lesssim \epsilon(\mathcal{W}_1 + \mathcal{Y}_2), \quad (6.159)$$

$$\left|\int_{D_{u,\underline{u}}} r^6 D_4 \alpha \nabla_4 tr\chi \rho^F \cdot \rho^F\right| \lesssim \epsilon^2(\mathcal{W}_1 + \mathcal{Y}_0 + \mathcal{Y}_1), \quad (6.160)$$

$$\left|\int_{D_{u,\underline{u}}} r^6 D_4 \alpha \alpha^F \cdot \widehat{\nabla}_4^2 \rho^F\right| \lesssim \epsilon(\mathcal{W}_1 + \mathcal{Y}_2), \quad (6.161)$$

$$\left|\int_{D_{u,\underline{u}}} r^6 D_4 \alpha \widehat{\nabla}_4 \underline{\alpha}^F \cdot \alpha^F tr\chi\right| \lesssim \epsilon(\mathcal{W}_1 + \mathcal{Y}_1), \quad (6.162)$$

$$\left|\int_{D_{u,\underline{u}}} r^6 D_4 \alpha \widehat{\nabla}_4 \underline{\alpha}^F \cdot \widehat{\nabla}_4 \alpha^F\right| \lesssim \epsilon(\mathcal{W}_1 + \mathcal{Y}_1 + \mathcal{Y}_2), \quad (6.163)$$

$$|\int_{D_{u,\underline{u}}} r^6 D_4\alpha \nabla_4 tr\chi \alpha^F \cdot \underline{\alpha}^F| \lesssim \epsilon^2(\mathcal{W}_1 + \mathcal{Y}_0), \tag{6.164}$$

$$|\int_{D_{u,\underline{u}}} r^6 D_4\alpha \underline{\alpha}^F \cdot \widehat{\nabla}_4 \widehat{\nabla}_4 \alpha^F| \lesssim \epsilon(\mathcal{W}_1 + \mathcal{Y}_2), \tag{6.165}$$

$$|\int_{D_{u,\underline{u}}} r^6 D_4\alpha(\eta + \underline{\eta})\widehat{\nabla}_4 \alpha^F \cdot \rho^F| \lesssim \epsilon^2(\mathcal{W}_1 + \mathcal{Y}_1), \tag{6.166}$$

$$|\int_{D_{u,\underline{u}}} r^6 D_4\alpha \nabla_4 \eta \alpha^F \cdot \rho^F| \lesssim \epsilon^2(\mathcal{W}_1 + \mathcal{Y}_0),$$

$$|\int_{D_{u,\underline{u}}} r^6 D_4\beta \nabla_4(tr\chi)\beta| \lesssim \epsilon(\mathcal{W}_0 + \mathcal{W}_1), |\int_{D_{u,\underline{u}}} r^6 D_4\beta(\nabla_4\omega)\beta| \lesssim \epsilon(\mathcal{W}_0 + \mathcal{W}_1), \tag{6.167}$$

$$|\int_{D_{u,\underline{u}}} r^6 \omega |D_4\beta|^2|\lesssim \epsilon \mathcal{W}_1, \ |\int_{D_{u,\underline{u}}} r^6 D_4\beta \nabla_4\eta \alpha| \lesssim \epsilon(\mathcal{W}_0 + \mathcal{W}_1), \tag{6.168}$$

$$|\int_{D_{u,\underline{u}}} r^6 D_4\beta \eta D_4\alpha| \lesssim \epsilon \mathcal{W}_1, \tag{6.169}$$

$$|\int_{D_{u,\underline{u}}} D_4\beta D_4\alpha(\eta,\underline{\eta})| \lesssim \epsilon \mathcal{W}_1, |\int_{D_{u,\underline{u}}} D_4\beta \alpha^F \cdot \widehat{\nabla}_4\alpha^F| \lesssim \epsilon(\mathcal{W}_1 + \mathcal{Y}_1), \tag{6.170}$$

$$|\int_{D_{u,\underline{u}}} D_4\beta(\rho^F, \sigma^F) \cdot \widehat{\nabla}_4\alpha^F| \lesssim \epsilon(\mathcal{W}_1 + \mathcal{Y}_1),$$

$$|\int_{D_{u,\underline{u}}} D_4\beta(\eta,\underline{\eta})\alpha^F \cdot \alpha^F| \lesssim \epsilon^2(\mathcal{W}_1 + \mathcal{Y}_0), \tag{6.171}$$

$$|\int_{D_{u,\underline{u}}} r^6 D_4\beta \widehat{\nabla}\alpha^F \cdot \widehat{\nabla}_4\alpha^F| \lesssim \epsilon(\mathcal{W}_1 + \mathcal{Y}_1), \tag{6.172}$$

$$|\int_{D_{u,\underline{u}}} r^6 D_4\beta \alpha^F \cdot \widehat{\nabla}\widehat{\nabla}_4\alpha^F| \lesssim \epsilon(\mathcal{W}_1 + \mathcal{Y}_2) \tag{6.173}$$

Here we estimate $\nabla_4 \omega$ employing the estimates from Propositions (6.4) and (6.5). The collection of all the terms yields the proof. □

*Remark 6.* Notice an important fact that we can not simply replace $D_4\beta$ ($D_3\underline{\beta}$ resp.) by its evolution equation since that would imply controlling $\nabla \alpha$ ($\nabla \underline{\alpha}$) on $\underline{H}$ ($H$) which is impossible. Therefore, we need to include $D_4\beta$ ($D_3\underline{\beta}$) in the initial data.

Now we move on to estimating the Yang–Mills curvature components on the null hypersurfaces. To accomplish this we need to establish the following proposition.

*Proposition 6.10.* Null Yang–Mills equations are manifestly hyperbolic.

*Remark 7.* Note that this is natural since Yang–Mills equations are derivable from a Lagrangian and as such it is equipped with a canonical stress-energy tensor.

*Proof.* The integration lemma can be utilized to prove the manifestly hyperbolic characteristics of the Yang–Mills equations while expressed in the double null coordinates. First consider the null triple $(\underline{\alpha}^F, \rho^F, \sigma^F)$ and recall their gauge covariant evolution equations

$$\widehat{\nabla}_4 \underline{\alpha}^F + \frac{1}{2} tr\chi \underline{\alpha}^F = -\widehat{\nabla}\rho^F + {}^*\widehat{\nabla}\sigma^F - 2{}^*\underline{\eta}\sigma^F - 2\underline{\eta}\rho^F + 2\omega\underline{\alpha}^F - \hat{\underline{\chi}} \cdot \alpha^F, \tag{6.174}$$

$$\widehat{\nabla}_3 \rho^F + tr\underline{\chi}\rho^F = -\widehat{\operatorname{div}}\underline{\alpha}^F + (\eta - \underline{\eta}) \cdot \underline{\alpha}^F, \tag{6.175}$$

$$\widehat{\nabla}_3 \sigma^F + tr\underline{\chi}\sigma^F = -\widehat{\operatorname{curl}}\underline{\alpha}^F + (\eta - \underline{\eta}) \cdot {}^*\underline{\alpha}^F. \tag{6.176}$$

Now define $f_1 := |\underline{\alpha}^F|^2_{\gamma,\delta}$, $f_2 = |\rho^F|^2_{\gamma,\delta}$, and $f_3 := |\sigma^F|^2_{\gamma,\delta}$. With these definitions in mind, let us apply the integration lemma to $f_1$, $f_2$, and $f_3$ to yield

$$\int_{D_{u,\underline{u}}} \nabla_4 f_1 + \int_{D_{u,\underline{u}}} \nabla_3 f_2 + \int_{D_{u,\underline{u}}} \nabla_3 f_3 = \int_{\underline{H}_u} f_1 + \int_{H_u} f_2 + \int_{H_u} f_3 - \int_{\underline{H}_{u_0}} f_1 \tag{6.177}$$

$$- \int_{H_{u_0}} f_2 - \int_{H_{u_0}} f_3 + \int_{D_{u,\underline{u}}} (2\omega - tr\chi)f_1 + \int_{D_{u,\underline{u}}} (2\underline{\omega} - tr\underline{\chi})(f_2 + f_3).$$

In order for these equations to exhibit a hyperbolic characteristic, the left-hand side should simplify to terms that are algebraic in $\underline{\alpha}^F, \rho^F$, and $\sigma^F$ upon using the null evolution equations. Now we note the most important point: $f_1$, $f_2$, and $f_3$ are gauge invariant objects and therefore we have the following as a consequence of the compatibility of the gauge covariant connection $\widehat{\nabla}$ with the metrics $\gamma$ and $\delta$ of the fibers ($\gamma$ is a Riemannian metric on the topological two-sphere $S_{u,\underline{u}}$ and $\delta$ is a Riemannian metric on the fibers of the associated gauge bundle)

$$\nabla_4 f_1 = 2\langle \underline{\alpha}^F, \widehat{\nabla}_4 \underline{\alpha}^F \rangle_{\gamma,\delta}, \quad \nabla_3 f_2 = 2\langle \rho^F, \widehat{\nabla}_3 \rho^F \rangle_{\gamma,\delta}, \quad \nabla_3 f_3 = 2\langle \sigma^F, \widehat{\nabla}_3 \sigma^F \rangle_{\gamma,\delta}. \tag{6.178}$$

Now we only focus on the principal terms for the hyperbolicity argument

$$\langle \underline{\alpha}^F, \widehat{\nabla}_4 \underline{\alpha}^F \rangle_{\gamma,\delta} = \langle \underline{\alpha}^F, -\widehat{\nabla}\rho^F + *\widehat{\nabla}\sigma^F + \cdots \rangle_{\gamma,\delta}$$
$$= -\mathrm{div}\langle \underline{\alpha}^F, \rho^F \rangle_{\gamma,\delta} + \langle \widehat{\mathrm{div}}\underline{\alpha}^F, \rho^F \rangle_{\gamma,\delta} + \mathrm{div}\; {}^*\langle \underline{\alpha}^F, \sigma^F \rangle_{\gamma,\delta} + \langle \widehat{\mathrm{curl}}\underline{\alpha}^F, \sigma^F \rangle + \mathrm{l.o.t}$$

$$\langle \rho^F, \widehat{\nabla}_3 \rho^F \rangle_{\gamma,\delta} = \langle \rho^F, -\widehat{\mathrm{div}}\underline{\alpha}^F + \cdots \rangle_{\gamma,\delta}$$
$$\langle \sigma^F, \widehat{\nabla}_3 \sigma^F \rangle_{\gamma,\delta} = \langle \sigma^F, -\widehat{\mathrm{curl}}\underline{\alpha}^F + \cdots \rangle_{\gamma,\delta}. \tag{6.180}$$

Now, in addition, we have

$$\langle \underline{\alpha}^F, \widehat{\nabla}_4 \underline{\alpha}^F \rangle_{\gamma,\delta} + \langle \rho^F, \widehat{\nabla}_3 \rho^F \rangle_{\gamma,\delta} + \langle \sigma^F, \widehat{\nabla}_3 \sigma^F \rangle_{\gamma,\delta} \tag{6.181}$$

$$= -\mathrm{div}\langle \underline{\alpha}^F, \rho^F \rangle_{\gamma,\delta} + \langle \widehat{\mathrm{div}}\underline{\alpha}^F, \rho^F \rangle_{\gamma,\delta} + \mathrm{div}\; {}^*\langle \underline{\alpha}^F, \sigma^F \rangle_{\gamma,\delta} + \langle \widehat{\mathrm{curl}}\underline{\alpha}^F, \sigma^F \rangle$$

$$- \langle \rho^F, \widehat{\mathrm{div}}\underline{\alpha}^F \rangle_{\gamma,\delta} - \langle \sigma^F, \widehat{\mathrm{curl}}\underline{\alpha}^F \rangle_{\gamma,\delta} + \mathrm{l.o.t}$$

$$= -\mathrm{div}\langle \underline{\alpha}^F, \rho^F \rangle_{\gamma,\delta} + \mathrm{div}\; {}^*\langle \underline{\alpha}^F, \sigma^F \rangle_{\gamma,\delta} + \mathrm{l.o.t}$$

which upon integration over the topological two–sphere yields terms that are algebraic in $\underline{\alpha}^F, \rho^F$, and $\sigma^F$. Here $\langle \underline{\alpha}^F, \rho^F \rangle_{\gamma,\delta} = (\underline{\alpha}^F)^P{}_{Qab}(\rho^F)^Q{}_P{}^b$ and ${}^*\langle \underline{\alpha}^F, \sigma^F \rangle_{\gamma,\delta} = \epsilon^{ca}(\underline{\alpha}^F)^P{}_{Qab}\rho^Q{}_P{}^b$. The most vital property that is utilized here is the compatibility of the connection $\widehat{\nabla}$ with the inner product $\langle , \rangle_{\gamma,\delta}$ induced by the fiber metrics $\gamma$ and $\delta$ together with the Hodge structure present in the null Yang–Mills equations. Notice that nowhere in the procedure do we need the information about the Yang–Mills connection one–form $A^P{}_{Q\mu}dx^\mu$. The remaining Yang–Mills null evolution equations may be utilized to obtain energy identities associated with the triple $\underline{\alpha}^F, \rho^F, \sigma^F$. This concludes the proof of the hyperbolic characteristics of the null Yang–Mills equations. □

*Lemma 6.7.* The following estimates hold true in the exterior region

$$\int_H r^2 |\underline{\alpha}^F|^2 + \int_{\underline{H}} r^2 (|\rho^F|^2 + |\sigma^F|^2) \lesssim \int_{\widehat{\Sigma}_0} r^2 |\underline{\alpha}^F|^2 + \int_{\widehat{\Sigma}_0} r^2 (|\rho^F|^2 + |\sigma^F|^2) + \epsilon \mathbf{Y},$$

$$\int_H r^4 |\widehat{\nabla}\underline{\alpha}^F|^2 + \int_{\underline{H}} r^4 (|\widehat{\nabla}\rho^F|^2 + |\widehat{\nabla}\sigma^F|^2) \lesssim \int_{\widehat{\Sigma}_0} r^4 |\widehat{\nabla}\underline{\alpha}^F|^2 + \int_{\widehat{\Sigma}_0} r^4 (|\widehat{\nabla}\rho^F|^2 + |\widehat{\nabla}\sigma^F|^2) + \epsilon(\mathbf{W} + \mathbf{Y}),$$

$$\int_H r^6 |\widehat{\nabla}^2 \underline{\alpha}^F|^2 + \int_{\underline{H}} r^6 (|\widehat{\nabla}^2 \rho^F|^2 + |\widehat{\nabla}^2 \sigma^F|^2) \lesssim \int_{\widehat{\Sigma}_0} r^6 |\widehat{\nabla}^2 \underline{\alpha}^F|^2 + \int_{\widehat{\Sigma}_0} r^6 (|\widehat{\nabla}^2 \rho^F|^2 + |\widehat{\nabla}^2 \sigma^F|^2)$$
$$+ \epsilon(\mathbf{W} + \mathbf{Y}),$$

$$\int_H r^8 |\widehat{\nabla}^3 \underline{\alpha}^F|^2 + \int_{\underline{H}} r^8 (|\widehat{\nabla}^3 \rho^F|^2 + |\widehat{\nabla}^3 \sigma^F|^2) \lesssim \int_{\widehat{\Sigma}_0} r^8 |\widehat{\nabla}^3 \underline{\alpha}^F|^2 + \int_{\widehat{\Sigma}_0} r^8 (|\widehat{\nabla}^3 \rho^F|^2 + |\widehat{\nabla}^3 \sigma^F|^2)$$
$$+ \epsilon(\mathbf{W} + \mathbf{Y})$$

*Proof.* We only focus on estimating the second derivatives. The remaining derivatives follow in an exactly similar way. First, note the following integration by parts identity

$$\int_H r^6 |\widehat{\nabla}^2 \underline{\alpha}^F|^2 + \int_{\underline{H}} r^6 (|\widehat{\nabla}^2 \rho^F|^2 + |\widehat{\nabla}^2 \sigma^F|^2) - \int_{D_{u,\underline{u}}} (3r^6 \Omega^{-1} \overline{\Omega tr\chi} - 3r^6 tr\underline{\chi}) |\widehat{\nabla}(\rho^F, \sigma^F)|^2$$

$$- \int_{D_{u,\underline{u}}} r^6 (3\Omega^{-1} \overline{\Omega tr\underline{\chi}} - 2tr\chi) |\widehat{\nabla}^2 \underline{\alpha}^F|^2 = \int_{\widehat{\Sigma}_0} r^6 |\widehat{\nabla}^2 \underline{\alpha}^F|^2 + \int_{\widehat{\Sigma}_0} r^6 (|\widehat{\nabla}^2 \rho^F|^2 + |\widehat{\nabla}^2 \sigma^F|^2)$$
$$+ \mathcal{E}^{\nabla^2 \underline{\alpha}^F, \nabla^2 (\rho^F, \sigma^F)},$$

where $\mathcal{E}^{\widehat{\nabla}^2 \alpha^F, \widehat{\nabla}^2(\rho^F, \sigma^F)}$ verifies

$$\mathcal{E}^{\widehat{\nabla}^2 \alpha^F, \widehat{\nabla}^2(\rho^F, \sigma^F)} \sim \int_{D_{u,\underline{u}}} 2r^6 \left\langle \widehat{\nabla}^2 \rho^F, -\widehat{\nabla}(\nabla(tr\chi)\rho^F) - \widehat{\nabla}^2((\eta - \underline{\eta}) \cdot \alpha^F) \right.$$

$$\left. -\widehat{\chi} \cdot \widehat{\nabla}^2 \rho^F + \mathcal{E}^{[\widehat{\nabla}_4, \widehat{\nabla}^2]\rho^F} + \mathcal{E}^{[\widehat{\nabla},\widehat{\nabla}^2](\rho^F,\sigma^F)} \right\rangle + \int_{D_{u,\underline{u}}} 2r^6 \left\langle \widehat{\nabla}^2 \sigma^F, -\widehat{\nabla}(\widehat{\nabla}(tr\chi)\sigma^F) + \widehat{\nabla}^2((\eta - \underline{\eta}) \cdot {}^\star \alpha^F) \right.$$

$$\left. -\widehat{\nabla}(\widehat{\chi} \cdot \widehat{\nabla}\sigma) + \mathcal{E}^{[\widehat{\nabla}_4, \widehat{\nabla}^2]\sigma^F} \right\rangle + \int_{D_{u,\underline{u}}} 2r^6 \left\langle \widehat{\nabla}^2 \alpha^F, -\frac{1}{2}\widehat{\nabla}(\widehat{\nabla}(tr\underline{\chi})\alpha^F) \right.$$

$$\left. -2\widehat{\nabla} \,{}^\star\widehat{\nabla}(\eta\sigma^F) + 2\widehat{\nabla}^2(\eta\rho^F) + 2\widehat{\nabla}^2(\underline{\omega}\alpha^F) - \widehat{\nabla}^2(\hat{\chi} \cdot \underline{\alpha}^F) + \mathcal{E}^{[\widehat{\nabla}_3, \widehat{\nabla}^2]\alpha^F} \right\rangle.$$

Now we control the error term. The proof relies on noting the presence of null-structure and appropriate scaling. The slowly decaying term $\underline{\omega}$ appears with $\alpha^F$ i.e.,

$$|\int_{D_{u,\underline{u}}} r^6 \underline{\omega} \widehat{\nabla}^2 \alpha^F \cdot \widehat{\nabla}^2 \alpha^F| \lesssim \sup_{D_{u,\underline{u}}}(r|u|\|\underline{\omega}\|) \left(\int_{u_0}^u |u'|^{-1} r^{-1} du'\right) \int_H r^6 |\widehat{\nabla}^2 \alpha^F|^2 \lesssim \epsilon \mathcal{Y}_2 \qquad (6.182)$$

$$|\int_{D_{u,\underline{u}}} r^6 \widehat{\nabla}^2 \alpha^F \cdot \widehat{\nabla} \alpha^F \nabla \omega| \qquad (6.183)$$

$$\lesssim \sup_{u,\underline{u}} \|r^{\frac{3}{2}}|u|\nabla\omega\|_{L^4(S_{u,\underline{u}})} \int_{u,\underline{u}} \left(\|r^3 \widehat{\nabla}^2 \alpha^F\|_{L^2(S_{u,\underline{u}})} \|r^{\frac{5}{2}} \widehat{\nabla} \alpha^F\|_{L^4(S_{u,\underline{u}})} |u|^{-1} r^{-1}\right)$$

$$\lesssim \epsilon \left(\int_{u_0}^u |u'|^{-1} r^{-1}\right) \int_H (r^4 |\widehat{\nabla} \alpha^F|^2 + r^6 |\widehat{\nabla}^2 \alpha^F|^2) \lesssim \epsilon(\mathcal{Y}_1 + \mathcal{Y}_2),$$

where we have made use of the weighted co-dimension 1 trace inequality (Proposition 3.1). The other connections are estimated in a similar way i.e.,

$$|\int_{D_{u,\underline{u}}} r^6 \widehat{\nabla}^2 \alpha^F \cdot \widehat{\nabla}((\rho^F, \sigma^F)\nabla\eta)| \lesssim \epsilon(\mathcal{Y}_1 + \mathcal{Y}_2), \qquad (6.184)$$

$$|\int_{D_{u,\underline{u}}} r^6 \widehat{\nabla}^2 \alpha^F \cdot \widehat{\nabla}(\widehat{\nabla}(\rho^F, \sigma^F)\eta)| \lesssim \epsilon(\mathcal{Y}_1 + \mathcal{Y}_2), \qquad (6.185)$$

$$|\int_{D_{u,\underline{u}}} r^6 \widehat{\nabla}^2 \alpha^F \cdot \widehat{\nabla}^2 \underline{\alpha}^F \hat{\chi}| \lesssim \sup_{D_{u,\underline{u}}}(r^2|\hat{\chi}|) \int_{D_{u,\underline{u}}} \left(\frac{r^6|\widehat{\nabla}^2 \alpha^F|^2}{u^2} + \frac{u^2 r^4|\widehat{\nabla}^2 \underline{\alpha}^F|^2}{r^2}\right) \qquad (6.186)$$

$$\lesssim \epsilon \left(\left(\int_{u_0}^u |u'|^{-2}\right) \int_H r^6|\widehat{\nabla}^2 \alpha^F|^2 + \left(\int_{\underline{u}_0}^{\underline{u}} |r|^{-2} d\underline{u}'\right) \int_{\underline{H}} u^2 r^4 |\widehat{\nabla}^2 \underline{\alpha}^F|^2\right) \lesssim \epsilon \mathcal{Y}_2.$$

The most vital term that we want to focus on is the error term $\mathcal{E}^{[\widehat{\nabla}_3, \widehat{\nabla}^2]\alpha^F}$ since this is where the *true nonlinearity* of the Yang–Mills equations shows up. We estimate this error term as follows

$$|\int_{D_{u,\underline{u}}} r^6 \widehat{\nabla}^2 \alpha^F \cdot \widehat{\nabla} \underline{\alpha}^F \alpha^F| \lesssim \sup_{u,\underline{u}} \||u|^{\frac{3}{2}} r^{\frac{3}{2}} \widehat{\nabla} \underline{\alpha}^F\|_{L^4(S_{u,\underline{u}})} \int_{u,\underline{u}} \|r^3 \widehat{\nabla}^2 \alpha^F\|_{L^2(S_{u,\underline{u}})} \|r^{\frac{3}{2}} \alpha^F\|_{L^4(S_{u,\underline{u}})} |u|^{-\frac{3}{2}}$$

$$\lesssim \epsilon \left(\int_{u_0}^u |u|^{-\frac{3}{2}}\right) \int_H (r^2 |\alpha^F|^2 + r^6 |\widehat{\nabla} \alpha^F|^2) \lesssim \epsilon(\mathcal{Y}_0 + \mathcal{Y}_2),$$

where we have utilized the $L^4 - H^1$ Sobolev inequality on the topological two–sphere. The Yang–Mills-gravity coupling term satisfies

$$|\int_{D_{u,\underline{u}}} r^6 \widehat{\nabla}^2 \alpha^F \cdot \alpha^F \nabla\underline{\beta}| \lesssim \sup_{D_{u,\underline{u}}} \left(r^{\frac{5}{2}}|\alpha^F|\right) \left(\int_{u_0}^u |u|^{-1} r^{-\frac{3}{2}} du'\right) \int_H (r^6|\widehat{\nabla}^2 \alpha^F|^2 + u^2 r^4 |\nabla\underline{\beta}|^2)$$

$$\lesssim \epsilon(\mathcal{W}_1 + \mathcal{Y}_2). \qquad (6.187)$$

Similarly, we obtain

$$|\int_{D_{u,\underline{u}}} r^6 \widehat{\nabla}^2 \alpha^F \cdot \widehat{\nabla}\widehat{\nabla}_3 \alpha^F(\eta, \underline{\eta})| \lesssim \epsilon^2(\mathcal{Y}_0 + \mathcal{Y}_1 + \mathcal{Y}_2). \qquad (6.188)$$

Once again, we only estimate the top order terms involving $\rho^F$ since $\sigma^F$ satisfies similar estimates. Once again note that the commutator terms are the manifestation of Yang–Mills non-linearity. We estimate the error terms as follows

$$|\int_{D_{u,\underline{u}}} r^6 \widehat{\nabla}^2 \rho^F \cdot \alpha^F \widehat{\nabla}\rho^F| \lesssim \sup_{D_{u,\underline{u}}} \left(r^{\frac{5}{2}}|\alpha^F|\right) \left(\int_{\underline{u}_0}^{\underline{u}} r^{-\frac{3}{2}} d\underline{u}'\right) \int_{\underline{H}} (r^6|\widehat{\nabla}^2 \rho^F|^2 + r^4|\widehat{\nabla}\rho^F|^2) \qquad (6.189)$$

$$\lesssim \epsilon(\mathcal{Y}_1 + \mathcal{Y}_2).$$

Similarly, the remaining terms are estimated as follows

$$|\int_{D_{u,\underline{u}}} r^6 \widehat{\nabla}^2 \rho^F \cdot \widehat{\nabla}(\rho^F \beta)| \lesssim \epsilon(\mathcal{W}_0 + \mathcal{W}_1 + \mathcal{Y}_1 + \mathcal{Y}_2), \tag{6.190}$$

$$|\int_{D_{u,\underline{u}}} r^6 \widehat{\nabla}^2 \rho^F \cdot \widehat{\nabla}(\widehat{\nabla}_4 \rho^F(\eta, \underline{\eta}))| \lesssim \epsilon^2(\mathcal{Y}_0 + \mathcal{Y}_1 + \mathcal{Y}_2), \tag{6.191}$$

$$|\int_{D_{u,\underline{u}}} r^6 \widehat{\nabla}^2 \rho^F \cdot \widehat{\nabla}(\rho^F \nabla tr\chi)| \lesssim \epsilon(\mathcal{Y}_0 + \mathcal{Y}_1 + \mathcal{Y}_2), \tag{6.192}$$

$$|\int_{D_{u,\underline{u}}} r^6 \widehat{\nabla}^2 \rho^F \cdot \widehat{\nabla}(\alpha^F \nabla(\eta, \underline{\eta}))| \lesssim \epsilon(\mathcal{Y}_0 + \mathcal{Y}_1 + \mathcal{Y}_2), \tag{6.193}$$

$$|\int_{D_{u,\underline{u}}} r^6 \widehat{\nabla}^2 \rho^F \cdot \widehat{\nabla}(\widehat{\nabla}\rho^F \hat{\chi})| \lesssim \epsilon(\mathcal{Y}_0 + \mathcal{Y}_1 + \mathcal{Y}_2). \tag{6.194}$$

Collecting all the terms yields

$$|\mathcal{E}^{\widehat{\nabla}^2 \alpha^F, \widehat{\nabla}^2(\rho^F, \sigma^F)}| \lesssim \epsilon(\mathcal{W}_0 + \mathcal{W}_1 + \mathcal{Y}_0 + \mathcal{Y}_1 + \mathcal{Y}_2). \tag{6.195}$$

The error estimates combining with the fact that $(3\Omega^{-1}\overline{\Omega tr\chi} - 3tr\chi) \lesssim \frac{\epsilon}{r^2}$, $(3\Omega^{-1}\overline{\Omega tr\underline{\chi}} - 2tr\underline{\chi}) < 0$ in the exterior domain due to the connection estimates, we conclude the proof. □

A similar energy estimate for the Yang–Mills curvature $\underline{\alpha}^F$ holds. The following lemma establishes such an estimate

*Lemma 6.8.*

$$\int_{\underline{H}} u^2 |\underline{\alpha}^F|^2 + \int_H u^2(|\rho^F|^2 + |\sigma^F|^2) \lesssim \int_{\widehat{\Sigma}_0} u^2 |\underline{\alpha}^F|^2 + \int_{\widehat{\Sigma}_0} u^2(|\rho^F|^2 + |\sigma^F|^2) + \epsilon \mathbf{Y},$$

$$\int_{\underline{H}} u^2 r^2 |\widehat{\nabla}\underline{\alpha}^F|^2 + \int_H u^2 r^2(|\widehat{\nabla}\rho^F|^2 + |\widehat{\nabla}\sigma^F|^2) \leq \int_{\widehat{\Sigma}_0} u^2 r^2 |\widehat{\nabla}\underline{\alpha}^F|^2 + \int_{\widehat{\Sigma}_0} u^2 r^2(|\widehat{\nabla}\rho^F|^2 + |\widehat{\nabla}\sigma^F|^2)$$
$$+ \epsilon(\mathbf{W} + \mathbf{Y}),$$

$$\int_{\underline{H}} u^2 r^4 |\widehat{\nabla}^2\underline{\alpha}^F|^2 + \int_H u^2 r^4(|\widehat{\nabla}^2\rho^F|^2 + |\widehat{\nabla}^2\sigma^F|^2)$$
$$\lesssim \int_{\widehat{\Sigma}_0} u^2 r^4 |\widehat{\nabla}^2\underline{\alpha}^F|^2 + \int_{\widehat{\Sigma}_0} u^2 r^4(|\widehat{\nabla}^2\rho^F|^2 + |\widehat{\nabla}^2\sigma^F|^2) + \epsilon(\mathbf{W} + \mathbf{Y}),$$

$$\int_{\underline{H}} u^2 r^6 |\widehat{\nabla}^3\underline{\alpha}^F|^2 + \int_H u^2 r^6(|\widehat{\nabla}^3\rho^F|^2 + |\widehat{\nabla}^3\sigma^F|^2) \lesssim \int_{\widehat{\Sigma}_0} u^2 r^6 |\widehat{\nabla}^3\underline{\alpha}^F|^2 + \int_{\widehat{\Sigma}_0} u^2 r^6(|\widehat{\nabla}^3\rho^F|^2 + |\widehat{\nabla}^3\sigma^F|^2)$$
$$+ \epsilon(\mathbf{W} + \mathbf{Y})$$

*Proof.* Consider the Null Yang–Mills pair $\underline{\alpha}^F$ and $(\rho^F, \sigma^F)$ as usual, use the integration Lemma (6.1) and write down the following identity

$$\int_{\underline{H}} u^2 r^4 |\widehat{\nabla}^2\underline{\alpha}^F|^2 + \int_H u^2 r^4(|\widehat{\nabla}^2\rho^F|^2 + |\widehat{\nabla}^2\sigma^F|^2) - 2\int_{D_{u,\underline{u}}} u^2 r^4 (\Omega^{-1}\overline{\Omega tr\chi} - tr\chi)|\widehat{\nabla}^2\underline{\alpha}^F|^2$$
$$- \int_{D_{u,\underline{u}}} (2\Omega^{-1} ur^2 + 2u^2 r^4 \Omega^{-1}\overline{\Omega tr\underline{\chi}} - 3u^2 r^4 tr\underline{\chi})|\widehat{\nabla}^2(\rho^F, \sigma^F)|^2$$
$$= \int_{\widehat{\Sigma}_0} u^2 r^4 |\widehat{\nabla}^2\underline{\alpha}^F|^2 + \int_{\widehat{\Sigma}_0} u^2 r^4(|\widehat{\nabla}^2\rho^F|^2 + |\widehat{\nabla}^2\sigma^F|^2) + \mathcal{E}^{\widehat{\nabla}^2\underline{\alpha}^F, \widehat{\nabla}^2(\rho^F, \sigma^F)},$$

where the error term is defined as follows

$$\mathcal{E}^{\widehat{\nabla}^2\underline{\alpha}^F, \widehat{\nabla}^2(\rho^F, \sigma^F)} \sim \int_{D_{u,\underline{u}}} 2u^2 r^4 \left\langle \widehat{\nabla}^2\underline{\alpha}^F, \widehat{\nabla}\left(\widehat{\nabla}\left(-\frac{1}{2}tr\underline{\chi}\right)\underline{\alpha}^F\right)\right.$$
$$\left. - \widehat{\nabla}^2(2 *\underline{\eta}\sigma^F - 2\underline{\eta}\rho^F + 2\omega\underline{\alpha}^F - \underline{\hat{\chi}} \cdot \alpha^F)) - \widehat{\nabla}(\underline{\hat{\chi}} \cdot \widehat{\nabla}\alpha^F) + \mathcal{E}^{[\widehat{\nabla}_4, \widehat{\nabla}^2]\underline{\alpha}^F} + \mathcal{E}^{[\widehat{\nabla}, \widehat{\nabla}^2](\rho^F, \sigma^F)}\right\rangle$$
$$+ \int_{D_{u,\underline{u}}} 2u^2 r^4 \Big( \langle \widehat{\nabla}^2\rho^F, -\widehat{\nabla}(\nabla(tr\underline{\chi})\rho^F) + \widehat{\nabla}^2((\eta - \underline{\eta}) \cdot \underline{\alpha}^F) - \widehat{\nabla}(\underline{\hat{\chi}} \cdot \widehat{\nabla}\rho^F)\rangle$$
$$+ \langle \widehat{\nabla}^2\sigma^F, -\widehat{\nabla}(\widehat{\nabla}(tr\underline{\chi})\sigma^F) + \widehat{\nabla}^2((\eta - \underline{\eta}) \cdot {}^*\underline{\alpha}^F) - \widehat{\nabla}(\underline{\hat{\chi}} \cdot \widehat{\nabla}\sigma^F) + \mathcal{E}^{[\widehat{\nabla}_3, \widehat{\nabla}^2](\sigma^F, \rho^F)}\rangle \Big)$$

$$|\int_{D_{u,\underline{u}}} u^2 r^4 \widehat{\nabla}^2\underline{\alpha}^F \cdot \alpha^F \widehat{\nabla}\underline{\alpha}^F| \lesssim \sup_{u,\underline{u}} \|r^2 \alpha^F\|_{L^4(S_{u,\underline{u}})} \int_{u,\underline{u}} \left( \|ur^2 \widehat{\nabla}^2\underline{\alpha}^F\|_{L^2(S_{u,\underline{u}})} \|u|r^{\frac{3}{2}}\widehat{\nabla}\underline{\alpha}^F\|_{L^4(S_{u,\underline{u}})} r^{-\frac{3}{2}} \right)$$
$$\lesssim \epsilon \left( \int_{\underline{H}} (u^2 r^2 |\widehat{\nabla}\underline{\alpha}^F|^2 + u^2 r^4 |\widehat{\nabla}^2\underline{\alpha}^F|^2) \right) \lesssim \epsilon(\mathcal{Y}_1 + \mathcal{Y}_2).$$

This is essentially the most dangerous term at the level of second derivative estimates of the Yang–Mills fields. Notice that all the derivative estimates are essentially the same since the appearance of a derivative is compensated by the scale factor $r$. Another important feature is the presence of the null structure. Notice that $\hat{\chi}$ appears with $|\widehat{\nabla}^2\underline{\alpha}^F|^2$ (recall $\hat{\underline{\chi}}$ appeared multiplied with $|\widehat{\nabla}\alpha^F|^2$ which verifies a weaker decay along the outgoing direction but such a property did not make a difference since $\alpha^F$ is only controlled on $H$) and satisfies a stronger decay along the outgoing null direction. More specifically, we have

$$\left|\int_{D_{u,\underline{u}}} u^2 r^4 \widehat{\nabla}^2\underline{\alpha}^F \cdot \widehat{\nabla}^2\underline{\alpha}^F \hat{\chi}\right| \lesssim \sup_{D_{u,\underline{u}}}(r^2|\hat{\chi}|)\left(\int_{\underline{u}_0}^{\underline{u}} r^{-2} d\underline{u}'\right)\int_{\underline{H}} u^2 r^4 |\widehat{\nabla}^2\underline{\alpha}^F|^2 \lesssim \epsilon \mathcal{Y}_2. \tag{6.197}$$

The remaining terms satisfy the following estimates (we use the available null evolution equations)

$$\left|\int_{D_{u,\underline{u}}} u^2 r^4 \widehat{\nabla}^2\underline{\alpha}^F \cdot \widehat{\nabla}(\underline{\alpha}^F \beta)\right| \lesssim \epsilon(\mathcal{W}_0 + \mathcal{W}_1 + \mathcal{Y}_1 + \mathcal{Y}_2)), \tag{6.198}$$

$$\left|\int_{D_{u,\underline{u}}} u^2 r^4 \widehat{\nabla}^2\underline{\alpha}^F \cdot \widehat{\nabla}(\underline{\alpha}^F \alpha^F \cdot (\rho^F, \sigma^F))\right| \lesssim \epsilon^2(\mathcal{Y}_0 + \mathcal{Y}_1 + \mathcal{Y}_2)), \tag{6.199}$$

$$\left|\int_{D_{u,\underline{u}}} u^2 r^4 \widehat{\nabla}^2\underline{\alpha}^F \cdot \widehat{\nabla}(\widehat{\nabla}_4\underline{\alpha}^F(\eta,\underline{\eta}))\right| \lesssim \epsilon(\mathcal{Y}_0 + \mathcal{Y}_1 + \mathcal{Y}_2), \tag{6.200}$$

$$\left|\int_{D_{u,\underline{u}}} u^2 r^4 \widehat{\nabla}^2\underline{\alpha}^F \cdot \widehat{\nabla}(\underline{\alpha}^F \nabla tr\chi)\right| \lesssim \epsilon(\mathcal{Y}_0 + \mathcal{Y}_1 + \mathcal{Y}_2), \tag{6.201}$$

$$\left|\int_{D_{u,\underline{u}}} u^2 r^4 \widehat{\nabla}^2\underline{\alpha}^F \cdot \widehat{\nabla}(\widehat{\nabla}(\sigma^F,\rho^F)\underline{\eta})\right| \lesssim \epsilon(\mathcal{Y}_1 + \mathcal{Y}_2) \tag{6.202}$$

$$\left|\int_{D_{u,\underline{u}}} u^2 r^4 \widehat{\nabla}^2\underline{\alpha}^F \cdot \widehat{\nabla}(\widehat{\nabla}\alpha^F \underline{\hat{\chi}})\right| \lesssim \epsilon(\mathcal{Y}_1 + \mathcal{Y}_2), \tag{6.203}$$

$$\left|\int_{D_{u,\underline{u}}} u^2 r^4 \widehat{\nabla}(\rho^F,\sigma^F) \cdot \widehat{\nabla}((\rho^F,\sigma^F)\nabla tr\underline{\chi})\right| \lesssim \epsilon(\mathcal{Y}_0 + \mathcal{Y}_1 + \mathcal{Y}_2), \tag{6.204}$$

$$\left|\int_{D_{u,\underline{u}}} u^2 r^4 \widehat{\nabla}(\rho^F,\sigma^F) \cdot \widehat{\nabla}((\rho^F,\sigma^F)\nabla(\eta,\underline{\eta}))\right| \lesssim \epsilon(\mathcal{Y}_0 + \mathcal{Y}_1 + \mathcal{Y}_2), \tag{6.205}$$

$$\left|\int_{D_{u,\underline{u}}} u^2 r^4 \widehat{\nabla}(\rho^F,\sigma^F) \cdot \widehat{\nabla}(\widehat{\nabla}(\rho^F,\sigma^F)(\eta,\underline{\eta}))\right| \lesssim \epsilon(\mathcal{Y}_1 + \mathcal{Y}_2), \tag{6.206}$$

$$\left|\int_{D_{u,\underline{u}}} u^2 r^4 \widehat{\nabla}(\rho^F,\sigma^F) \cdot \widehat{\nabla}(\widehat{\nabla}(\rho^F,\sigma^F)\hat{\chi})\right| \lesssim \epsilon(\mathcal{Y}_1 + \mathcal{Y}_2), \tag{6.207}$$

$$\left|\int_{D_{u,\underline{u}}} u^2 r^4 \widehat{\nabla}(\rho^F,\sigma^F) \cdot \mathcal{E}^{[\widehat{\nabla}_3,\widehat{\nabla}^2](\sigma^F,\rho^F)}\right| \lesssim \epsilon(\mathcal{Y}_0 + \mathcal{Y}_1 + \mathcal{Y}_2)). \tag{6.208}$$

Collecting all the terms yields the desired estimate

$$\left|\mathcal{E}^{\nabla^2\underline{\alpha}^F,\nabla^2(\rho^F,\sigma^F)}\right| \lesssim \epsilon(\mathcal{W}_0 + \mathcal{W}_1 + \mathcal{Y}_0 + \mathcal{Y}_1 + \mathcal{Y}_2). \tag{6.209}$$

Now $|(\Omega^{-1}\overline{\Omega tr\underline{\chi}} - tr\underline{\chi})| \lesssim \frac{\epsilon}{r^2}$ and $2\Omega^{-1}u^{-1} + 2\Omega^{-1}\overline{\Omega tr\underline{\chi}} - 3tr\underline{\chi} < 0$ due to an application of the connection estimates finishes the proof. □

*Remark 8.* We need the third derivative estimates for the connections (lemma) in estimating the top-most derivatives of the Yang–Mills curvature.

In an exact similar way but commuting with a combination of $\widehat{\nabla}_4$ and $\widehat{\nabla}$, we obtain the following lemma.

*Lemma 6.9.* *The following estimates hold for the mixed derivatives of the Yang–Mills curvature components*

$$\int_H r^4|\widehat{\nabla}_4\alpha^F|^2 + \int_{\underline{H}} r^4(|\widehat{\nabla}_4\rho^F|^2 + |\widehat{\nabla}_4\sigma^F|^2) \lesssim \int_{\widehat{\Sigma}_0} r^4|\widehat{\nabla}_4\alpha^F|^2 + \int_{\widehat{\Sigma}_0} r^4(|\widehat{\nabla}_4\rho^F|^2 + |\widehat{\nabla}_4\sigma^F|^2)$$
$$+ \int_{\widehat{\Sigma}_0} r^4|\widehat{\nabla}\alpha^F|^2 + \epsilon(\mathbf{W}+\mathbf{Y}),$$

$$\int_{\underline{H}} u^4|\widehat{\nabla}_3\underline{\alpha}^F|^2 + \int_H u^4(|\widehat{\nabla}_3\rho^F|^2 + |\widehat{\nabla}_3\sigma^F|^2) \lesssim \int_{\widehat{\Sigma}_0} u^4|\widehat{\nabla}_3\underline{\alpha}^F|^2 + \int_{\widehat{\Sigma}_0} u^4(|\widehat{\nabla}_3\rho^F|^2 + |\widehat{\nabla}_3\sigma^F|^2)$$
$$+\epsilon(\mathbf{W}+\mathbf{Y}),$$

$$\int_{\underline{H}} r^6 |\widehat{\nabla} \widehat{\nabla}_4 \alpha^F|^2 + \int_{\underline{H}} r^6 (|\widehat{\nabla} \widehat{\nabla}_4 \rho^F|^2 + |\widehat{\nabla} \widehat{\nabla}_4 \sigma^F|^2)$$

$$\lesssim \int_{\widehat{\Sigma}_0} r^6 |\widehat{\nabla} \widehat{\nabla}_4 \alpha^F|^2 + \int_{\widehat{\Sigma}_0} r^6 (|\widehat{\nabla} \widehat{\nabla}_4 \rho^F|^2 + |\widehat{\nabla} \widehat{\nabla}_4 \sigma^F|^2) + \int_{\widehat{\Sigma}_0} r^6 |\widehat{\nabla}^2 \alpha^F|^2 + \epsilon(\mathbf{W} + \mathbf{Y}),$$

$$\int_{\underline{H}} u^4 r^2 |\widehat{\nabla} \widehat{\nabla}_3 \underline{\alpha}^F|^2 + \int_{\underline{H}} u^4 r^2 (|\widehat{\nabla} \widehat{\nabla}_3 \rho^F|^2 + |\widehat{\nabla} \widehat{\nabla}_3 \sigma^F|^2)$$

$$\lesssim \int_{\widehat{\Sigma}_0} u^4 r^2 |\widehat{\nabla} \widehat{\nabla}_3 \underline{\alpha}^F|^2 + \int_{\widehat{\Sigma}_0} u^4 r^2 (|\widehat{\nabla} \widehat{\nabla}_3 \rho^F|^2 + |\widehat{\nabla} \widehat{\nabla}_3 \sigma^F|^2) + \epsilon(\mathbf{W} + \mathbf{Y}),$$

$$\int_{\underline{H}} r^6 |\widehat{\nabla}_4 \widehat{\nabla}_4 \alpha^F|^2 + \int_{\underline{H}} r^6 (|\widehat{\nabla}_4 \widehat{\nabla}_4 \rho^F|^2 + |\widehat{\nabla}_4 \widehat{\nabla}_4 \sigma^F|^2)$$

$$\lesssim \int_{\widehat{\Sigma}_0} r^6 |\widehat{\nabla}_4 \widehat{\nabla}_4 \alpha^F|^2 + \int_{\widehat{\Sigma}_0} r^6 (|\widehat{\nabla}_4 \widehat{\nabla}_4 \rho^F|^2 + |\widehat{\nabla}_4 \widehat{\nabla}_4 \sigma^F|^2) + \int_{\widehat{\Sigma}_0} r^6 |\widehat{\nabla} \widehat{\nabla} \alpha^F|^2 + \epsilon(\mathbf{W} + \mathbf{Y}),$$

$$\int_{\underline{H}} u^6 |\widehat{\nabla}_3 \widehat{\nabla}_3 \underline{\alpha}^F|^2 + \int_{\underline{H}} u^6 (|\widehat{\nabla}_3 \widehat{\nabla}_3 \rho^F|^2 + |\widehat{\nabla}_3 \widehat{\nabla}_3 \sigma^F|^2)$$

$$\lesssim \int_{\widehat{\Sigma}_0} u^4 r^2 |\widehat{\nabla}_3 \widehat{\nabla}_3 \underline{\alpha}^F|^2 + \int_{\widehat{\Sigma}_0} u^4 r^2 (|\widehat{\nabla}_3 \widehat{\nabla}_3 \rho^F|^2 + |\widehat{\nabla}_3 \widehat{\nabla}_3 \sigma^F|^2) + \epsilon(\mathbf{W} + \mathbf{Y}).$$

*Proof.* The proof is exactly similar to the Einstein–Maxwell case except for the fact that one needs to keep track of the non-linear terms that appear as a result of the commutation of the gauge-covariant derivatives. These true non-linear terms (a feature of the Yang–Mills theory) are estimated as follows

$$\left| \int_{D_{u,\underline{u}}} r^4 \widehat{\nabla}_4 \alpha^F \cdot \rho^F \alpha^F \right| \lesssim \sup_{u,\underline{u}} \||u|r\rho^F\|_{L^4(S_{u,\underline{u}})} \int_{u,\underline{u}} \|r^2 \widehat{\nabla}_4 \alpha^F\|_{L^2(S_{u,\underline{u}})} \|r^{\frac{3}{2}} \alpha^F\|_{L^4(S_{u,\underline{u}})} |u|^{-1} r^{-\frac{1}{2}}$$

$$\lesssim \epsilon \left( \int_{u_0}^{u} |u|^{-1} r^{-\frac{1}{2}} \right) \int_H (r^2|\alpha^F|^2 + r^4|\widehat{\nabla}\alpha^F|^2 + r^4|\widehat{\nabla}_4 \alpha^F|^2) \lesssim \epsilon(\mathbf{W} + \mathbf{Y}),$$

$$\left| \int_{D_{u,\underline{u}}} u^4 \widehat{\nabla}_3 \underline{\alpha}^F \cdot \rho^F \underline{\alpha}^F \right| \lesssim \sup_{u,\underline{u}} \|r^2 \rho^F\|_{L^4(S_{u,\underline{u}})} \int_{u,\underline{u}} \left( \|u^2 \widehat{\nabla}_3 \underline{\alpha}^F\|_{L^2(S_{u,\underline{u}})} \||u|r^{\frac{1}{2}} \underline{\alpha}^F\|_{L^4(S_{u,\underline{u}})} |u|r^{-\frac{5}{2}} \right)$$

$$\lesssim \epsilon \left( \int_{\underline{H}} (u^2|\underline{\alpha}^F|^2 + u^2 r^2 |\widehat{\nabla}\underline{\alpha}^F|^2 + u^4|\widehat{\nabla}\underline{\alpha}^F|^2) \right) \lesssim \epsilon(\mathbf{W} + \mathbf{Y}),$$

$$\left| \int_{D_{u,\underline{u}}} r^6 \widehat{\nabla}_4^2 \alpha^F \cdot \widehat{\nabla}_4 \rho^F \alpha^F \right| \lesssim \sup_{u,\underline{u}} \||u|r^2 \widehat{\nabla}_4 \rho^F\|_{L^4(S_{u,\underline{u}})} \int_{u,\underline{u}} \|r^3 \widehat{\nabla}_4^2 \alpha^F\|_{L^2(S_{u,\underline{u}})} \|r^{\frac{3}{2}} \alpha^F\|_{L^4(S_{u,\underline{u}})} |u|^{-1} r^{-\frac{1}{2}}$$

$$\lesssim \epsilon \left( \int_{u_0}^{u} |u|^{-1} r^{-\frac{1}{2}} \right) \int_H (r^2|\alpha^F|^2 + r^4|\widehat{\nabla}\alpha^F|^2 + r^6|\widehat{\nabla}_4^2 \alpha^F|^2) \lesssim \epsilon(\mathbf{W} + \mathbf{Y}),$$

$$\left| \int_{D_{u,\underline{u}}} u^6 \widehat{\nabla}_3^2 \underline{\alpha}^F \cdot \widehat{\nabla}_3 \rho^F \underline{\alpha}^F \right| \lesssim \sup_{u,\underline{u}} \|r^2|u|\widehat{\nabla}_3 \rho^F\|_{L^4(S_{u,\underline{u}})} \int_{u,\underline{u}} \left( \|u^3 \widehat{\nabla}_3^2 \underline{\alpha}^F\|_{L^2(S_{u,\underline{u}})} \||u|r^{\frac{1}{2}} \underline{\alpha}^F\|_{L^4(S_{u,\underline{u}})} |u|r^{-\frac{5}{2}} \right)$$

$$\lesssim \epsilon \left( \int_{\underline{H}} (u^2|\underline{\alpha}^F|^2 + u^2 r^2 |\widehat{\nabla}\underline{\alpha}^F|^2 + u^6|\widehat{\nabla}_3^2 \underline{\alpha}^F|^2) \right) \lesssim \epsilon(\mathbf{W} + \mathbf{Y})$$

and similar other terms. Collecting all these terms yields the desired estimates. □

*Remark 9.* In estimating these terms, we need the estimates for $\nabla_4 \omega, \nabla \nabla_4 \omega, \nabla^2 \nabla_4 \omega, \nabla_3 \underline{\omega}, \nabla \nabla_3 \underline{\omega},$ and $\nabla^2 \nabla_3 \underline{\omega}$.

## VII. PROOF OF THE MAIN THEOREM

Once we have obtained the energy estimates for the Weyl and Yang–Mills curvature, the proof of the remainder of the main theorem follows straightforwardly. The following lemma essentially completes the proof.

*Lemma 7.1.* The total energy verifies the following uniform estimate uniform in $u$ and $\underline{u}$

$$\mathbf{W} + \mathbf{Y} \lesssim \mathbf{W}_0 + \mathbf{Y}_0. \tag{7.1}$$

*Proof.* Collecting all the estimates proved in Lemmas (6.2)–(6.14) yields

$$\mathbf{W} + \mathbf{Y} \leq \mathbf{W}_0 + \mathbf{Y}_0 + C\epsilon(\mathbf{W} + \mathbf{Y}), \tag{7.2}$$

for a constant $C$ independent of $u, \underline{u}$ and $\epsilon$. Therefore for sufficiently small $\epsilon$, we obtain

$$\mathbf{W} + \mathbf{Y} \leq C(\mathbf{W}_0 + \mathbf{Y}_0) \tag{7.3}$$

for $C$ independent of $u$ and $\underline{u}$. Now we can choose the initial data $\mathbf{W}_0 + \mathbf{Y}_0$ to be sufficiently small such that $C(\mathbf{W}_0 + \mathbf{Y}_0) \leq \frac{\epsilon}{2}$ and therefore $\mathbf{W} + \mathbf{Y} \leq \frac{\epsilon}{2}$. This closes the bootstrap assumption. □

*Remark 10.* Since the connection norm, $\mathcal{O}$ verifies $\mathcal{O} \lesssim \mathbf{W} + \mathbf{Y}$, this closes the bootstrap assumption on the connections too.

By employing the gauge-invariant global Sobolev inequalities, we obtain the point-wise bound on the Weyl and Yang–Mills curvature components.

*Corollary 7.1.* The following gauge-invariant point-wise estimates of the Weyl and Yang–Mills curvature hold in the exterior domain

$$|\alpha| \lesssim r^{-\frac{7}{2}}, \qquad |\underline{\alpha}| \lesssim r^{-1}|u|^{-\frac{5}{2}}, \qquad |\beta| \lesssim r^{-\frac{7}{2}}, \qquad |\rho, \sigma| \lesssim r^{-3}|u|^{-\frac{1}{2}}, \tag{7.4}$$

$$|\alpha^F| \lesssim r^{-\frac{5}{2}}, \qquad |\underline{\alpha}^F| \lesssim r^{-1}|u|^{-\frac{3}{2}}, \qquad |\rho^F, \sigma^F| \lesssim r^{-2}|u|^{-\frac{1}{2}}. \tag{7.5}$$

Even though, in the context of the apriori estimates of the main theorem, we do not need to address the Cauchy problem, we do so briefly. Using these gauge-invariant estimates and standard arguments for the Cauchy problem up to a suitably defined "time" $\tau = u + \underline{u}$, one can prove the existence of a solution to the Einstein–Yang–Mills equations through a continuity argument in space-time harmonic generalized Coloumb/temporal gauge. We omit such detail since they are standard (see Sec. 6 of Ref. 28 for example for vacuum case, Yang–Mills equations in temporal gauge are symmetric hyperbolic and so follows similarly). For the moment, if we consider the metric in the ADM form (suitable for the Cauchy problem) $^{1+3}g = -N^2 dt \otimes dt + g_{ij}(dx^i + X^i dt) \otimes (dx^j + X^j dt)$, where $\{x^i\}_{i=0}^3 = (t, x^1, x^2, x^3)$ is local chart, $N$ is the lapse function, and $X$ is the shift vector field. Also, let $\mathbf{n} := \frac{1}{N}(\partial_t - X)$ be the $t$ = constant hypersurface orthogonal future directed unit normal field. In the framework of a Cauchy problem, the Einstein and gauge covariant Yang–Mills equations read[31]

$$\partial_t g_{ij} = -2Nk_{ij} + L_X g_{ij} \tag{7.6}$$

$$\partial_t k_{ij} = -\nabla_i \nabla_j N + N\Big\{R_{ij} + tr_g k k_{ij} - 2k_{ik} k^k_j + \mathcal{E}^a_{bi} \mathcal{E}^b_{aj}$$

$$- g^{kl} F^a_{bik} F^b_{akl} - \frac{g^{kl} \mathcal{E}^a_{bk} \mathcal{E}^b_{al}}{n-1} g_{ij} + \frac{g^{km} g^{ln} F^a_{bkl} F^b_{amn}}{2(n-1)} g_{ij}\Big\} + L_X k_{ij},$$

$$R(g) - |k|^2 + (tr_g k)^2 = g^{ij} \mathcal{E}^a_{bi} \mathcal{E}^b_{aj} + \frac{1}{2} g^{ik} g^{jl} F^a_{bij} F^b_{akl}, \tag{7.7}$$

$$\nabla_j k^j_i - \nabla_i tr_g k = g^{jk} F^a_{bij} \mathcal{E}^b_{ak}, \tag{7.8}$$

$$g^{ij} \widehat{\nabla}_i \mathcal{E}^a_{bj} = 0, \tag{7.9}$$

$$\widehat{\mathcal{L}}_{\partial_t} \mathcal{E}^a_{bi} = \widehat{\mathcal{L}}_X \mathcal{E}^a_{bi} - N\epsilon_i{}^{jk} \widehat{\nabla}_j H^a_{bk} - 2NK_i{}^j \mathcal{E}^a_{bj} + N tr_g K \mathcal{E}^a_{bi} - \epsilon_i{}^{jk} \nabla_j N H^a_{bk}, \tag{7.10}$$

$$\widehat{\mathcal{L}}_{\partial_t} H^a_{bi} = \widehat{\mathcal{L}}_X H^a_{bi} + N\epsilon_i{}^{jk} \widehat{\nabla}_j \mathcal{E}^a_{bk} - 2NK_i{}^j H^a_{bj} + N tr_g K H^a_{bi} + \epsilon_i{}^{jk} \nabla_j N \mathcal{E}^a_{bk}, \tag{7.11}$$

where $\mathcal{E}_i := F(\mathbf{n}, \partial_i), H_i = {}^*F(\mathbf{n}, e_i)$ are the chromo-electric and chromo-magnetic fields, respectively. Here $\widehat{\mathcal{L}}$ denotes the gauge-covariant Lie derivative operator. More explicitly, $\widehat{\mathcal{L}}_{\partial_t} := \partial_t + [A_0, \cdot]$ and $\widehat{\mathcal{L}}_X \mathcal{E}_i := \mathcal{L}_X \mathcal{E}_i + X^j [A_j, \mathcal{E}_i]$. The Yang–Mills equations are symmetric hyperbolic and one may obtain a local well-posedness result in the temporal gauge ($A_0 = 0$). Similarly, one may obtain a local well-posedness result for Einstein's equation in spacetime harmonic or wave gauge. In light of the gauge invariant estimates for the Weyl and Yang–Mills curvature, one may obtain the suitable $H^4 \times H^3 \times H^3 \times H^3$ estimates for $(g - {}^3\eta, k, \mathcal{E}, \mathcal{H})$, where ${}^3\eta$ is the spatial part of the Minkowski metric i.e., in the usual rectangular coordinates ${}^3\eta = \text{diag}(1,1,1)$.

## VIII. CONCLUDING REMARKS

We established the exterior global stability of the trivial flat solutions of the Einstein–Yang–Mills system. The main challenge is to obtain suitable gauge invariant estimates for the physical degrees of freedom. In the coupled Einstein–Yang–Mills system, the total *energy* is contributed by the Weyl curvature (pure gravity) and the Yang–Mills curvature. The idea is to perturb the trivial solution of the Minkowski space (and flat Yang–Mills solutions) and study whether the energy associated with these perturbations escapes to null infinity or can the non-linearities associated with the coupled equations concentrate the energy to yield the incompleteness of the spacetime. This incompleteness can be understood in terms of a trapped surface formation (which by Penrose's theorem would imply a null geodesic incompleteness[32]) or a naked singularity. To this end, it is natural to work in a double null framework where one may explicitly study the energy dispersion along the outgoing cone and energy concentration along the incoming cone. Since both the Einstein and Yang–Mills equations possess the same characteristics, it is relatively easy to study the radiation problem in the double-null framework. Our result confirms that the geometric decay is just sufficient to yield a global existence result for the coupled theory. The point-wise decay of the Yang–Mills curvature is weaker than that of the Weyl curvature components. On the other hand, Yang–Mills non-linearities are weaker than that of gravity (while the former is semi-linear the latter is quasi-linear). Therefore, one can close the argument with less decay. In summary, the main idea of the proof essentially relies on the geometric scaling of different curvature components. Once the curvature components (Weyl and Yang–Mills) are estimated uniformly, the total degrees of freedom are exhausted and as a consequence, the global well-posedness of the Cauchy problem follows.

Estimation of the gravitational radiation at the null infinity is of fundamental importance in general relativity for a detailed understanding of the gravitational waves. A double null foliation seems to be a natural choice for such studies as one can explicitly study the energy dispersion along the outgoing null cone. Our results suggest that if one disturbs an exterior domain of the Minkowski space by means of coupled Einstein–Yang–Mills perturbations, the perturbations escape to null infinity, and as such the resulting spacetime is close to the Minkowski space in a global sense. It is known that the passing of gravitational radiation causes permanent displacement of the test particles (Chirstodoulou's non-linear memory effect[33]). Therefore, one would like to understand the impact of the coupled gravitational–Yang–Mills radiation on the memory effect. Reference [34] proved that the electromagnetic field contributes at the highest order to the nonlinear memory effect of gravitational waves, enlarging the permanent displacement of test masses. One would expect that a similar feature is observed in the Einstein–Yang–Mills case. In particular, we want to understand the limits of several masses (e.g., Wang-Yau quasi-local mass, Hawking mass, etc.) associated with the topological two–spheres foliating the future null infinity $\mathcal{I}^+$. Can we discern the information about the gauge group by studying the Yang–Mills contribution to the memory effect? We would like to investigate such an issue in the future.

Lastly, we want to mention the fact that our estimates associated with the Yang–Mills curvature components are completely gauge-invariant. In particular, we obtain estimates for the fully gauge covariant angular derivatives. This essentially hints at an apparent similarity with the Maxwell and Yang–Mills theory even though the latter is a fully non-linear theory because the gauge covariant derivative *hides* the information of the connection and the non-linear coupling shows up only as the commutator of the full gauge covariant derivatives (this non-linear term is the most dangerous term among all the other non-linear terms present in the Einstein–Yang–Mills theory and we note that this term causes serious problem in case of the stability of Milne universe under coupled gravity-Yang–Mills perturbations). This does not cause a problem in the current context of obtaining estimates since all the associated inequalities are formulated in terms of the gauge covariant derivatives. The linear decay is just sufficient to control the nonlinear terms (and the estimates required for a local existence theory for coupled Yang–Mills equations can be obtained in a gauge-invariant way; note that at the end one ought to choose a gauge and work with the equation for connection) indicating a scattering property of the coupled Einstein–Yang–Mills system. Our preferred choice of gauge is the temporal gauge/Coulomb gauge where the Yang–Mills equations take the form of a symmetric hyperbolic/elliptic-hyperbolic system and the spatial connection can be determined in terms of the gauge invariant norms of the Yang–Mills curvature). There is of course a physical motivation behind this. Since the double null framework in some sense encodes the information about the *physical* nature of the Yang–Mills fields, the choice of the gauge should not matter and one should expect to obtain gauge-invariant estimates as is done in the current context. In this framework of the gauge-invariant estimates, therefore, the exterior stability of the Minkowski space under coupled gravity-Yang–Mills perturbations is proven to hold as a consequence of the main theorem. The problem that remains open is the study of the interior domain (even in pure vacuum the double null framework does not work in the interior domain). One hope is to use the Eardley-Moncrief[35] or Klainerman-Rodnianski[36] type argument. This is under investigation.

## ACKNOWLEDGMENTS

We thank the referee for a thorough review. This work was supported by the Center of Mathematical Sciences and Applications (CMSA) at Harvard University.

## AUTHOR DECLARATIONS

### Conflict of Interest

The authors have no conflicts to disclose.

### Author Contributions

**Puskar Mondal**: Conceptualization (lead); Formal analysis (lead); Investigation (equal); Writing – original draft (lead); Writing – review & editing (lead). **Shing-Tung Yau**: Conceptualization (supporting); Formal analysis (supporting); Investigation (equal); Project administration (lead); Supervision (lead).

## DATA AVAILABILITY

Data sharing is not applicable to this article as no new data were created or analyzed in this study.

## REFERENCES

[1] R. Schoen and S. T. Yau, "On the proof of the positive mass conjecture in general relativity," Commun. Math. Phys. **65**, 45–76 (1979).

[2] R. Schoen and S. T. Yau, "Proof of the positive mass theorem. II," Commun. Math. Phys. **79**, 231–260 (1981).

[3] M. T. Wang and S. T. Yau, "Quasilocal mass in general relativity," Phys. Rev. Lett. **102**, 021101 (2009).

[4] M. T. Wang and S. T. Yau, "Isometric embeddings into the Minkowski space and new quasi-local mass," Commun. Math. Phys. **288**, 919–942 (2009).

[5] P. N. Chen, M. T. Wang, and S. T. Yau, "Evaluating small sphere limit of the Wang–Yau quasi-local energy," Commun. Math. Phys. **357**, 731–774 (2018).

[6] D. Christodoulou and S. Klainerman, "The global nonlinear stability of the Minkowski space," in *Séminaire Équations aux dérivées partielles (Polytechnique) dit aussi" Séminaire Goulaouic-Schwartz* (Numdam, 1993), Vol. 1–29.

[7] H. Lindblad and I. Rodnianski, "Global existence for the Einstein vacuum equations in wave coordinates," Commun. Math. Phys. **256**, 43–110 (2005).

[8] Y. Foures-Bruhat, "Théorème d'existence pour certains systèmes d'équations aux dérivées partielles non linéaires," Acta Math. **88**, 141–225 (1952).

[9] H. Lindblad and I. Rodnianski, "The global stability of Minkowski space-time in harmonic gauge," Ann. Math. **171**, 1401–1477 (2010).

[10] L. Bieri, "An extension of the stability theorem of the Minkowski space in general relativity," J. Differ. Geom. **86**, 17–70 (2010).

[11] S. Klainerman and F. Nicoló, *The Evolution Problem in General Relativity* (Springer Science, 2012), Vol. 25.

[12] N. Zipser, *The Global Nonlinear Stability of the Trivial Solution of the Einstein-Maxwell Equations* (Harvard University, 2000).

[13] P. G. LeFloch and Y. Ma, "The global nonlinear stability of Minkowski space for self-gravitating massive fields," Commun. Math. Phys. **346**, 603–665 (2016).

[14] P. G. LeFloch and Y. Ma, "The Euclidian-hyperboloidal foliation method and the nonlinear stability of Minkowski spacetime," arXiv:1712.10048 (2017).

[15] D. Fajman, J. Joudioux, and J. Smulevici, "The stability of the Minkowski space for the Einstein–Vlasov system," Anal. PDE **14**, 425–531 (2021).

[16] L. Bigorgne, D. Fajman, J. Joudioux, J. Smulevici, and M. Thaller, "Asymptotic stability of Minkowski space-time with non-compactly supported massless Vlasov matter," Arch. Ration. Mech. Anal. **242**, 1–147 (2021).

[17] M. Taylor, "The global nonlinear stability of Minkowski space for the massless Einstein–Vlasov system," Ann. PDE **3**, 9 (2017).

[18] H. Friedrich, "On the global existence and the asymptotic behavior of solutions to the Einstein–Maxwell–Yang–Mills equations," J. Differ. Geom. **34**, 275–345 (1991).

[19] C. Liu, T. Oliynyk, and J. Wang, "Global existence and stability of de Sitter-like solutions to the Einstein–Yang–Mills equations in spacetime dimensions $n \geq 4$," arXiv:2202.05432 (2022).

[20] P. Griggs and P. Mondal, "On the global well-posedness of the Einstein–Yang–Mills system," J. Math. Phys. **64**, 062501 (2023).

[21] P. Mondal and S.-T. Yau, "Einstein–Yang–Mills equations in the double null framework," arXiv:2205.01101 (2022).

[22] V. Moncrief, "Gribov degeneracies: Coulomb gauge conditions and initial value constraints," J. Math. Phys. **20**, 579–585 (1979).

[23] I. M. Singer, "Some remarks on the Gribov ambiguity," Commun. Math. Phys. **60**, 7–12 (1978).

[24] O. Babelon and C. M. Viallet, "The Riemannian geometry of the configuration space of gauge theories," Commun. Math. Phys. **81**, 515–525 (1981).

[25] Y. Choquet-Bruhat, *General Relativity and the Einstein Equations* (Oxford University Press, 2009).

[26] R. Bartnik, J. Isenberg, R. Bartnik, and J. Isenberg, "The constraint equations," in *The Einstein Equations and the Large Scale Behavior of Gravitational Fields*, edited by P. T. Chruściel and H. Friedrich (Birkhäuser-Verlag, Basel, 2004), pp. 1–38.

[27] L. Andersson and V. Moncrief, "Future complete vacuum spacetimes," in *The Einstein Equations and the Large Scale Behavior of Gravitational Fields* (Springer, 2004), pp. 299–330.

[28] J. Luk, "On the local existence for the characteristic initial value problem in general relativity," Int. Math. Res. Not. **2012**, 4625–4678.

[29] $A \hat{\otimes} B := (A \otimes B + B \otimes A - A \cdot B\gamma)$, $(A \wedge B) := \epsilon^{ac}\gamma^{bd}A_{ab}B_{cd}$, $(\text{curl } A)_{a_1 \cdots a_n} := \epsilon^{cd}\nabla_c A_{da_1 \cdots a_n}$.

[30] P. Mondal and S.-T. Yau, "Aspects of quasilocal energy for gravity coupled to gauge fields," Phys. Rev. D **105**, 104068 (2022).

[31] These coupled equations are essentially the first-order formulation of the wave equation for the Yang–Mills curvature $\widehat{D}^\alpha \widehat{D}_\alpha F^{\hat{a}}{}_{\hat{b}\mu\nu} = 2F^{\hat{a}}{}_{\hat{c}\mu\beta}F^{\hat{c}}{}_{\hat{b}\nu}{}^\beta - 2F^{\hat{a}}{}_{\hat{c}\nu\beta}F^{\hat{c}}{}_{\hat{b}\mu}{}^\beta - R^\gamma{}_{\beta\mu\nu}F^{\hat{a}}{}_{\hat{b}\gamma}{}^\beta - R^\gamma{}_\nu F^{\hat{a}}{}_{\hat{b}\gamma\mu} - R^\gamma{}_\mu F^{\hat{a}}{}_{\hat{b}\nu\gamma}$.

[32] R. Penrose, "Gravitational collapse and space-time singularities," Phys. Rev. Lett. **14**(3), 57 (1965).

[33] D. Christodoulou, "Nonlinear nature of gravitation and gravitational-wave experiments," Phys. Rev. Lett. **67**, 1486 (1991).

[34] L. Bieri, P. N. Chen, and S.-T. Yau, "The electromagnetic Christodoulou memory effect and its application to neutron star binary mergers," Classical Quantum Gravity **29**(21), 215003 (2012).

[35] D. M. Eardley and V. Moncrief, "The global existence of Yang–Mills–Higgs fields in 4-dimensional Minkowski space: II. Completion of proof," Commun. Math. Phys. **83**, 193–212 (1982).

[36] S. Klainerman and I. Rodnianski, "A Kirchoff–Sobolev parametrix for the wave equation and applications," J. Hyperbolic Differ. Equ. **04**, 401–433 (2007).